\documentclass[aps,prx,twocolumn,amsmath,amssymb,nofootinbib,superscriptaddress,floatfix,reprint,longbibliography]{revtex4-1}
\usepackage[dvips]{graphicx}
\usepackage{latexsym}
\usepackage{amsmath}
\usepackage{amsfonts}
\usepackage{amssymb}
\usepackage{bm}
\usepackage{color}
\usepackage{xcolor}
\usepackage{txfonts}
\usepackage{float}
\usepackage{braket}
\usepackage{url}

\usepackage[colorlinks=true, allcolors=blue]{hyperref}
\usepackage{ulem}
\usepackage{cleveref}

\newcommand{\rmd}{{\rm d}}

	\newcommand{\fig}[2]{\includegraphics[width=#1]{#2}}
	
	\def\bk{{\mathbf{k}}}
	\def\bK{{\mathbf{K}}}

	\def\br{{\mathbf{r}}}
	
	\def\bq{{\mathbf{q}}}

	\def\bG{{\mathbf{G}}}

    \newcommand{\fvec}[1]{\boldsymbol{#1}}
    \newcommand{\pd}{\partial}
    \newcommand{\mtr}{\mathrm{Tr}}

\begin{document}

\title{Contribution of remote bands to orbital magnetization in twisted bilayer graphene}

\author{Pinzhuo Li}
\affiliation{School of Physical Science and Technology, ShanghaiTech University, Shanghai 201210, China}
\affiliation{State Key Laboratory of Quantum Functional Materials, School of Physical Science and Technology, ShanghaiTech University, Shanghai 201210, China}

\author{Kun Jiang}
\affiliation{Beijing National Laboratory for Condensed Matter Physics and Institute of Physics,
	Chinese Academy of Sciences, Beijing 100190, China}
\affiliation{University of Chinese Academy of Sciences, Beijing 100190, China}

\author{Ziqiang Wang}
\email{ziqiang.wang@bc.edu}
\affiliation{Department of Physics, Boston College, Chestnut Hill, MA 02467, USA}

\author{Jian Kang}
\email{kangjian@shanghaitech.edu.cn}
\affiliation{School of Physical Science and Technology, ShanghaiTech University, Shanghai 201210, China}

\author{Yi Zhang}
\email{zhangyi821@shu.edu.cn}
\affiliation{Department of Physics and Institute for Quantum Science and Technology, Shanghai University, Shanghai 200444, China}
\affiliation{Shanghai Key Laboratory of High Temperature Superconductors, Shanghai University, Shanghai 200444, China}


\begin{abstract}
Motivated by recent theoretical and experimental studies of orbital magnetization $M_{\mathrm{orb}}$ in interacting systems, we develop a gauge-invariant framework to compute $M_{\mathrm{orb}}$ for correlated phases of magic-angle twisted bilayer graphene within a self-consistent Hartree--Fock approximation. Based on the projector formulation of the theory of orbital magnetization, we evaluate both $M_{\mathrm{orb}}$ and the self-rotation contribution $m_{\mathrm{SR}}$ directly from the Hartree--Fock Hamiltonian. We demonstrate that both $M_{\mathrm{orb}}$ and $m_{\mathrm{SR}}$ receive substantial contributions from remote bands, and thus require careful convergence with respect to the number of included remote bands.  Applying this approach to correlated phases at integer fillings, we obtain converged $M_{\mathrm{orb}}$ and $m_{\mathrm{SR}}$ for time-reversal-symmetry-broken Chern insulating states at $\nu=\pm3$ and for competing correlated phases at other integer fillings.  Our results establish a systematic and controlled approach for evaluating orbital magnetization in correlated moir\'e systems and clarify the crucial role of remote bands in determining their magnetic response.
\end{abstract}

\maketitle

\section{Introduction}
The discovery of correlated insulating states~\cite{cao_2018_1,jiang_2019,kerelsky_2019,xie_2019,choi_2019,xie2021,sharpe_2019,serlin_2020,wus_2020,das_2020,nuk_2020,ste_2021,feldman_2022,lu_2019,yank_2019} and unconventional superconductivity~\cite{cao_2018_2,lu_2019,yank_2019,code_2019} in twisted bilayer graphene (TBG) has stimulated intense interest in correlated moir\'e systems. Near the magic angle, the isolated low-energy bands are extremely flat~\cite{mac_2011} and possess nontrivial topology~\cite{tar_2019,liu_2019,song_2019,song_2020,ahn_2019,po_2019,po_2018,po_2018_1}, so electron-electron interactions are expected to drive a rich variety of symmetry-broken and topological phases. Indeed, theoretical studies have predicted several strongly correlated states at integer fillings~\cite{kang_2019,kang_2020,zhang_2019,xu_2018,xie_2020,liuc_2018,
huang_2019,liu_2021,lius_2021,bul1_2020,bul_2020,wu_2020,chen_2020,lu_2020,
liao_2021,lian_2020,xief_2020, hejazi_2020,sb_2019,ra_2019,gu_2018,seo_2019,bernevig_2020,kang_2021,zhang_2020,wang2024,wang_2022}.
Among these phases, time-reversal symmetry breaking (TRSB) states are particularly interesting because they can become Chern insulators and exhibit integer or fractional quantum anomalous Hall (QAH) effects, as observed experimentally in TBG~\cite{sharpe_2019,serlin_2020,wus_2020,das_2020,nuk_2020,ste_2021,feldman_2022,xie2021} and other moir\'e materials~\cite{chen2020,li2021,xu2023,park2023,lu2024}. In particular, the ferromagnetic QAH state at $\nu=3$ was found to exhibit a strong magnetic response to external dc currents~\cite{sharpe_2019,serlin_2020}, pointing to a large orbital magnetization associated with TRSB~\cite{he_2020,su_2020,huang_2021,lee_2020,zhangs_2020}. Orbital magnetization in TBG has also been probed directly by high-resolution magnetometry~\cite{tsch2021,grover2022}.

Transport and scanning tunneling microscopy experiments further showed that weak to moderate out-of-plane magnetic fields can drive the states at $\nu=\pm1,\pm2,\pm3$ into correlation-induced Chern insulators in magic-angle TBG~\cite{das_2020,nuk_2020}. These observations highlight orbital magnetization as a key quantity governing the coupling of correlated moir\'e states to external magnetic fields. A quantitative understanding of orbital magnetization is therefore essential for elucidating the microscopic origin of Chern-insulating phases and field-induced transitions between competing states in TBG~\cite{das_2020,nuk_2020,ste_2021,feldman_2022}. 

On the theoretical side, the Berry curvature and orbital magnetic moments of moir\'e bands have been extensively analyzed within the non-interacting Bistritzer–MacDonald (BM) continuum model~\cite{liu_2019_1,he_2020,review_tbg,zhu_2020}. These studies established that the flat bands of TBG can host nontrivial topology and substantial orbital magnetization per unit cell largely arising from sizable Berry curvature. However, the flat-band nature of magic-angle TBG implies that electron–electron interactions are comparable to or even larger than the bandwidth, and self-consistent Hartree–Fock (HF) calculations have demonstrated that interactions tremendously reconstruct the band structure and stabilize correlated Chern insulating states at integer fillings~\cite{kang_2019,kang_2020,zhang_2019,zhang_2020,xu_2018,xie_2020,liuc_2018,
huang_2019,liu_2021,lius_2021,bul1_2020,bul_2020,wu_2020,chen_2020,lu_2020,
liao_2021,bernevig_2020,lian_2020,xief_2020, hejazi_2020,sb_2019,ra_2019,gu_2018,seo_2019,kang_2021,wang2024,wang_2022}. Despite this progress, a systematic calculation of the bulk orbital magnetization in such interaction-driven states remains lacking. In particular, it is not clear how interaction-induced band reconstruction, valley polarization, and remote-band contributions quantitatively modify the orbital magnetization beyond the non-interacting predictions, nor how these effects relate to experimentally observed field-induced switching between competing correlated phases.

In this work, we combine the recently developed projector formulation of orbital magnetization~\cite{kang_2025,zhu_2025,liu_2025,ye_2026,song_2026} with self-consistent 
HF theory to construct a gauge-invariant framework for correlated states of magic-angle TBG~\cite{cere_2006}. Starting from the BM continuum model with Coulomb interactions projected onto the relevant bands, we determine symmetry-broken ground states at integer fillings and evaluate the orbital magnetization using the modern theory expressed in terms of  projection operators. This formulation avoids gauge ambiguities and enables a controlled treatment of the large number of remote bands inherent to the continuum description. 

Using this framework, we compute both the total orbital magnetization $M_{\mathrm{orb}}$ and the self-rotation contribution $m_{\mathrm{SR}}$ relevant for magneto-optical probes such as magnetic circular dichroism (MCD)~\cite{xia_2026}. We additionally show that, unlike topological invariants such as the Chern number, the orbital magnetization is strongly sensitive to remote-band contributions and therefore requires careful convergence.
The calculated orbital magnetization respects the particle-hole symmetry that relates the fillings of $\nu$ and $-\nu$~\cite{bernevig_2020} and reproduces the characteristic linear dependence of magnetization on chemical potential within the insulating gap, with slope proportional to the Chern number. We further analyze the sign structure of the magnetization and its relation to changes in Hall conductivity. At $\nu=\pm 1$, we compare competing Chern and intervalley-coherent states and show that Chern insulating phases possess substantially larger orbital magnetization, providing a microscopic explanation for their stabilization under external magnetic fields.
Our results establish a systematic approach for evaluating orbital magnetization in correlated moir\'e systems and elucidate the important role of interactions as well as the critical contribution of remote bands in shaping orbital ferromagnetism in twisted bilayer graphene.

\section{Non-interacting model}
\begin{figure}
	\begin{center}
		\fig{3.4in}{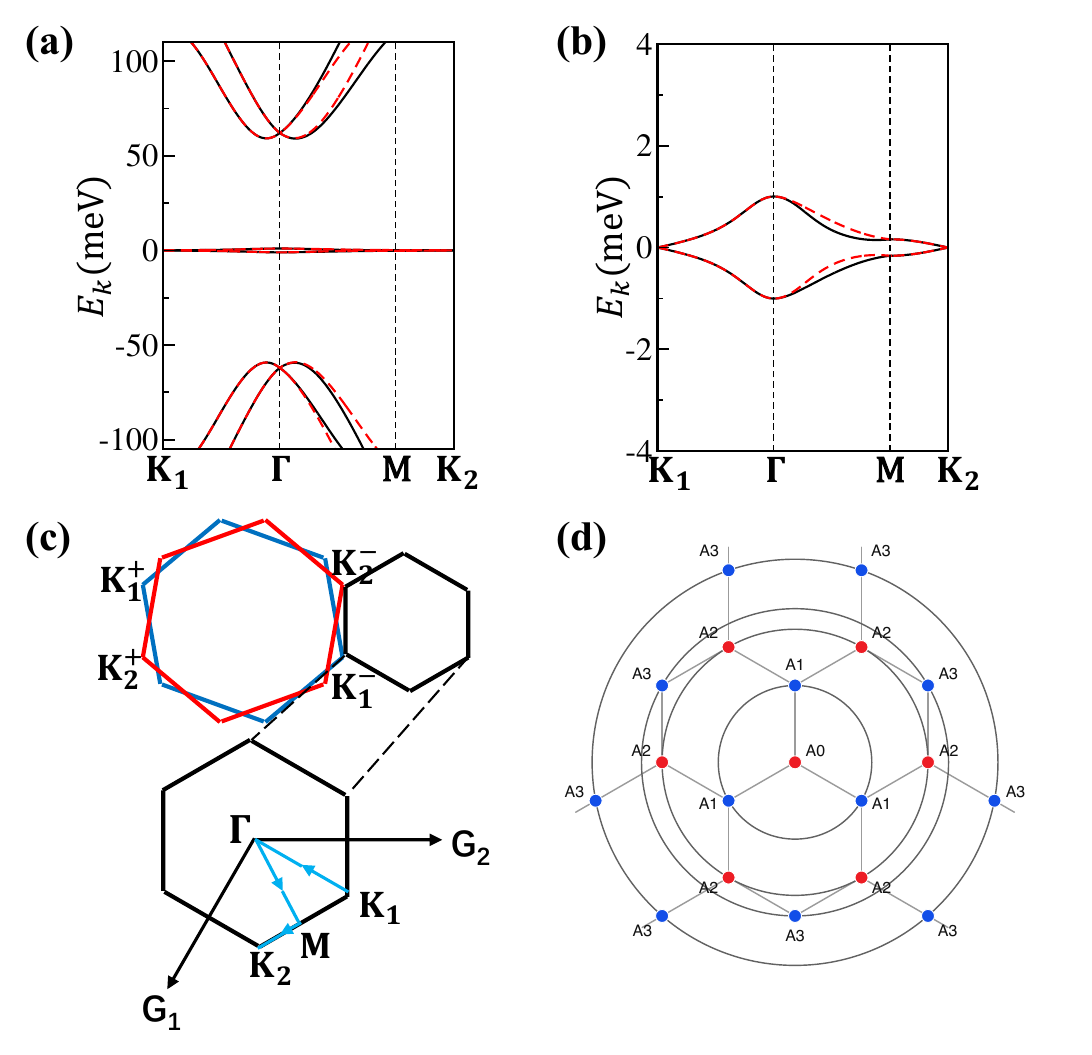}\caption{(a) Low-energy band structure of non-interacting TBG calculated using the BM model. The black solid and red dashed curves correspond to the $K^{\prime}$ and $K$ valleys, respectively. (b) Zoomed-in view of the isolated flat bands near charge neutrality, highlighting their separation from the higher-energy remote bands. (c) Schematic illustration of Brillouin zone folding in TBG. The small hexagon denotes the moir\'e Brillouin zone (mBZ), reciprocal to the moir\'e superlattice. The cyan arrows indicate the momentum path in the enlarged moir\'e Brillouin zone along which the band dispersions in panels (a) and (b) are plotted. (d) Tight-binding representation of the BM continuum model on a momentum-space honeycomb lattice. The red and blue sites denote plane-wave Dirac states associated with the two graphene layers, while the gray bonds represent the interlayer tunneling amplitudes $u_0$ and $u_1$ connecting nearest-neighbor momentum sites. The labels $A_0$--$A_3$ identify successive graph-distance shells measured from the central site at $\fvec G=0$; these shells contain 1, 3, 6, and 9 sites, respectively.
			\label{fig:fig1}}
	\end{center}
	\vskip-0.5cm
\end{figure}

The moir\'e pattern in TBG is formed by twisting the top and bottom layers of an
aligned bilayer graphene by angles $\theta/2$ and $-\theta/2$, which forms a
periodic superlattice structure of a triangular lattice with the lattice vectors
$\mathbf{L_{M1}}=(0,-1)L_M$, $\mathbf{L_{M2}}=(\frac{\sqrt{3}}{2},-\frac{1}{2})L_M$ and
$L_M=a_0/(2\sin(\theta/2))$, where $a_0$=0.246 nm, is the lattice
constant of the monolayer graphene.
The corresponding reciprocal lattice vectors of the moir\'e lattice are
$\bG_1=(-\frac{2\pi}{\sqrt{3} L_M},-\frac{2\pi}{L_M})$ and
$\bG_2=(\frac{4\pi}{\sqrt{3}L_M},0)$.
The non-interacting physics of TBG can be described by the continuum model introduced by
Bistritzer and MacDonald~\cite{mac_2011,lop_2012}. The Hamiltonian of the continuum model for each valley is given by,
\begin{equation}
	H^{\tau}_{BM}(\hat{\mathbf{k}})=\left(\begin{array}{cc}
		-\hbar v_F (\hat{\bk}-\bK_1^{\tau}) \cdot \pmb{\sigma}^{\tau}  &  U  \\
		U^+  & -\hbar v_F (\hat{\bk}-\bK_2^{\tau}) \cdot \pmb{\sigma}^{\tau}
	\end{array}\right)
	\label{eq:model}
\end{equation}
where $\hat{\bk}=-i\partial_{\br}$, $\tau=\pm$ is the valley index, and $\pmb{\sigma}^{\tau}=(\tau \sigma_x,\sigma_y)$  the Pauli matrices acting in the A,B sublattice space.
In Eq.~(\ref{eq:model}), $\bK^{\tau}_1$ and $\bK^{\tau}_2$ are the Dirac points of valley $\tau$ in the bottom and top layers twisted by angles $\mp\frac{\theta}{2}$. They are equivalent to $-\tau\bK_1$ and $-\tau\bK_2$ in the moir\'e Brillouin zone (mBZ), as shown in Fig.~\ref{fig:fig1}(c).
The interlayer tunneling between the Dirac states in the two layers is described by the matrix
\begin{equation}
	\begin{split}
		U=&
		\left(\begin{array}{cc}
			u_0  & u_1 \\
			u_1 & u_0
		\end{array}\right)
		+   \left(\begin{array}{cc}
			u_0  & u_1 e^{-i\tau \frac{2\pi}{3}} \\
			u_1 e^{i\tau \frac{2\pi}{3}} & u_0
		\end{array}\right)e^{-i \tau \bG_1 \cdot \br}\\
		&+   \left(\begin{array}{cc}
			u_0  & u_1 e^{i\tau \frac{2\pi}{3}} \\
			u_1 e^{-i\tau \frac{2\pi}{3}} & u_0
		\end{array}\right)e^{-i \tau(\bG_1+\bG_2) \cdot \br}
	\end{split}
\end{equation}
where $u_0$ and $u_1$ are the intra- and inter-sublattice interlayer tunneling amplitudes. We adopt realistic parameters for the continuum model: $\hbar v_F/a_0 = 2.365$ eV, inter-sublattice interlayer coupling $u_1 = 0.11$ eV, and intra-sublattice interlayer coupling $u_0 = 0.06$ eV~\cite{zhang_2020}. The resulting ratio $u_0/u_1\simeq 0.55$ incorporates the suppression of intra-sublattice tunneling arising from lattice relaxation~\cite{nam_2017,kang_2023,kang_2025_1}, as well as Coulomb interaction induced renormalization from higher energy states at remote bands~\cite{kang_2020_1}.  The twist angle is fixed at the magic value $\theta$=1.086$^{\circ}$, at which the low-energy moir\'e bands become perfectly flat in the limit $u_0 \to 0$~\cite{tar_2019}.
The low-energy flat bands exhibit a bandwidth of approximately 2 meV and are well separated from the higher-energy remote bands by an energy gap of about 50 meV as shown in Fig.~\ref{fig:fig1}.
The eigenstates of the continuum Hamiltonian $H_{BM}^{\tau}$  have the Bloch wavefunction form
\begin{equation}
	\psi_{m,\tau,\bk}^X(\br)=\sum_{\bG}u_{m,\tau;\bG,X }(\bk) e^{i(\bk+\bG)\cdot \br}
	\label{eq:bloch}
\end{equation}
where $X=\{A_1, B_1, A_2, B_2\}$ is the layer and sublattice index with the eigen-band energy $\epsilon_{m\bk\tau}$ and $\bG$ is the general reciprocal lattice vector in the mBZ.
Here, $m$ and $\tau$ are the band and valley indices, and the spin
index $s$ is omitted since the Hamiltonian is spin independent.
In the calculation, we take 121 $\bG$ vectors as expressed as $\bG=n_1 \bG_1+n_2 \bG_2$ with $n_1, n_2 \in [-5,5]$, which is enough to produce the flat bands. 

It is useful to view the BM continuum model as a tight-binding model on a momentum-space lattice~\cite{sm, bern_2021}. Each momentum site, labeled by a reciprocal lattice vector $\fvec G$, is associated with the intralayer Dirac Hamiltonian $\hbar v_F \fvec{\sigma}\cdot(\fvec k-\fvec G)$ and therefore has a characteristic energy scale $\hbar v_F|\fvec k-\fvec G|$. The interlayer tunneling amplitudes $u_0$ and $u_1$ play the role of hopping amplitudes between neighboring momentum space sites. As depicted in Fig.~\ref{fig:fig1}(d), these sites can be organized into momentum shells according to their graph distance, measured by the number of hopping from the central site at $\fvec G = 0$. Sites in higher shells generally have larger kinetic energy scales of order $\hbar v_F|\fvec G|$. For the BM parameters employed in our calculations, the wave functions of the active band states are strongly localized in momentum space, with most of their spectral weight concentrated within the three innermost shells~\cite{sm}.

\section{Interacting case}
The full Hamiltonian is written as $H=H_0+H_{int}$, with the non-interacting part given by
\begin{equation}
	H_0=\sum_{\tau} H_{BM}^{\tau}=\sum_{m\bk s\tau}\epsilon_{m\bk\tau}d_{\bk,m,s,\tau}^{\dagger}d_{\bk,m,s,\tau},
\end{equation}
Here, $d^{\dagger}_{\bk,m,s,\tau}$ and $d_{\bk,m,s,\tau}$ are the creation and annihilation operators in the eigen-band basis, respectively. To incorporate interactions, we project the single-gate-screened Coulomb interaction onto the two lowest-energy flat bands for each spin and valley flavor, and solve the resulting problem self-consistently within a standard HF approximation in momentum space. During the HF approximation, interaction effects are retained only in the active flat-band subspace, while the remote bands are kept fixed. Correspondingly, the interaction is chosen to be in the form of
\begin{equation}
	H_{int}=\frac{1}{2 S}\sum_{\bq\neq0}V(\bq)\delta\rho_{\bq}\delta\rho_{-\bq} \ ,
	\label{eq:coulomb}
\end{equation}
where $S$ is the total area and $\delta\rho_{\bq} = \rho_{\bq}-\bar{\rho}_{\bq}$ denotes the density fluctuation relative to the  density $\bar{\rho}_{\bq}$ of charge neutrality of the decoupled graphene layers. In contrast to normal ordering the density-density interaction within the active-band subspace, this subtraction prescription incorporates the renormalization effects associated with the remote bands~\cite{kang_2020_1}.

The detailed construction of the projected interaction, the form factors, and the resulting HF Hamiltonian is presented in the Supplemental Material~\cite{sm}, following the standard formulations of Refs.~\cite{bernevig_2020,kang_2021,zhang_2020}. In the numerical calculations, we adopt a dielectric constant $\varepsilon=7$ and a gate distance $d_s=40$ nm for the single gate screened Coulomb interaction. The results presented here are obtained using a $30\times30$ momentum mesh, corresponding to $N_k = 900$ kpoints in the mBZ. Calculations performed on finer momentum meshes are presented in the Supplemental Material~\cite{sm} to demonstrate the numerical convergence of the orbital magnetization.

It is well established that electron–electron Coulomb interactions in twisted bilayer graphene can drive spontaneous symmetry breaking, giving rise to a series of correlated insulating states at commensurate integer fillings. Among these, states that break the combined twofold-rotation and time-reversal symmetry $C_2\mathcal{T}$, where $\mathcal{T}$ denotes time reversal, are energetically competitive. In particular, at fillings $\nu = \pm 3$, the TRSB Chern insulating states can emerge as ground states even in the absence of an explicit staggered sublattice potential. Within this projected-basis HF framework, we systematically identify the correlated insulating states at all integer fillings~\cite{zhang_2020} and compute the orbital magnetization of these interaction-driven phases.

\section{Orbital magnetization}
The topology of the correlated electronic states and their response to an external magnetic field are governed by the Berry curvature and orbital magnetization of the interaction-renormalized bands. In our approach, the Coulomb interaction is projected onto the lowest two flat bands per spin and per valley. As a consequence, only those HF renormalized flat bands that break the combined $C_2\mathcal{T}$ symmetry can acquire nontrivial Berry curvature and hence finite Chern numbers.
Since the Berry curvature depends only on the periodic part of the Bloch wave functions, it can be directly computed from the HF eigenstates as
\begin{equation}
	\label{eq:berry}
	{\bf\Omega}_n(\bk)=i\sum_{\mu\nu} \epsilon_{\mu\nu} \left\langle \pd_{\mu} u_n^{HF}(\fvec k) \middle| \pd_{\nu} u_n^{HF}(\fvec k) \right\rangle ,
\end{equation}
where $n$ denotes the band index, $\epsilon_{\mu\nu}$ is the antisymmetric tensor, $\mu,\nu$ are Cartesian indices, and $\partial_{\mu} \equiv \partial/\partial k_{\mu}$. For all phases considered in this work, the remote bands do not contribute to the total Chern number. In the presence of $C_2 \mathcal{T}$ symmetry, their Chern number is constrained to vanish. When $C_2\mathcal{T}$ is weakly broken, the total Chern number of remote bands remains zero as long as the relevant band gaps remain open, since a Chern number cannot change under a continuous deformation without a gap closing. Consequently, the total Chern number is determined entirely by the active bands and is evaluated using the gauge-invariant discretized BZ method introduced in Ref.~\cite{fukui_2005}.

For the orbital magnetization, it has been recently proved that both the total orbital magnetization $M_{\mathrm{orb}}$~\cite{kang_2025} and the self-rotation contribution $m_{\mathrm{SR}}$ can be evaluated using the HF Hamiltonian and its eigenfunctions~\cite{souza_2008,resta_2020,wang_2025}: 
\begin{widetext}
\begin{equation}
    M_{orb} = \frac{e}{2i \hbar } \sum_{n\fvec k} \sum_{\mu\nu} \epsilon_{\mu\nu} \left\langle \pd_{\mu} u_n(\fvec k) \left| \left( H^{\rm HF}(\fvec k) + E_n^{\rm HF}(\fvec k) - 2\mu \right) \right| \pd_{\nu} u_n(\fvec k) \right\rangle n_F\left( E_n^{\rm HF}(\fvec k) - \mu \right)
    \label{eq:Morb}
\end{equation}  
\begin{equation}
    m_{SR}  = \frac{e}{2i \hbar } \sum_{n, \alpha} \sum_{\fvec k} \sum_{\mu\nu} \epsilon_{\mu\nu} \left\langle \pd_{\mu} u_n(\fvec k) \middle| u_{\alpha}(\fvec k) \right\rangle \left( E^{\rm HF}_{\alpha}(\fvec k) - E_n^{\rm HF}(\fvec k) \right) \left\langle u_{\alpha}(\fvec k) \middle| \pd_{\nu} u_n(\fvec k) \right\rangle n_F \left( E_n^{\rm HF}(\fvec k) - \mu \right)  \left(1 - n_F\left( E_{\alpha}^{\rm HF}(\fvec k) - \mu \right) \right)
    \label{eq:msr}
\end{equation}
\end{widetext}
where $n_F(E-\mu) = \Theta(\mu - E)$ is the Fermi–Dirac distribution function at $T = 0$ and $\mu$ is the chemical potential. 
These quantities can be reformulated in terms of projection operators.

Both $M_{\mathrm{orb}}$ and $m_{\mathrm{SR}}$ characterize TRSB in the orbital channel and therefore vanish in a time reversal symmetric state. They nevertheless represent distinct physical quantities. The total orbital magnetization is defined thermodynamically as the response to an external magnetic field, 
\begin{align}
    M_{\mathrm{orb}} & =-\left( \frac{\pd F}{\pd B}\right)_{T,N} = - \left( \frac{\pd \Omega}{\pd B}\right)_{T,\mu} \ ,
\end{align}
where $F$ and $\Omega$ are the free energy and grand potential, respectively. By contrast, $m_{\mathrm{SR}}$ is the self-rotation component of the orbital magnetization and describes the internal circulation of an electronic wave packet~\cite{xiao_2010}, and thus related with the orbital angular momentum of electrons. In noninteracting electron systems, $m_{\mathrm{SR}}$ is related through the dichroic $f$-sum rule to the frequency integrated MCD spectrum~\cite{souza_2008,resta_2020}. In interacting systems, however, the optically excited particle hole states are renormalized by interactions. Establishing a quantitative relation between the integrated MCD signal~\cite{xia_2026} and $m_{\mathrm{SR}}$ therefore generally requires an explicit treatment of many-body effects, which lies beyond the scope of the present work.

For orbital magnetization, it is important to emphasize that the HF Hamiltonian $H^{\mathrm{HF}}(\mathbf{k})$ entering Eqs.~\ref{eq:Morb} and \ref{eq:msr} is, strictly speaking, defined in the full Hilbert space, including the active flat-band sector together with both the occupied and empty remote bands. The derivation in Ref.~\cite{kang_2025} assumes that the HF problem is solved self-consistently in this full Hilbert space. In the present work, however, the HF calculation is carried out only within the active flat-band subspace, while the remote bands are kept fixed. To partially account for interaction effects associated with the remote bands, we express the interaction in terms of the density fluctuation $\delta\rho$ and adopt renormalized BM parameters rather than their bare values~\cite{kang_2020_1}. Although this treatment captures part of the influence of the interacting remote bands, the frozen remote band approximation may still introduce quantitative uncertainties into the calculated orbital magnetization.  

To estimate this uncertainty, we additionally perform self-consistent HF calculations including up to four remote conduction bands and four remote valence bands for each valley and each spin~\cite{sm}. We find that both $M_{\rm orb}$ and $m_{\rm SR}$ vary with the number of remote bands included in self-consistent HF calculation. The values for $m_{\rm SR}$ can differ by approximately $20\%-30\%$ from those obtained within the frozen-remote-band approximation. 
A fully self-consistent zero-field HF calculation including a substantially larger number of remote bands, together with a systematic convergence analysis of the orbital magnetization, is left for future work. 

Even within this approximation, Eqs.~\ref{eq:Morb} and \ref{eq:msr} still involve sums over infinitely many occupied and empty remote states. To make the calculation tractable, we evaluate $M_{\mathrm{orb}}$ and $m_{\mathrm{SR}}$ in a truncated Hilbert space. For clarity of notation, we label the HF Bloch states $|u_n^{A/R+/R-}(\mathbf{k})\rangle$ by the superscripts $A$, $R+$, and $R-$, which denote the active bands, the empty remote bands, and the occupied remote bands, respectively. The remote bands are indexed outward from charge neutrality, such that the states $|u_n^{R+}(\mathbf{k})\rangle$ and $|u_n^{R-}(-\mathbf{k})\rangle$ are related by particle-hole symmetry~\cite{bernevig_2020}. On this basis, we introduce two projection operators onto the truncated occupied and empty sectors:
\begin{align}
    P_{n_{\rm cut}}(\mathbf{k}) & = \sum_{n \in \mathrm{occupied}} |u^{\rm A}_n(\mathbf{k})\rangle\langle u^{\rm A}_n(\mathbf{k})| + \sum_{n \le n_{\rm cut}} |u^{R-}_n(\mathbf{k})\rangle \langle u^{R-}_n(\mathbf{k})|  \label{eq:Pdef} \\
    Q_{n_{\rm cut}}(\mathbf{k}) & = \sum_{n \in \mathrm{empty}} |u^{\rm A}_n(\mathbf{k})\rangle\langle u^{\rm A}_n(\mathbf{k})| + \sum_{n \le n_{\rm cut}} |u^{R+}_n(\mathbf{k})\rangle \langle u^{R+}_n(\mathbf{k})| \label{eq:Qdef} ,
\end{align}
These two projection operators for truncated occupied and empty states explicitly depend on the choice of $n_{\rm cut}$, and thus $P_{n_{\rm cut}}(k) + Q_{n_{\rm cut}}(k) = I$ only when $n_{\rm cut} \rightarrow \infty$. In this representation, we introduce the orbital magnetization for the truncated Hilbert space that is expressed as~\cite{sm} 
\begin{equation}
    M^{\rm trun}_{orb}= - \frac{e}{\hbar} \sum_{\fvec k} \mathrm{Im}\left[ W^{\rm trun}_{xy}(\fvec k) - N^{\rm trun}_{xy}(\fvec k)  \right]
    \label{eq:Morb1}
\end{equation}
\begin{equation}
    m^{\rm trun}_{SR}= \frac{e}{\hbar} \sum_{\fvec k} \mathrm{Im}\left[ W^{\rm trun}_{xy}(\fvec k) + N^{\rm trun}_{xy}(\fvec k)  \right] ,
    \label{eq:msr1}
\end{equation}
where 
\begin{equation}
    W^{\rm trun}_{\mu\nu}(\fvec k)  = \mtr\left[ P_{n_{cut}}(\fvec k) \big( \pd_{\mu} P_{n_{cut}}(\fvec k) \big) \big( \pd_{\nu} Q_{n_{cut}}(\fvec k) \big) \big( H^{HF}(\fvec k) - \mu \big)  \right]  \label{Eqn:WmunuTrunc}
\end{equation}
and
\begin{equation}
    N^{\rm trun}_{\mu\nu}(\fvec k)  = \mtr\left[ Q_{n_{cut}}(\fvec k) \big( \pd_{\mu} P_{n_{cut}}(\fvec k) \big) \big( \pd_{\nu} Q_{n_{cut}}(\fvec k) \big) \big( H^{HF}(\fvec k) - \mu \big)  \right] .  \label{Eqn:NmunuTrunc}
\end{equation}
Because the projection operators $P_{n_{cut}}(\mathbf{k})$ and $Q_{n_{cut}}(\mathbf{k})$ are manifestly gauge invariant, this formulation avoids the gauge ambiguities that typically arise in direct evaluations of orbital magnetization based on $\fvec k$-derivatives of Bloch wave functions. Moreover, it circumvents the singularities associated with band degeneracy points. Specifically, even in the presence of band crossings or near degeneracies within the occupied or unoccupied states, the derivatives of individual Bloch wavefunctions may be ill-defined, whereas the projectors $P_{n_{\rm cut}}(\mathbf{k})$ and $Q_{n_{\rm cut}}(\mathbf{k})$ remain smooth and possess well-defined momentum derivatives.
The choice $n_{\rm cut}=0$ retains only the active-band contributions, whereas physical quantities are defined in the limit $n_{\mathrm{cut}} \to \infty$. Nevertheless, we anticipate rapid convergence with increasing $n_{\mathrm{cut}}$, enabling reliable numerical evaluation within a finite truncated Hilbert space.

\subsection{Valley magnetization for the BM model}
\begin{figure}
	\begin{center}
		\fig{3.4in}{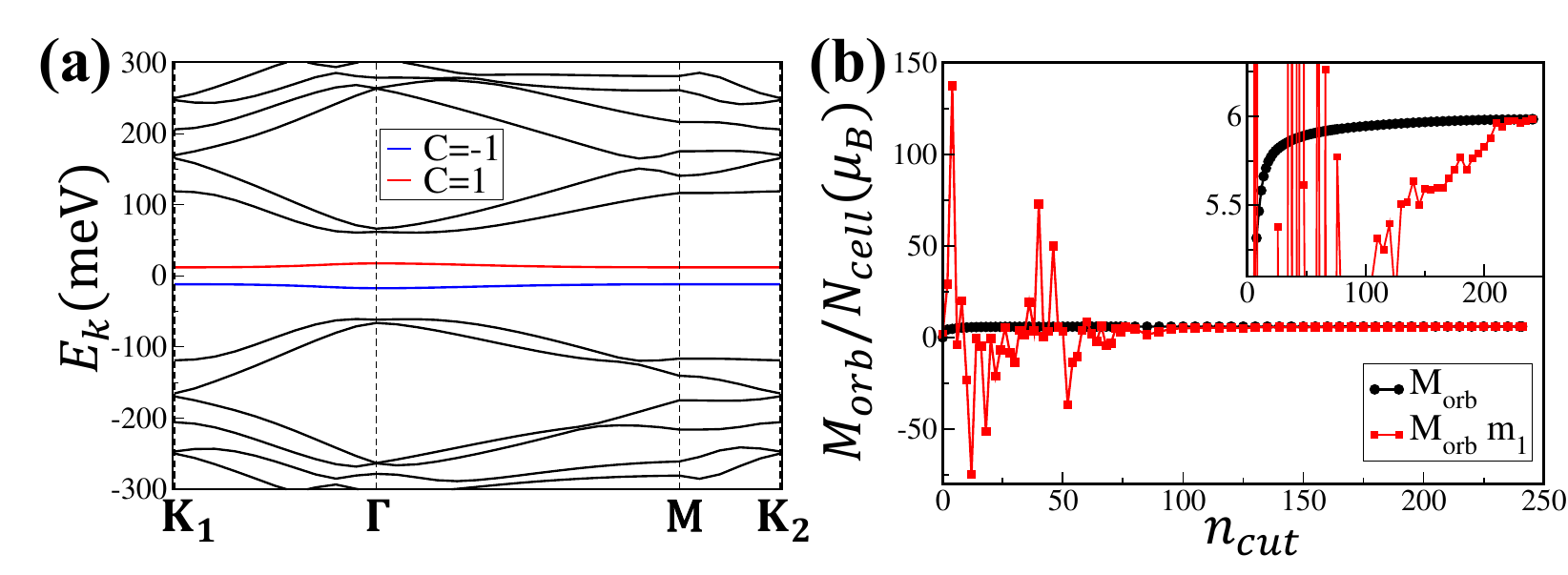}\caption{(a) Band structure for valley ($\tau = -$) in the non-interacting BM model with a staggered potential $\Delta_{C2B}=20$ meV. The gapped low-energy flat bands carrying nontrivial Chern numbers are highlighted in red ($C = 1$) and blue ($C = -1$). (b) Calculated total orbital magnetization $M_{\mathrm{orb}}$ per moir\'e cell for the valley $-$ state shown in (a), obtained using two truncation schemes within the projection-matrix formalism and plotted as functions of the number of remote-band pairs $n_{\mathrm{cut}}$. When evaluating $M_{\mathrm{orb}}$, the chemical potential is fixed at $\mu=-10$ meV, inside the gap. The inset shows the enlarged view of panel (b), highlighting the oscillatory behavior of scheme m$_1$.
			\label{fig:om_bm}}
	\end{center}
	\vskip-0.5cm
\end{figure}

We first apply the projection-matrix formalism to the non-interacting BM model. To generate a finite valley magnetization, we introduce a staggered potential through an additional term $\Delta_{C2B}\sigma_z$ in the BM Hamiltonian. This term breaks $C_2$ symmetry and thus gaps the Dirac points, as shown in Fig.~\ref{fig:om_bm}(a). Because the staggered potential preserves time-reversal symmetry, the orbital magnetizations in the two valleys remain equal in magnitude and opposite in sign, such that the total orbital magnetization vanishes. We therefore restrict our analysis to the $\tau = -$ valley and evaluate $M_{\mathrm{orb}}$ within this sector.  The self-rotation contribution $m_{\mathrm{SR}}$ is not evaluated in this case, as it is logarithmically divergent in the ultraviolet limit unless an additional high-energy cutoff is introduced~\cite{sm}.

To highlight the importance of the truncation procedure, we compare our scheme with an alternative prescription, denoted m$_1$, in which only the occupied projection matrix $P_{n_{\rm cut}}(\mathbf{k})$ is truncated, while the empty-state projection matrix is taken as $Q_{n_{\rm cut}}^{\prime}(\mathbf{k}) = I - P_{n_{\rm cut}}(\mathbf{k})$~\cite{sm}. As shown in Fig.~\ref{fig:om_bm}(b), truncating both projection matrices yields smooth and rapid convergence of $M_{\mathrm{orb}}$ with increasing $n_{\mathrm{cut}}$ (black curves). In contrast, as analyzed in the Supplemental Material~\cite{sm}, the $m_1$ scheme contains small energy differences in the denominators of individual terms, resulting in pronounced numerical oscillations. Convergence is recovered only when essentially all remote bands are included in the occupied projection matrix, where the two prescriptions become equivalent. The origin of this oscillatory behavior and the numerical stability of the symmetric truncation scheme are discussed in the Supplemental Material~\cite{sm}.

\subsection{$\nu = \pm 3$}
\begin{figure}
	\begin{center}
      \fig{3.4in}{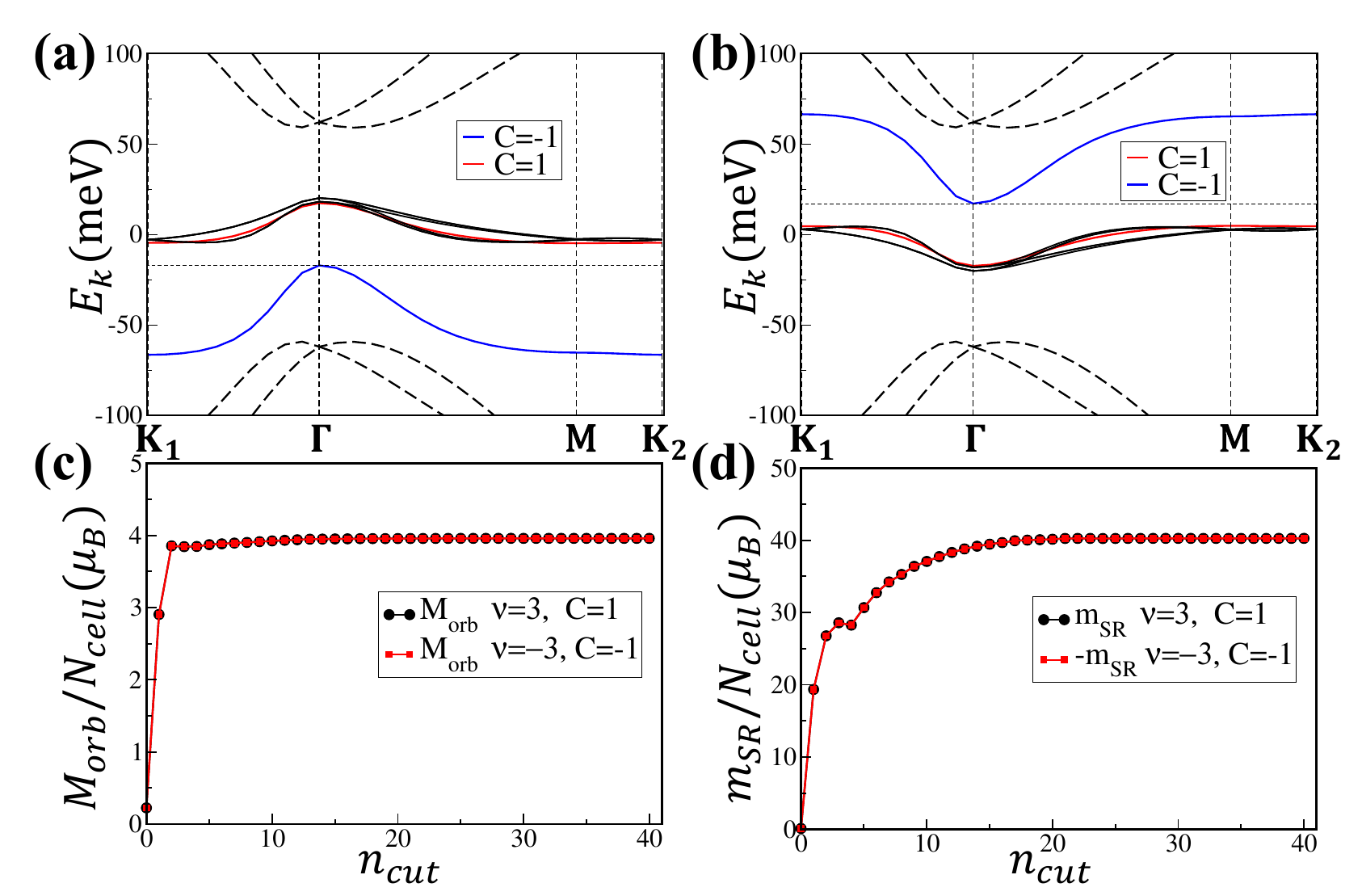}\caption{(a,b) Band dispersions of the Chern insulating states at fillings $\nu = -3$ (a) and $\nu = 3$ (b), obtained within the HF approximation. The solid curves represent the eight interaction-renormalized HF bands, while the dashed curves denote the remote bands that are excluded from the self-consistent HF calculation. Bands carrying nontrivial Chern numbers are highlighted in red ($C = 1$) and blue ($C = -1$), corresponding to the experimentally observed Chern insulating states with $C = \pm 1$ at $\nu = \pm 3$. (c,d) Calculated total orbital magnetization $M_{\mathrm{orb}}$ and self-rotation contribution $m_{\mathrm{SR}}$ per moir\'e cell for the states shown in (a,b), plotted as functions of the number of remote-band pairs ($n_{\mathrm{cut}}$) included in the projection-matrix formalism. Both quantities exhibit clear convergence for $n_{\mathrm{cut}} \approx 20$. When evaluating $M_{\mathrm{orb}}$, the chemical potential $\mu$ is fixed at $-17.08\,\mathrm{meV}$ the top of the valence band for $\nu = -3$ and  at $17.08 \,\mathrm{meV}$ the bottom of the conduction band for $\nu = 3$, as indicated by the horizontal dashed lines in panels (a) and (b).  			\label{fig:fig3}}
	\end{center}
	\vskip-0.5cm
\end{figure}

With the truncation scheme validated in the non-interacting BM model, we now apply the same formalism to the interacting Hartree-Fock states, beginning with the Chern insulating phases at $\nu=\pm3$.
The self-consistent Hartree–Fock calculation yields a spin–valley polarized Chern insulating ground state characterized by spontaneous breaking of the combined $C_2\mathcal{T}$ symmetry. The corresponding band dispersions, shown in Fig.~\ref{fig:fig3}(a) and (b), indicate that among the eight active bands, a single pair associated with one spin–valley flavor develops a gap due to $C_2\mathcal{T}$ symmetry breaking and acquires a finite Chern number. The remaining three pairs of bands, corresponding to the other spin–valley flavors (shown in black), remain topologically trivial. The symmetry breaking generates an indirect energy gap of approximately 13 meV between the occupied Chern band and the higher-energy bands.
These states correspond to the experimentally observed quantum anomalous Hall phases with Chern number $C = 1$ ($C = -1$) at fillings $\nu = 3$ ($\nu = -3$)~\cite{sharpe_2019,serlin_2020,nuk_2020}. 

We next evaluate the orbital magnetization of these states using Eqs.~\ref{eq:Morb1} and \ref{eq:msr1}. 
As shown in Fig.~\ref{fig:fig3}(c,d), both the total orbital magnetization $M_{\mathrm{orb}}$ and the self-rotation contribution $m_{\mathrm{SR}}$ converge by $n_{\mathrm{cut}} \approx 20$. The results also respect the particle-hole symmetry relating fillings $\nu = 3$ and $\nu = -3$, with $M_{\mathrm{orb}}(\nu=3) = M_{\mathrm{orb}}(\nu=-3)$ and $m_{\mathrm{SR}}(\nu=3) = -m_{\mathrm{SR}}(\nu=-3)$~\cite{sm}, consistent with the experimentally observed Landau-fan structure~\cite{das_2020,nuk_2020}. Since convergence is already achieved for $n_{\mathrm{cut}} \approx 20$, we use $n_{\mathrm{cut}} = 30$ in the following calculations to ensure numerical stability.

One might be tempted to conclude that both the orbital magnetization $M_{\mathrm{orb}}$ and the self-rotation contribution $m_{\mathrm{SR}}$ arise exclusively from the occupied Chern bands that explicitly break time-reversal symmetry, and that only these bands need to be considered. Such reasoning, however, is generally incorrect. Although the occupied states of the remote bands, $|u_n(\mathbf{k})\rangle$, remain invariant under time-reversal symmetry, the remote and active band wavefunctions, when expanded as in Eq.~\eqref{eq:bloch}, can carry substantial weight on the same momentum space site labelled by $\fvec G$. Consequently, their momentum derivatives can exhibit sizable interband matrix elements. Once time reversal symmetry is broken in the active band sector, the mixed active-remote contributions are no longer constrained to vanish and can contribute substantially to both $M_{\mathrm{orb}}$ and $m_{\mathrm{SR}}$. 
The need to include $20$ remote bands on each side of the charge-neutrality point (CNP) to reach numerical convergence provides direct evidence that these contributions are substantial.

We further find that the required remote band cutoff $n_{\rm cut}$ is governed primarily by the momentum space extension of the active band wavefunctions, or equivalently, by the number of reciprocal lattice vectors $\fvec G$ in Eq.~\eqref{eq:bloch} that carry appreciable wavefunction weight.  Consequently, $n_{\mathrm{cut}}$ depends only weakly on the interaction parameters and is controlled mainly by the BM model parameters $u_{0,1}/(\hbar v_F |\fvec K| \theta)$, as discussed in the Supplemental Material~\cite{sm}. For the parameters used here, the active band wave functions are concentrated predominantly within the three innermost momentum shells, which together contain $18$ momentum sites~\cite{sm}. With the momentum space hopping interpretation of the BM model, this suggests that approximately $18$ remote bands on each side of the charge-neutrality point should be retained, which is consistent with the numerical convergence obtained at $n_{\mathrm{cut}}\approx20$.

\begin{figure}
	\begin{center}
		\fig{3.4in}{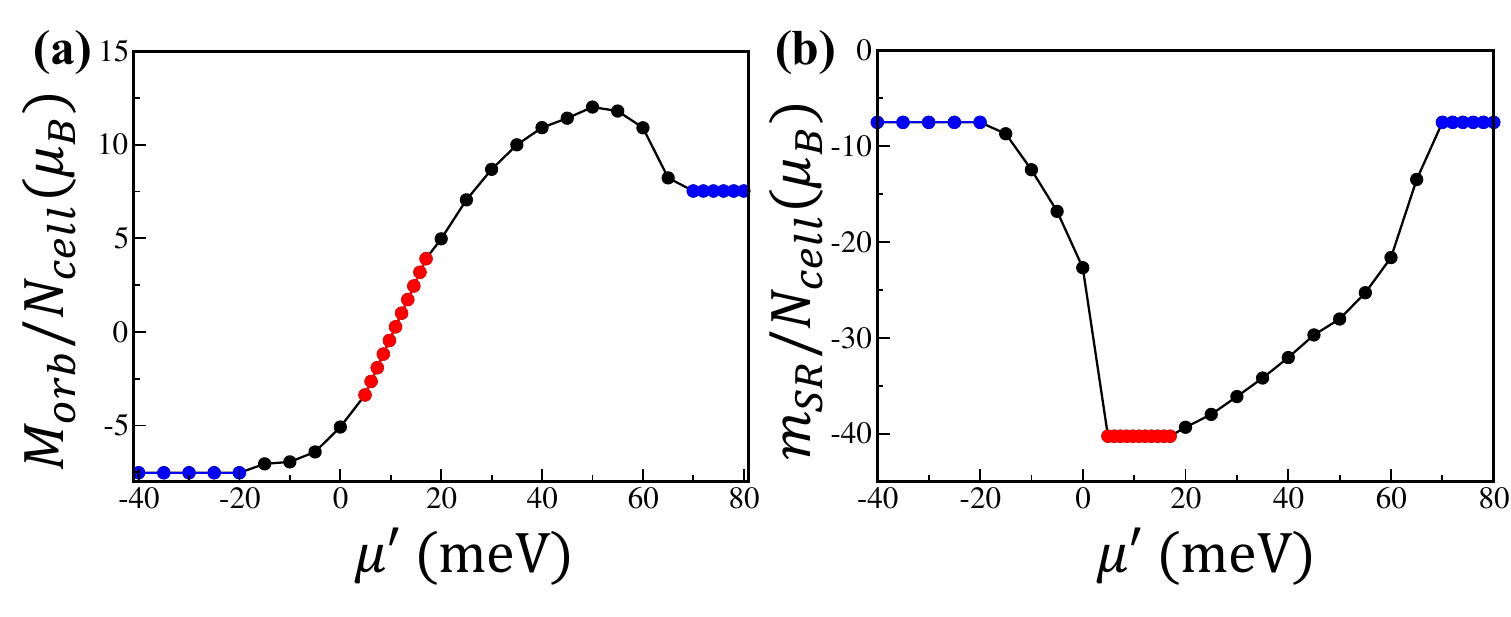}\caption{Calculated total orbital magnetization $M_{\mathrm{orb}}$ (a) and self-rotation contribution $m_{\mathrm{SR}}$ (b) per moir\'e cell as functions of the energy level $\mu^{\prime}$ for the Chern insulating state at $\nu = 3$ inside the insulating gap, whose band dispersion is shown in Fig.~\ref{fig:fig3}(b).
        The red dots denote the values within the Chern insulating gap, where $M_{\mathrm{orb}}$ shows a linear dependence with $\mu^{\prime}$($\mu$) and changes sign within the gap while $m_{\mathrm{SR}}$ remains constant. The blue dots represent the contribution from the remote bands arising from the time-reversal symmetry breaking of the HF Hamiltonian $H^{\mathrm{HF}}$. 
        The number of remote band pairs $n_{cut}$ is set to 30 in the calculation. 
			\label{fig:fig4}}
	\end{center}
	\vskip-0.5cm
\end{figure}

To make the role of the remote bands more transparent, we fix the HF band structure obtained at $\nu = 3$ and introduce an auxiliary energy level $\mu^{\prime}$ in place of the physical chemical potential $\mu$. We emphasize that varying $\mu'$ neither changes the physical filling nor modifies the self-consistent HF band structure and eigenstates. Instead, it changes the partition between occupied and empty states encoded in the projectors $P$ and $Q$ defined in Eqs.~\eqref{eq:Pdef} and \eqref{eq:Qdef}, and replace $\mu$ in Eqs.~\eqref{Eqn:WmunuTrunc} and \eqref{Eqn:NmunuTrunc} by $\mu'$. The resulting quantities should therefore be regarded as a diagnostic decomposition of the orbital magnetization rather than as self-consistent results at different physical fillings. We vary $\mu^{\prime}$ rigidly from $-40$ meV, well below the active HF bands, to $80$ meV, well above them, and compute both $M_{\mathrm{orb}}$ and $m_{\mathrm{SR}}$ as functions of $\mu^{\prime}$. For a fixed HF spectrum, the projector $P(\mu')$ includes the negative-energy remote bands together with all active bands below $\mu'$. For $\mu'\lesssim -20$~meV, $P(\mu')$ contains only the occupied remote bands. Consequently, the blue plateaus in $M_{\mathrm{orb}}$ and $m_{\mathrm{SR}}$ shown in Figs.~\ref{fig:fig4}(a) and \ref{fig:fig4}(b) arise entirely from the mixed pair consisting of negative energy remote bands and active bands. 
Their magnitudes reach approximately $7.5\,\mu_B$ per moir\'e unit cell, directly demonstrating the sizable contribution of the remote bands to the orbital magnetization. In this regime, $M_{\mathrm{orb}} = m_{\mathrm{SR}}$ because the frozen remote bands preserve $C_2\mathcal{T}$ symmetry and therefore has a vanishing Berry curvature contribution.
In the Supplemental Material, we further decompose the orbital magnetization according to the occupied-empty band pairs entering its evaluation~\cite{sm}. This analysis confirms that the purely remote contribution, involving pairs of occupied and empty remote bands, vanishes. The finite remote band contribution therefore originates from the mixed active remote sectors.

More physically relevant is the behavior of $M_{\mathrm{orb}}$ and $m_{\mathrm{SR}}$ within the Chern insulating gap, indicated by the red dots in Fig.~\ref{fig:fig4}(a,b). Inside the Chern gap, the auxiliary energy level $\mu^{\prime}$ is equivalent to the true chemical potential $\mu$ of the Chern insulating state. The total orbital magnetization $M_{\mathrm{orb}}$ exhibits the expected linear dependence on $\mu$.
Thermodynamics gives the relation $\partial M_{orb} / \partial \mu = \partial N / \partial B$, consistent with the Diophantine equation that characterizes quantum Hall states.
Taking the derivative of Eq.\ref{eq:Morb} with respect to $\mu$ yields
\begin{equation}
    \frac{\partial M_{orb}}{\partial \mu} = S \frac{e}{h}  C  ,  \label{Eq:dMdmu}
\end{equation}
where $C$ is the Chern number.
The linear dependence leads to a sign reversal of $M_{\mathrm{orb}}$ across the gap, corresponding to magnetization reversal previously analyzed in the non-interacting case~\cite{zhu_2020}.
The in-gap plateau of $m_{\mathrm{SR}}$ demonstrates that the Chern insulating state possesses a large self-rotation contribution of approximately $40\mu_B$. Such a large self-rotation contribution may give rise to a sizable magneto-optical signature, for example in frequency integrated MCD measurements.

To further quantify the uncertainty associated with the frozen-remote-band approximation, we enlarge the self-consistent HF subspace to include up to four remote bands above and below the CNP for each spin and valley~\cite{sm}. Both the HF gap within the active bands and the overall energy spectrum evolve as the number of self-consistently treated remote bands, $n_{\mathrm{R}}^{\mathrm{HF}}$, is increased. The self-rotation contribution $m_{\mathrm{SR}}$ grows systematically and begins to approach convergence for $n_{\mathrm{R}}^{\mathrm{HF}}\gtrsim3$. In particular, $|m_{\mathrm{SR}}|$ increases from approximately $40\mu_{\mathrm{B}}$ when all remote bands are frozen ($n_{\mathrm{R}}^{\mathrm{HF}}=0$) to approximately $50\mu_{\mathrm{B}}$ for $n_{\mathrm{R}}^{\mathrm{HF}}=4$. This comparison indicates an uncertainty of order $25\%$ in $m_{\mathrm{SR}}$ within the frozen-remote-band approximation. A similar convergence trend is observed for $M_{\mathrm{orb}}$~\cite{sm}. A direct comparison of $M_{\mathrm{orb}}$ between $n_{\rm R}^{\mathrm{HF}}=0$ and $n_{\mathrm{R}}^{\mathrm{HF}}=4$ is, however, complicated by the strong dependence of the full HF spectrum on $n_{\mathrm{R}}^{\mathrm{HF}}$. As follows from Eq.~\eqref{Eq:dMdmu}, $M_{\mathrm{orb}}$ varies with the chemical potential $\mu$ within the HF gap even at fixed particle number. Nevertheless, when $\mu$ is consistently chosen at the bottom of the conduction band at $\nu = 3$, $M_{\mathrm{orb}}$ exhibits only a rather weak dependence on $n_{\mathrm{R}}^{\mathrm{HF}}$. A more detailed comparison is provided in the Supplemental Material~\cite{sm}.

Importantly, for all values of $n_{\rm R}^{\rm HF}$ considered, both $M_{\rm orb}$ and $m_{\rm SR}$ converge with respect to the remote-band cutoff entering the orbital-magnetization formulas at essentially the same value, $n_{\rm cut}\simeq20$. This observation indicates that the value of $n_{\rm cut}$ required for convergence is largely insensitive to the interaction-induced reconstruction and energetics of the active bands. Instead, it is primarily controlled by the wave-function geometry and the resulting matrix elements between the active and remote sectors.

\subsection{Competing states at $\nu = \pm 1$}
\begin{figure}
	\begin{center}
		\fig{3.4in}{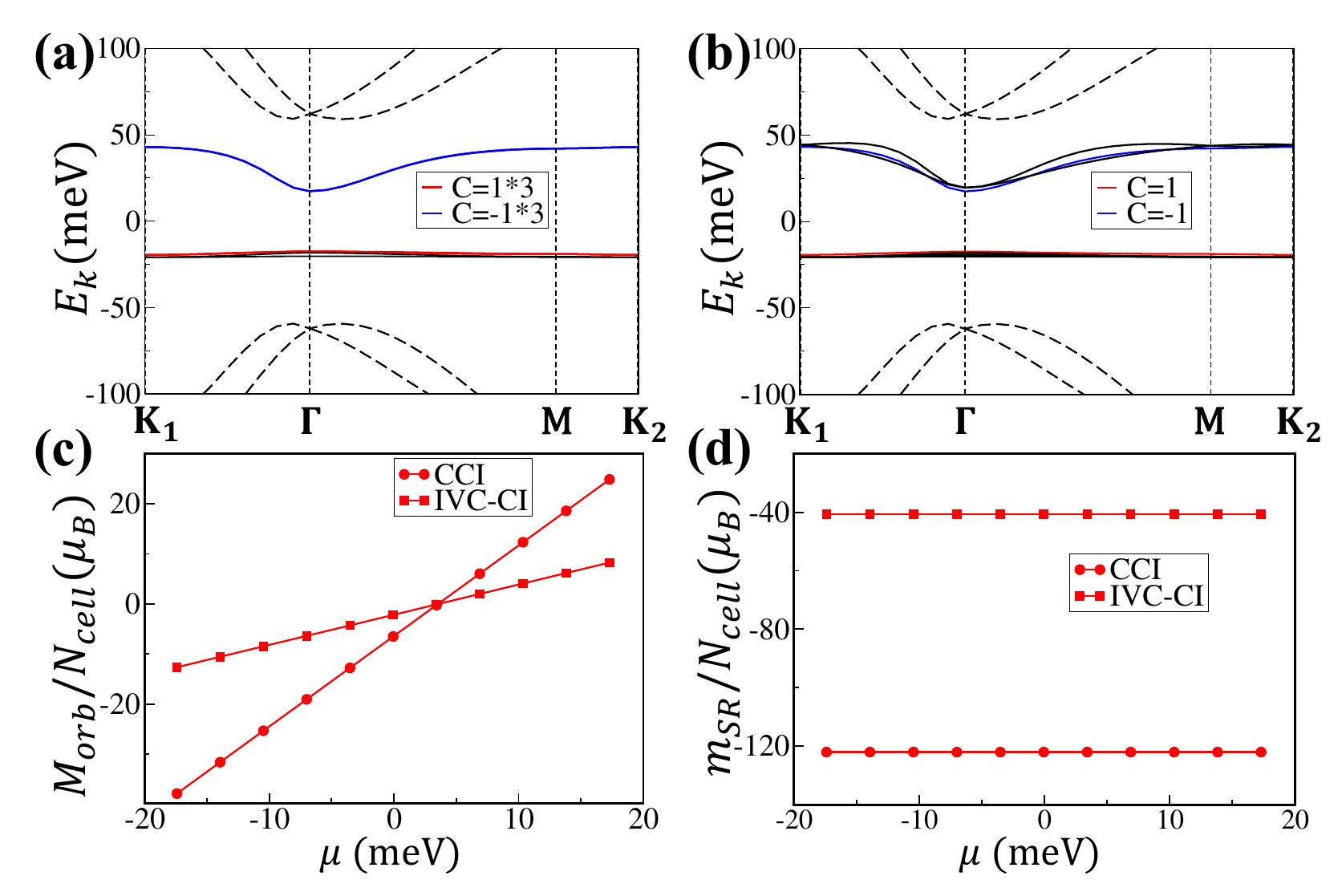}\caption{(a,b) HF band dispersions of the correlated Chern insulating state (CCI) (a) and the mixed intervalley-coherent and Chern insulating (IVC-CI) state (b) at filling $\nu=1$. The solid curves represent the eight interaction-renormalized HF bands, while the dashed curves denote the remote bands excluded from the self-consistent HF calculation. Bands carrying nontrivial Chern numbers are highlighted in red ($C=1$) and blue ($C=-1$). For the CCI, the Chern bands are triply degenerate, yielding a total Chern number $C=3$.
        (c,d) Total orbital magnetization $M_{\mathrm{orb}}$ (c) and self-rotation contribution $m_{\mathrm{SR}}$ (d) per moir\'e cell as functions of the chemical potential $\mu$ inside the insulating gap for the two states shown in (a) and (b). Solid circles correspond to the CCI, and solid squares to the mixed IVC-CI state. $M_{\mathrm{orb}}$ varies linearly with $\mu$ and changes sign across the gap, while $m_{\mathrm{SR}}$ remains constant. The number of remote-band pairs is fixed at $n_{\mathrm{cut}}=30$ in the calculation. 
			\label{fig:fig5}}
	\end{center}
	\vskip-0.5cm
\end{figure}

We next consider the correlated insulating states at fillings $\nu = \pm 1$. Owing to the particle–hole symmetry relating $\nu$ and $-\nu$, we focus on the case $\nu = 1$. The HF calculation yields two types of nearly degenerate correlated insulating states. 
The lowest-energy state is a mixed phase exhibiting both Kramers intervalley-coherent (K-IVC) order~\cite{bul_2020} and $C_2 \mathcal{T}$ symmetry breaking. Its band dispersion is shown in Fig.~\ref{fig:fig5}(b). In this state, the five valence bands consist of two bands associated with a spin-polarized IVC sector carrying zero Chern number, one $C_2\mathcal{T}$-broken band with finite Chern number, and an additional pair of ungapped bands from a single spin-valley flavor, yielding a total Chern number $C=1$~\cite{zhang_2020}. This state is lower in energy by approximately 1.52 meV per moir\'e cell than a family of states in which three valence bands break $C_2\mathcal{T}$ symmetry and each carry a finite Chern number. We refer to the former state as the mixed IVC-CI state, and the latter as correlated Chern insulators (CCIs). Among these degenerate CCI solutions, we present the state with the largest Chern number, namely $C=3$, whose dispersion is shown in Fig.~\ref{fig:fig5}(a).

We then evaluate the orbital magnetization for both states. As shown in Fig.~\ref{fig:fig5}(c), the CCI exhibits a larger $M_{\mathrm{orb}}$ than the mixed IVC-CI state. Moreover, the slope of $M_{\mathrm{orb}}$ as a function of chemical potential within the insulating gap is steeper in the CCI, reflecting its larger Chern number. In the mixed IVC-CI state, both the Chern number and the orbital magnetization arise exclusively from the single $C_2\mathcal{T}$-broken band with $C = 1$. Although the two bands in the spin-polarized K-IVC sector break both $C_2\mathcal{T}$ and physical time-reversal symmetry $\mathcal{T}$, this sector remains invariant under the antiunitary symmetry $\widetilde{T} = \tau_z \mathcal{T}$, where $\tau_z$ acts in valley space. Because both $M_{\mathrm{orb}}$ and $m_{\mathrm{SR}}$ are odd under $\widetilde{\mathcal{T}}$, the K-IVC sector does not contribute to either quantity. Consequently, the mixed IVC-CI state exhibits substantially smaller $M_{\mathrm{orb}}$ and $m_{\mathrm{SR}}$ than the CCI state with  $C=3$, in which three $C_2\mathcal{T}$-broken bands with unit Chern number contribute to the orbital magnetization.

As a consequence, the CCI is energetically favored in the presence of a finite magnetic field, consistent with experimental observations. The self-rotation contribution $m_{\mathrm{SR}}$, shown in Fig.~\ref{fig:fig5}(d), is likewise significantly larger in magnitude for the CCI than for the mixed IVC-CI state.
Using the difference in orbital magnetization at the center of the gap, $\delta M_{\mathrm{orb}}=4.31\,\mu_B$ per moir\'e cell, together with the energy difference between the two insulating states, $\delta E=1.52$ meV per moir\'e cell, we estimate a critical magnetic field for the transition from the mixed IVC-CI state to the CCI of $B_c \sim \delta E/\delta M_{\mathrm{orb}} \approx 6.1$ T. This estimate depends on the microscopic parameters employed in our calculations, including the BM model parameters, and interaction parameters such as dielectric constant and gate distance, and is therefore not universal. Nevertheless, the estimated field lies within an experimentally accessible range, indicating that the orbital magnetization effect is sufficiently strong to drive a magnetic-field-induced transition between a Chern insulating state and competing insulating phases.

We also note a sign-convention subtlety in comparison with experiment. In Landau-fan analyses based on the Streda formula, the Chern number is extracted from the slope of the carrier density with respect to the applied magnetic field. Because the charge carriers are electrons and therefore carry negative charge, the physical Chern number is the negative of this slope, in agreement with the convention used in our theoretical analysis. Thus, for direct comparison with the experimentally reported values, one should consider the time-reversed partner of the ground state obtained in our calculation, under which both the Chern number and the associated orbital magnetization change sign. 

Although the HF band structures and orbital magnetizations at $\nu=\pm 1$ differ substantially from those at $\nu=\pm3$, the same cutoff, $n_{\mathrm{cut}}\approx20$, is sufficient to achieve numerical convergence at both sets of fillings. This common convergence scale reflects the fact that $n_{\mathrm{cut}}$ is controlled primarily by the momentum space size of the underlying active band Hilbert space and is only weakly dependent on the detailed interaction parameters.

\section{Conclusion}
In this work, we developed a practical and gauge-invariant framework to compute orbital magnetization in correlated states of magic-angle twisted bilayer graphene within self-consistent Hartree–Fock theory. Building on the recently developed theory of orbital magnetization in interacting systems~\cite{kang_2025}, we implemented a projector formulation with symmetric truncations of the occupied and unoccupied remote band sectors. This formulation enables us to evaluate both the total orbital magnetization $M_{\mathrm{orb}}$ and the self-rotation contribution $m_{\mathrm{SR}}$ directly from the HF Hamiltonian and its eigenstates, thereby avoiding gauge ambiguities associated with Bloch-wavefunction derivatives. A key advantage of this approach is that it enables controlled inclusion of remote-band effects through the projection matrices. We demonstrated that, unlike the Chern number, the orbital magnetization is highly sensitive to remote bands and requires careful convergence with respect to the number of remote band pairs $n_{\mathrm{cut}}$, which saturates for $n_{\mathrm{cut}}\gtrsim 20$ in our calculations. We further showed that the required cutoff is governed primarily by the momentum-space extent of the active-band wavefunctions and depends only weakly on the interaction parameters.

While the projector-based formulation is broadly applicable to multiband systems, we focused here on interacting TBG at integer fillings. We obtained converged orbital magnetization for the TRSB Chern insulating states at $\nu=\pm3$ and for competing correlated phases at $\nu=1$. Our results ensure the thermodynamic linear dependence of $M_{\mathrm{orb}}$ on the chemical potential inside the insulating gaps with the slope determined by the Chern number, while yielding a sizable and constant in-gap $m_{\mathrm{SR}}$ relevant for magneto-optical probes. More broadly, the gauge-invariant projector formalism introduced here provides a systematic route to quantify orbital magnetization in interaction-driven moir\'e phases and can be readily extended to other moir\'e materials.

\section*{ACKNOWLEDGMENTS}
We thank Hui-Ke Jin,  Jie Wang, Minxuan Wang, Zhida Song, Wen-Yu He, and especially Oskar Vafek for helpful discussions. J.~K.~acknowledges the support from the NSFC (Grant No.~12574159 and No.~12074276), the Double First-Class Initiative Fund of ShanghaiTech University, and the start-up grant of ShanghaiTech University.
YZ is supported in part by the Shanghai Science and Technology Innovation Action Plan (Grant No. 24LZ1400800) and National Natural Science Foundation of China (NSFC) Grants No.~12274279 and No.~12674186.
ZW is supported by the U.S. Department of Energy, Basic Energy Sciences, Grant No. DE FG02-99ER45747.

\bibliography{reference} 

\begin{thebibliography}{87}%
\makeatletter
\providecommand \@ifxundefined [1]{%
 \@ifx{#1\undefined}
}%
\providecommand \@ifnum [1]{%
 \ifnum #1\expandafter \@firstoftwo
 \else \expandafter \@secondoftwo
 \fi
}%
\providecommand \@ifx [1]{%
 \ifx #1\expandafter \@firstoftwo
 \else \expandafter \@secondoftwo
 \fi
}%
\providecommand \natexlab [1]{#1}%
\providecommand \enquote  [1]{``#1''}%
\providecommand \bibnamefont  [1]{#1}%
\providecommand \bibfnamefont [1]{#1}%
\providecommand \citenamefont [1]{#1}%
\providecommand \href@noop [0]{\@secondoftwo}%
\providecommand \href [0]{\begingroup \@sanitize@url \@href}%
\providecommand \@href[1]{\@@startlink{#1}\@@href}%
\providecommand \@@href[1]{\endgroup#1\@@endlink}%
\providecommand \@sanitize@url [0]{\catcode `\\12\catcode `\$12\catcode
  `\&12\catcode `\#12\catcode `\^12\catcode `\_12\catcode `\%12\relax}%
\providecommand \@@startlink[1]{}%
\providecommand \@@endlink[0]{}%
\providecommand \url  [0]{\begingroup\@sanitize@url \@url }%
\providecommand \@url [1]{\endgroup\@href {#1}{\urlprefix }}%
\providecommand \urlprefix  [0]{URL }%
\providecommand \Eprint [0]{\href }%
\providecommand \doibase [0]{http://dx.doi.org/}%
\providecommand \selectlanguage [0]{\@gobble}%
\providecommand \bibinfo  [0]{\@secondoftwo}%
\providecommand \bibfield  [0]{\@secondoftwo}%
\providecommand \translation [1]{[#1]}%
\providecommand \BibitemOpen [0]{}%
\providecommand \bibitemStop [0]{}%
\providecommand \bibitemNoStop [0]{.\EOS\space}%
\providecommand \EOS [0]{\spacefactor3000\relax}%
\providecommand \BibitemShut  [1]{\csname bibitem#1\endcsname}%
\let\auto@bib@innerbib\@empty
\bibitem [{\citenamefont {Cao}\ \emph {et~al.}(2018{\natexlab{a}})\citenamefont
  {Cao}, \citenamefont {Fatemi}, \citenamefont {Demir}, \citenamefont {Fang},
  \citenamefont {Tomarken}, \citenamefont {Luo}, \citenamefont
  {Sanchez-Yamagishi}, \citenamefont {Watanabe}, \citenamefont {Taniguchi},
  \citenamefont {Kaxiras}, \citenamefont {Ashoori},\ and\ \citenamefont
  {Jarillo-Herrero}}]{cao_2018_1}%
  \BibitemOpen
  \bibfield  {author} {\bibinfo {author} {\bibfnamefont {Yuan}\ \bibnamefont
  {Cao}}, \bibinfo {author} {\bibfnamefont {Valla}\ \bibnamefont {Fatemi}},
  \bibinfo {author} {\bibfnamefont {Ahmet}\ \bibnamefont {Demir}}, \bibinfo
  {author} {\bibfnamefont {Shiang}\ \bibnamefont {Fang}}, \bibinfo {author}
  {\bibfnamefont {Spencer~L.}\ \bibnamefont {Tomarken}}, \bibinfo {author}
  {\bibfnamefont {Jason~Y.}\ \bibnamefont {Luo}}, \bibinfo {author}
  {\bibfnamefont {Javier~D.}\ \bibnamefont {Sanchez-Yamagishi}}, \bibinfo
  {author} {\bibfnamefont {Kenji}\ \bibnamefont {Watanabe}}, \bibinfo {author}
  {\bibfnamefont {Takashi}\ \bibnamefont {Taniguchi}}, \bibinfo {author}
  {\bibfnamefont {Efthimios}\ \bibnamefont {Kaxiras}}, \bibinfo {author}
  {\bibfnamefont {Ray~C.}\ \bibnamefont {Ashoori}}, \ and\ \bibinfo {author}
  {\bibfnamefont {Pablo}\ \bibnamefont {Jarillo-Herrero}},\ }\bibfield  {title}
  {\enquote {\bibinfo {title} {Correlated insulator behaviour at half-filling
  in magic-angle graphene superlattices},}\ }\href {\doibase
  10.1038/nature26154} {\bibfield  {journal} {\bibinfo  {journal} {Nature}\
  }\textbf {\bibinfo {volume} {556}},\ \bibinfo {pages} {80--84} (\bibinfo
  {year} {2018}{\natexlab{a}})}\BibitemShut {NoStop}%
\bibitem [{\citenamefont {Jiang}\ \emph {et~al.}(2019)\citenamefont {Jiang},
  \citenamefont {Lai}, \citenamefont {Watanabe}, \citenamefont {Taniguchi},
  \citenamefont {Haule}, \citenamefont {Mao},\ and\ \citenamefont
  {Andrei}}]{jiang_2019}%
  \BibitemOpen
  \bibfield  {author} {\bibinfo {author} {\bibfnamefont {Yuhang}\ \bibnamefont
  {Jiang}}, \bibinfo {author} {\bibfnamefont {Xinyuan}\ \bibnamefont {Lai}},
  \bibinfo {author} {\bibfnamefont {Kenji}\ \bibnamefont {Watanabe}}, \bibinfo
  {author} {\bibfnamefont {Takashi}\ \bibnamefont {Taniguchi}}, \bibinfo
  {author} {\bibfnamefont {Kristjan}\ \bibnamefont {Haule}}, \bibinfo {author}
  {\bibfnamefont {Jinhai}\ \bibnamefont {Mao}}, \ and\ \bibinfo {author}
  {\bibfnamefont {Eva~Y.}\ \bibnamefont {Andrei}},\ }\bibfield  {title}
  {\enquote {\bibinfo {title} {Charge order and broken rotational symmetry in
  magic-angle twisted bilayer graphene},}\ }\href {\doibase
  10.1038/s41586-019-1460-4} {\bibfield  {journal} {\bibinfo  {journal}
  {Nature}\ }\textbf {\bibinfo {volume} {573}},\ \bibinfo {pages} {91--95}
  (\bibinfo {year} {2019})}\BibitemShut {NoStop}%
\bibitem [{\citenamefont {Kerelsky}\ \emph {et~al.}(2019)\citenamefont
  {Kerelsky}, \citenamefont {McGilly}, \citenamefont {Kennes}, \citenamefont
  {Xian}, \citenamefont {Yankowitz}, \citenamefont {Chen}, \citenamefont
  {Watanabe}, \citenamefont {Taniguchi}, \citenamefont {Hone}, \citenamefont
  {Dean}, \citenamefont {Rubio},\ and\ \citenamefont
  {Pasupathy}}]{kerelsky_2019}%
  \BibitemOpen
  \bibfield  {author} {\bibinfo {author} {\bibfnamefont {Alexander}\
  \bibnamefont {Kerelsky}}, \bibinfo {author} {\bibfnamefont {Leo~J.}\
  \bibnamefont {McGilly}}, \bibinfo {author} {\bibfnamefont {Dante~M.}\
  \bibnamefont {Kennes}}, \bibinfo {author} {\bibfnamefont {Lede}\ \bibnamefont
  {Xian}}, \bibinfo {author} {\bibfnamefont {Matthew}\ \bibnamefont
  {Yankowitz}}, \bibinfo {author} {\bibfnamefont {Shaowen}\ \bibnamefont
  {Chen}}, \bibinfo {author} {\bibfnamefont {K.}~\bibnamefont {Watanabe}},
  \bibinfo {author} {\bibfnamefont {T.}~\bibnamefont {Taniguchi}}, \bibinfo
  {author} {\bibfnamefont {James}\ \bibnamefont {Hone}}, \bibinfo {author}
  {\bibfnamefont {Cory}\ \bibnamefont {Dean}}, \bibinfo {author} {\bibfnamefont
  {Angel}\ \bibnamefont {Rubio}}, \ and\ \bibinfo {author} {\bibfnamefont
  {Abhay~N.}\ \bibnamefont {Pasupathy}},\ }\bibfield  {title} {\enquote
  {\bibinfo {title} {Maximized electron interactions at the magic angle in
  twisted bilayer graphene},}\ }\href {\doibase 10.1038/s41586-019-1431-9}
  {\bibfield  {journal} {\bibinfo  {journal} {Nature}\ }\textbf {\bibinfo
  {volume} {572}},\ \bibinfo {pages} {95--100} (\bibinfo {year}
  {2019})}\BibitemShut {NoStop}%
\bibitem [{\citenamefont {Xie}\ \emph {et~al.}(2019)\citenamefont {Xie},
  \citenamefont {Lian}, \citenamefont {J{\"a}ck}, \citenamefont {Liu},
  \citenamefont {Chiu}, \citenamefont {Watanabe}, \citenamefont {Taniguchi},
  \citenamefont {Bernevig},\ and\ \citenamefont {Yazdani}}]{xie_2019}%
  \BibitemOpen
  \bibfield  {author} {\bibinfo {author} {\bibfnamefont {Yonglong}\
  \bibnamefont {Xie}}, \bibinfo {author} {\bibfnamefont {Biao}\ \bibnamefont
  {Lian}}, \bibinfo {author} {\bibfnamefont {Berthold}\ \bibnamefont
  {J{\"a}ck}}, \bibinfo {author} {\bibfnamefont {Xiaomeng}\ \bibnamefont
  {Liu}}, \bibinfo {author} {\bibfnamefont {Cheng-Li}\ \bibnamefont {Chiu}},
  \bibinfo {author} {\bibfnamefont {Kenji}\ \bibnamefont {Watanabe}}, \bibinfo
  {author} {\bibfnamefont {Takashi}\ \bibnamefont {Taniguchi}}, \bibinfo
  {author} {\bibfnamefont {B.~Andrei}\ \bibnamefont {Bernevig}}, \ and\
  \bibinfo {author} {\bibfnamefont {Ali}\ \bibnamefont {Yazdani}},\ }\bibfield
  {title} {\enquote {\bibinfo {title} {Spectroscopic signatures of many-body
  correlations in magic-angle twisted bilayer graphene},}\ }\href {\doibase
  10.1038/s41586-019-1422-x} {\bibfield  {journal} {\bibinfo  {journal}
  {Nature}\ }\textbf {\bibinfo {volume} {572}},\ \bibinfo {pages} {101--105}
  (\bibinfo {year} {2019})}\BibitemShut {NoStop}%
\bibitem [{\citenamefont {Choi}\ \emph {et~al.}(2019)\citenamefont {Choi},
  \citenamefont {Kemmer}, \citenamefont {Peng}, \citenamefont {Thomson},
  \citenamefont {Arora}, \citenamefont {Polski}, \citenamefont {Zhang},
  \citenamefont {Ren}, \citenamefont {Alicea}, \citenamefont {Refael},
  \citenamefont {von Oppen}, \citenamefont {Watanabe}, \citenamefont
  {Taniguchi},\ and\ \citenamefont {Nadj-Perge}}]{choi_2019}%
  \BibitemOpen
  \bibfield  {author} {\bibinfo {author} {\bibfnamefont {Youngjoon}\
  \bibnamefont {Choi}}, \bibinfo {author} {\bibfnamefont {Jeannette}\
  \bibnamefont {Kemmer}}, \bibinfo {author} {\bibfnamefont {Yang}\ \bibnamefont
  {Peng}}, \bibinfo {author} {\bibfnamefont {Alex}\ \bibnamefont {Thomson}},
  \bibinfo {author} {\bibfnamefont {Harpreet}\ \bibnamefont {Arora}}, \bibinfo
  {author} {\bibfnamefont {Robert}\ \bibnamefont {Polski}}, \bibinfo {author}
  {\bibfnamefont {Yiran}\ \bibnamefont {Zhang}}, \bibinfo {author}
  {\bibfnamefont {Hechen}\ \bibnamefont {Ren}}, \bibinfo {author}
  {\bibfnamefont {Jason}\ \bibnamefont {Alicea}}, \bibinfo {author}
  {\bibfnamefont {Gil}\ \bibnamefont {Refael}}, \bibinfo {author}
  {\bibfnamefont {Felix}\ \bibnamefont {von Oppen}}, \bibinfo {author}
  {\bibfnamefont {Kenji}\ \bibnamefont {Watanabe}}, \bibinfo {author}
  {\bibfnamefont {Takashi}\ \bibnamefont {Taniguchi}}, \ and\ \bibinfo {author}
  {\bibfnamefont {Stevan}\ \bibnamefont {Nadj-Perge}},\ }\bibfield  {title}
  {\enquote {\bibinfo {title} {Electronic correlations in twisted bilayer
  graphene near the magic angle},}\ }\href {\doibase 10.1038/s41567-019-0606-5}
  {\bibfield  {journal} {\bibinfo  {journal} {Nature Physics}\ }\textbf
  {\bibinfo {volume} {15}},\ \bibinfo {pages} {1174--1180} (\bibinfo {year}
  {2019})}\BibitemShut {NoStop}%
\bibitem [{\citenamefont {Xie}\ \emph {et~al.}(2021{\natexlab{a}})\citenamefont
  {Xie}, \citenamefont {Pierce}, \citenamefont {Park}, \citenamefont {Parker},
  \citenamefont {Khalaf}, \citenamefont {Ledwith}, \citenamefont {Cao},
  \citenamefont {Lee}, \citenamefont {Chen}, \citenamefont {Forrester},
  \citenamefont {Watanabe}, \citenamefont {Taniguchi}, \citenamefont
  {Vishwanath}, \citenamefont {Jarillo-Herrero},\ and\ \citenamefont
  {Yacoby}}]{xie2021}%
  \BibitemOpen
  \bibfield  {author} {\bibinfo {author} {\bibfnamefont {Yonglong}\
  \bibnamefont {Xie}}, \bibinfo {author} {\bibfnamefont {Andrew~T.}\
  \bibnamefont {Pierce}}, \bibinfo {author} {\bibfnamefont {Jeong~Min}\
  \bibnamefont {Park}}, \bibinfo {author} {\bibfnamefont {Daniel~E.}\
  \bibnamefont {Parker}}, \bibinfo {author} {\bibfnamefont {Eslam}\
  \bibnamefont {Khalaf}}, \bibinfo {author} {\bibfnamefont {Patrick}\
  \bibnamefont {Ledwith}}, \bibinfo {author} {\bibfnamefont {Yuan}\
  \bibnamefont {Cao}}, \bibinfo {author} {\bibfnamefont {Seung~Hwan}\
  \bibnamefont {Lee}}, \bibinfo {author} {\bibfnamefont {Shaowen}\ \bibnamefont
  {Chen}}, \bibinfo {author} {\bibfnamefont {Patrick~R.}\ \bibnamefont
  {Forrester}}, \bibinfo {author} {\bibfnamefont {Kenji}\ \bibnamefont
  {Watanabe}}, \bibinfo {author} {\bibfnamefont {Takashi}\ \bibnamefont
  {Taniguchi}}, \bibinfo {author} {\bibfnamefont {Ashvin}\ \bibnamefont
  {Vishwanath}}, \bibinfo {author} {\bibfnamefont {Pablo}\ \bibnamefont
  {Jarillo-Herrero}}, \ and\ \bibinfo {author} {\bibfnamefont {Amir}\
  \bibnamefont {Yacoby}},\ }\bibfield  {title} {\enquote {\bibinfo {title}
  {Fractional chern insulators in magic-angle twisted bilayer graphene},}\
  }\href {\doibase 10.1038/s41586-021-04002-3} {\bibfield  {journal} {\bibinfo
  {journal} {Nature}\ }\textbf {\bibinfo {volume} {600}},\ \bibinfo {pages}
  {439--443} (\bibinfo {year} {2021}{\natexlab{a}})}\BibitemShut {NoStop}%
\bibitem [{\citenamefont {Sharpe}\ \emph {et~al.}(2019)\citenamefont {Sharpe},
  \citenamefont {Fox}, \citenamefont {Barnard}, \citenamefont {Finney},
  \citenamefont {Watanabe}, \citenamefont {Taniguchi}, \citenamefont
  {Kastner},\ and\ \citenamefont {Goldhaber-Gordon}}]{sharpe_2019}%
  \BibitemOpen
  \bibfield  {author} {\bibinfo {author} {\bibfnamefont {Aaron~L.}\
  \bibnamefont {Sharpe}}, \bibinfo {author} {\bibfnamefont {Eli~J.}\
  \bibnamefont {Fox}}, \bibinfo {author} {\bibfnamefont {Arthur~W.}\
  \bibnamefont {Barnard}}, \bibinfo {author} {\bibfnamefont {Joe}\ \bibnamefont
  {Finney}}, \bibinfo {author} {\bibfnamefont {Kenji}\ \bibnamefont
  {Watanabe}}, \bibinfo {author} {\bibfnamefont {Takashi}\ \bibnamefont
  {Taniguchi}}, \bibinfo {author} {\bibfnamefont {M.~A.}\ \bibnamefont
  {Kastner}}, \ and\ \bibinfo {author} {\bibfnamefont {David}\ \bibnamefont
  {Goldhaber-Gordon}},\ }\bibfield  {title} {\enquote {\bibinfo {title}
  {Emergent ferromagnetism near three-quarters filling in twisted bilayer
  graphene},}\ }\href {\doibase 10.1126/science.aaw3780} {\bibfield  {journal}
  {\bibinfo  {journal} {Science}\ }\textbf {\bibinfo {volume} {365}},\ \bibinfo
  {pages} {605--608} (\bibinfo {year} {2019})}\BibitemShut {NoStop}%
\bibitem [{\citenamefont {Serlin}\ \emph {et~al.}(2020)\citenamefont {Serlin},
  \citenamefont {Tschirhart}, \citenamefont {Polshyn}, \citenamefont {Zhang},
  \citenamefont {Zhu}, \citenamefont {Watanabe}, \citenamefont {Taniguchi},
  \citenamefont {Balents},\ and\ \citenamefont {Young}}]{serlin_2020}%
  \BibitemOpen
  \bibfield  {author} {\bibinfo {author} {\bibfnamefont {M.}~\bibnamefont
  {Serlin}}, \bibinfo {author} {\bibfnamefont {C.~L.}\ \bibnamefont
  {Tschirhart}}, \bibinfo {author} {\bibfnamefont {H.}~\bibnamefont {Polshyn}},
  \bibinfo {author} {\bibfnamefont {Y.}~\bibnamefont {Zhang}}, \bibinfo
  {author} {\bibfnamefont {J.}~\bibnamefont {Zhu}}, \bibinfo {author}
  {\bibfnamefont {K.}~\bibnamefont {Watanabe}}, \bibinfo {author}
  {\bibfnamefont {T.}~\bibnamefont {Taniguchi}}, \bibinfo {author}
  {\bibfnamefont {L.}~\bibnamefont {Balents}}, \ and\ \bibinfo {author}
  {\bibfnamefont {A.~F.}\ \bibnamefont {Young}},\ }\bibfield  {title} {\enquote
  {\bibinfo {title} {Intrinsic quantized anomalous hall effect in a moiré
  heterostructure},}\ }\href {\doibase 10.1126/science.aay5533} {\bibfield
  {journal} {\bibinfo  {journal} {Science}\ }\textbf {\bibinfo {volume}
  {367}},\ \bibinfo {pages} {900--903} (\bibinfo {year} {2020})}\BibitemShut
  {NoStop}%
\bibitem [{\citenamefont {Wu}\ \emph {et~al.}(2021)\citenamefont {Wu},
  \citenamefont {Zhang}, \citenamefont {Watanabe}, \citenamefont {Taniguchi},\
  and\ \citenamefont {Andrei}}]{wus_2020}%
  \BibitemOpen
  \bibfield  {author} {\bibinfo {author} {\bibfnamefont {Shuang}\ \bibnamefont
  {Wu}}, \bibinfo {author} {\bibfnamefont {Zhenyuan}\ \bibnamefont {Zhang}},
  \bibinfo {author} {\bibfnamefont {K.}~\bibnamefont {Watanabe}}, \bibinfo
  {author} {\bibfnamefont {T.}~\bibnamefont {Taniguchi}}, \ and\ \bibinfo
  {author} {\bibfnamefont {Eva~Y.}\ \bibnamefont {Andrei}},\ }\bibfield
  {title} {\enquote {\bibinfo {title} {Chern insulators, van hove singularities
  and topological flat bands in magic-angle twisted bilayer graphene},}\ }\href
  {\doibase 10.1038/s41563-020-00911-2} {\bibfield  {journal} {\bibinfo
  {journal} {Nature Materials}\ }\textbf {\bibinfo {volume} {20}},\ \bibinfo
  {pages} {488--494} (\bibinfo {year} {2021})}\BibitemShut {NoStop}%
\bibitem [{\citenamefont {Das}\ \emph {et~al.}(2021)\citenamefont {Das},
  \citenamefont {Lu}, \citenamefont {Herzog-Arbeitman}, \citenamefont {Song},
  \citenamefont {Watanabe}, \citenamefont {Taniguchi}, \citenamefont
  {Bernevig},\ and\ \citenamefont {Efetov}}]{das_2020}%
  \BibitemOpen
  \bibfield  {author} {\bibinfo {author} {\bibfnamefont {Ipsita}\ \bibnamefont
  {Das}}, \bibinfo {author} {\bibfnamefont {Xiaobo}\ \bibnamefont {Lu}},
  \bibinfo {author} {\bibfnamefont {Jonah}\ \bibnamefont {Herzog-Arbeitman}},
  \bibinfo {author} {\bibfnamefont {Zhi-Da}\ \bibnamefont {Song}}, \bibinfo
  {author} {\bibfnamefont {Kenji}\ \bibnamefont {Watanabe}}, \bibinfo {author}
  {\bibfnamefont {Takashi}\ \bibnamefont {Taniguchi}}, \bibinfo {author}
  {\bibfnamefont {B.~Andrei}\ \bibnamefont {Bernevig}}, \ and\ \bibinfo
  {author} {\bibfnamefont {Dmitri~K.}\ \bibnamefont {Efetov}},\ }\bibfield
  {title} {\enquote {\bibinfo {title} {Symmetry-broken chern insulators and
  rashba-like landau-level crossings in magic-angle bilayer graphene},}\ }\href
  {\doibase 10.1038/s41567-021-01186-3} {\bibfield  {journal} {\bibinfo
  {journal} {Nature Physics}\ }\textbf {\bibinfo {volume} {17}},\ \bibinfo
  {pages} {710--714} (\bibinfo {year} {2021})}\BibitemShut {NoStop}%
\bibitem [{\citenamefont {Nuckolls}\ \emph {et~al.}(2020)\citenamefont
  {Nuckolls}, \citenamefont {Oh}, \citenamefont {Wong}, \citenamefont {Lian},
  \citenamefont {Watanabe}, \citenamefont {Taniguchi}, \citenamefont
  {Bernevig},\ and\ \citenamefont {Yazdani}}]{nuk_2020}%
  \BibitemOpen
  \bibfield  {author} {\bibinfo {author} {\bibfnamefont {Kevin~P.}\
  \bibnamefont {Nuckolls}}, \bibinfo {author} {\bibfnamefont {Myungchul}\
  \bibnamefont {Oh}}, \bibinfo {author} {\bibfnamefont {Dillon}\ \bibnamefont
  {Wong}}, \bibinfo {author} {\bibfnamefont {Biao}\ \bibnamefont {Lian}},
  \bibinfo {author} {\bibfnamefont {Kenji}\ \bibnamefont {Watanabe}}, \bibinfo
  {author} {\bibfnamefont {Takashi}\ \bibnamefont {Taniguchi}}, \bibinfo
  {author} {\bibfnamefont {B.~Andrei}\ \bibnamefont {Bernevig}}, \ and\
  \bibinfo {author} {\bibfnamefont {Ali}\ \bibnamefont {Yazdani}},\ }\bibfield
  {title} {\enquote {\bibinfo {title} {Strongly correlated chern insulators in
  magic-angle twisted bilayer graphene},}\ }\href {\doibase
  10.1038/s41586-020-3028-8} {\bibfield  {journal} {\bibinfo  {journal}
  {Nature}\ }\textbf {\bibinfo {volume} {588}},\ \bibinfo {pages} {610--615}
  (\bibinfo {year} {2020})}\BibitemShut {NoStop}%
\bibitem [{\citenamefont {Stepanov}\ \emph {et~al.}(2021)\citenamefont
  {Stepanov}, \citenamefont {Xie}, \citenamefont {Taniguchi}, \citenamefont
  {Watanabe}, \citenamefont {Lu}, \citenamefont {MacDonald}, \citenamefont
  {Bernevig},\ and\ \citenamefont {Efetov}}]{ste_2021}%
  \BibitemOpen
  \bibfield  {author} {\bibinfo {author} {\bibfnamefont {Petr}\ \bibnamefont
  {Stepanov}}, \bibinfo {author} {\bibfnamefont {Ming}\ \bibnamefont {Xie}},
  \bibinfo {author} {\bibfnamefont {Takashi}\ \bibnamefont {Taniguchi}},
  \bibinfo {author} {\bibfnamefont {Kenji}\ \bibnamefont {Watanabe}}, \bibinfo
  {author} {\bibfnamefont {Xiaobo}\ \bibnamefont {Lu}}, \bibinfo {author}
  {\bibfnamefont {Allan~H.}\ \bibnamefont {MacDonald}}, \bibinfo {author}
  {\bibfnamefont {B.~Andrei}\ \bibnamefont {Bernevig}}, \ and\ \bibinfo
  {author} {\bibfnamefont {Dmitri~K.}\ \bibnamefont {Efetov}},\ }\bibfield
  {title} {\enquote {\bibinfo {title} {Competing zero-field chern insulators in
  superconducting twisted bilayer graphene},}\ }\href {\doibase
  10.1103/PhysRevLett.127.197701} {\bibfield  {journal} {\bibinfo  {journal}
  {Phys. Rev. Lett.}\ }\textbf {\bibinfo {volume} {127}},\ \bibinfo {pages}
  {197701} (\bibinfo {year} {2021})}\BibitemShut {NoStop}%
\bibitem [{\citenamefont {Yu}\ \emph {et~al.}(2022)\citenamefont {Yu},
  \citenamefont {Foutty}, \citenamefont {Han}, \citenamefont {Barber},
  \citenamefont {Schattner}, \citenamefont {Watanabe}, \citenamefont
  {Taniguchi}, \citenamefont {Phillips}, \citenamefont {Shen}, \citenamefont
  {Kivelson},\ and\ \citenamefont {Feldman}}]{feldman_2022}%
  \BibitemOpen
  \bibfield  {author} {\bibinfo {author} {\bibfnamefont {Jiachen}\ \bibnamefont
  {Yu}}, \bibinfo {author} {\bibfnamefont {Benjamin~A.}\ \bibnamefont
  {Foutty}}, \bibinfo {author} {\bibfnamefont {Zhaoyu}\ \bibnamefont {Han}},
  \bibinfo {author} {\bibfnamefont {Mark~E.}\ \bibnamefont {Barber}}, \bibinfo
  {author} {\bibfnamefont {Yoni}\ \bibnamefont {Schattner}}, \bibinfo {author}
  {\bibfnamefont {Kenji}\ \bibnamefont {Watanabe}}, \bibinfo {author}
  {\bibfnamefont {Takashi}\ \bibnamefont {Taniguchi}}, \bibinfo {author}
  {\bibfnamefont {Philip}\ \bibnamefont {Phillips}}, \bibinfo {author}
  {\bibfnamefont {Zhi-Xun}\ \bibnamefont {Shen}}, \bibinfo {author}
  {\bibfnamefont {Steven~A.}\ \bibnamefont {Kivelson}}, \ and\ \bibinfo
  {author} {\bibfnamefont {Benjamin~E.}\ \bibnamefont {Feldman}},\ }\bibfield
  {title} {\enquote {\bibinfo {title} {Correlated hofstadter spectrum and
  flavour phase diagram in magic-angle twisted bilayer graphene},}\ }\href
  {\doibase 10.1038/s41567-022-01589-w} {\bibfield  {journal} {\bibinfo
  {journal} {Nature Physics}\ }\textbf {\bibinfo {volume} {18}},\ \bibinfo
  {pages} {825--831} (\bibinfo {year} {2022})}\BibitemShut {NoStop}%
\bibitem [{\citenamefont {Lu}\ \emph {et~al.}(2019)\citenamefont {Lu},
  \citenamefont {Stepanov}, \citenamefont {Yang}, \citenamefont {Xie},
  \citenamefont {Aamir}, \citenamefont {Das}, \citenamefont {Urgell},
  \citenamefont {Watanabe}, \citenamefont {Taniguchi}, \citenamefont {Zhang},
  \citenamefont {Bachtold}, \citenamefont {MacDonald},\ and\ \citenamefont
  {Efetov}}]{lu_2019}%
  \BibitemOpen
  \bibfield  {author} {\bibinfo {author} {\bibfnamefont {Xiaobo}\ \bibnamefont
  {Lu}}, \bibinfo {author} {\bibfnamefont {Petr}\ \bibnamefont {Stepanov}},
  \bibinfo {author} {\bibfnamefont {Wei}\ \bibnamefont {Yang}}, \bibinfo
  {author} {\bibfnamefont {Ming}\ \bibnamefont {Xie}}, \bibinfo {author}
  {\bibfnamefont {Mohammed~Ali}\ \bibnamefont {Aamir}}, \bibinfo {author}
  {\bibfnamefont {Ipsita}\ \bibnamefont {Das}}, \bibinfo {author}
  {\bibfnamefont {Carles}\ \bibnamefont {Urgell}}, \bibinfo {author}
  {\bibfnamefont {Kenji}\ \bibnamefont {Watanabe}}, \bibinfo {author}
  {\bibfnamefont {Takashi}\ \bibnamefont {Taniguchi}}, \bibinfo {author}
  {\bibfnamefont {Guangyu}\ \bibnamefont {Zhang}}, \bibinfo {author}
  {\bibfnamefont {Adrian}\ \bibnamefont {Bachtold}}, \bibinfo {author}
  {\bibfnamefont {Allan~H.}\ \bibnamefont {MacDonald}}, \ and\ \bibinfo
  {author} {\bibfnamefont {Dmitri~K.}\ \bibnamefont {Efetov}},\ }\bibfield
  {title} {\enquote {\bibinfo {title} {Superconductors, orbital magnets and
  correlated states in magic-angle bilayer graphene},}\ }\href {\doibase
  10.1038/s41586-019-1695-0} {\bibfield  {journal} {\bibinfo  {journal}
  {Nature}\ }\textbf {\bibinfo {volume} {574}},\ \bibinfo {pages} {653--657}
  (\bibinfo {year} {2019})}\BibitemShut {NoStop}%
\bibitem [{\citenamefont {Yankowitz}\ \emph {et~al.}(2019)\citenamefont
  {Yankowitz}, \citenamefont {Chen}, \citenamefont {Polshyn}, \citenamefont
  {Zhang}, \citenamefont {Watanabe}, \citenamefont {Taniguchi}, \citenamefont
  {Graf}, \citenamefont {Young},\ and\ \citenamefont {Dean}}]{yank_2019}%
  \BibitemOpen
  \bibfield  {author} {\bibinfo {author} {\bibfnamefont {Matthew}\ \bibnamefont
  {Yankowitz}}, \bibinfo {author} {\bibfnamefont {Shaowen}\ \bibnamefont
  {Chen}}, \bibinfo {author} {\bibfnamefont {Hryhoriy}\ \bibnamefont
  {Polshyn}}, \bibinfo {author} {\bibfnamefont {Yuxuan}\ \bibnamefont {Zhang}},
  \bibinfo {author} {\bibfnamefont {K.}~\bibnamefont {Watanabe}}, \bibinfo
  {author} {\bibfnamefont {T.}~\bibnamefont {Taniguchi}}, \bibinfo {author}
  {\bibfnamefont {David}\ \bibnamefont {Graf}}, \bibinfo {author}
  {\bibfnamefont {Andrea~F.}\ \bibnamefont {Young}}, \ and\ \bibinfo {author}
  {\bibfnamefont {Cory~R.}\ \bibnamefont {Dean}},\ }\bibfield  {title}
  {\enquote {\bibinfo {title} {Tuning superconductivity in twisted bilayer
  graphene},}\ }\href {\doibase 10.1126/science.aav1910} {\bibfield  {journal}
  {\bibinfo  {journal} {Science}\ }\textbf {\bibinfo {volume} {363}},\ \bibinfo
  {pages} {1059--1064} (\bibinfo {year} {2019})}\BibitemShut {NoStop}%
\bibitem [{\citenamefont {Cao}\ \emph {et~al.}(2018{\natexlab{b}})\citenamefont
  {Cao}, \citenamefont {Fatemi}, \citenamefont {Fang}, \citenamefont
  {Watanabe}, \citenamefont {Taniguchi}, \citenamefont {Kaxiras},\ and\
  \citenamefont {Jarillo-Herrero}}]{cao_2018_2}%
  \BibitemOpen
  \bibfield  {author} {\bibinfo {author} {\bibfnamefont {Yuan}\ \bibnamefont
  {Cao}}, \bibinfo {author} {\bibfnamefont {Valla}\ \bibnamefont {Fatemi}},
  \bibinfo {author} {\bibfnamefont {Shiang}\ \bibnamefont {Fang}}, \bibinfo
  {author} {\bibfnamefont {Kenji}\ \bibnamefont {Watanabe}}, \bibinfo {author}
  {\bibfnamefont {Takashi}\ \bibnamefont {Taniguchi}}, \bibinfo {author}
  {\bibfnamefont {Efthimios}\ \bibnamefont {Kaxiras}}, \ and\ \bibinfo {author}
  {\bibfnamefont {Pablo}\ \bibnamefont {Jarillo-Herrero}},\ }\bibfield  {title}
  {\enquote {\bibinfo {title} {Unconventional superconductivity in magic-angle
  graphene superlattices},}\ }\href {\doibase 10.1038/nature26160} {\bibfield
  {journal} {\bibinfo  {journal} {Nature}\ }\textbf {\bibinfo {volume} {556}},\
  \bibinfo {pages} {43--50} (\bibinfo {year} {2018}{\natexlab{b}})}\BibitemShut
  {NoStop}%
\bibitem [{\citenamefont {Codecido}\ \emph {et~al.}(2019)\citenamefont
  {Codecido}, \citenamefont {Wang}, \citenamefont {Koester}, \citenamefont
  {Che}, \citenamefont {Tian}, \citenamefont {Lv}, \citenamefont {Tran},
  \citenamefont {Watanabe}, \citenamefont {Taniguchi}, \citenamefont {Zhang},
  \citenamefont {Bockrath},\ and\ \citenamefont {Lau}}]{code_2019}%
  \BibitemOpen
  \bibfield  {author} {\bibinfo {author} {\bibfnamefont {Emilio}\ \bibnamefont
  {Codecido}}, \bibinfo {author} {\bibfnamefont {Qiyue}\ \bibnamefont {Wang}},
  \bibinfo {author} {\bibfnamefont {Ryan}\ \bibnamefont {Koester}}, \bibinfo
  {author} {\bibfnamefont {Shi}\ \bibnamefont {Che}}, \bibinfo {author}
  {\bibfnamefont {Haidong}\ \bibnamefont {Tian}}, \bibinfo {author}
  {\bibfnamefont {Rui}\ \bibnamefont {Lv}}, \bibinfo {author} {\bibfnamefont
  {Son}\ \bibnamefont {Tran}}, \bibinfo {author} {\bibfnamefont {Kenji}\
  \bibnamefont {Watanabe}}, \bibinfo {author} {\bibfnamefont {Takashi}\
  \bibnamefont {Taniguchi}}, \bibinfo {author} {\bibfnamefont {Fan}\
  \bibnamefont {Zhang}}, \bibinfo {author} {\bibfnamefont {Marc}\ \bibnamefont
  {Bockrath}}, \ and\ \bibinfo {author} {\bibfnamefont {Chun~Ning}\
  \bibnamefont {Lau}},\ }\bibfield  {title} {\enquote {\bibinfo {title}
  {Correlated insulating and superconducting states in twisted bilayer graphene
  below the magic angle},}\ }\href {\doibase 10.1126/sciadv.aaw9770} {\bibfield
   {journal} {\bibinfo  {journal} {Science Advances}\ }\textbf {\bibinfo
  {volume} {5}},\ \bibinfo {pages} {eaaw9770} (\bibinfo {year}
  {2019})}\BibitemShut {NoStop}%
\bibitem [{\citenamefont {Bistritzer}\ and\ \citenamefont
  {MacDonald}(2011)}]{mac_2011}%
  \BibitemOpen
  \bibfield  {author} {\bibinfo {author} {\bibfnamefont {Rafi}\ \bibnamefont
  {Bistritzer}}\ and\ \bibinfo {author} {\bibfnamefont {Allan~H.}\ \bibnamefont
  {MacDonald}},\ }\bibfield  {title} {\enquote {\bibinfo {title} {Moiré bands
  in twisted double-layer graphene},}\ }\href {\doibase
  10.1073/pnas.1108174108} {\bibfield  {journal} {\bibinfo  {journal}
  {Proceedings of the National Academy of Sciences}\ }\textbf {\bibinfo
  {volume} {108}},\ \bibinfo {pages} {12233--12237} (\bibinfo {year}
  {2011})}\BibitemShut {NoStop}%
\bibitem [{\citenamefont {Tarnopolsky}\ \emph {et~al.}(2019)\citenamefont
  {Tarnopolsky}, \citenamefont {Kruchkov},\ and\ \citenamefont
  {Vishwanath}}]{tar_2019}%
  \BibitemOpen
  \bibfield  {author} {\bibinfo {author} {\bibfnamefont {Grigory}\ \bibnamefont
  {Tarnopolsky}}, \bibinfo {author} {\bibfnamefont {Alex~Jura}\ \bibnamefont
  {Kruchkov}}, \ and\ \bibinfo {author} {\bibfnamefont {Ashvin}\ \bibnamefont
  {Vishwanath}},\ }\bibfield  {title} {\enquote {\bibinfo {title} {Origin of
  magic angles in twisted bilayer graphene},}\ }\href {\doibase
  10.1103/PhysRevLett.122.106405} {\bibfield  {journal} {\bibinfo  {journal}
  {Phys. Rev. Lett.}\ }\textbf {\bibinfo {volume} {122}},\ \bibinfo {pages}
  {106405} (\bibinfo {year} {2019})}\BibitemShut {NoStop}%
\bibitem [{\citenamefont {Liu}\ \emph {et~al.}(2019{\natexlab{a}})\citenamefont
  {Liu}, \citenamefont {Liu},\ and\ \citenamefont {Dai}}]{liu_2019}%
  \BibitemOpen
  \bibfield  {author} {\bibinfo {author} {\bibfnamefont {Jianpeng}\
  \bibnamefont {Liu}}, \bibinfo {author} {\bibfnamefont {Junwei}\ \bibnamefont
  {Liu}}, \ and\ \bibinfo {author} {\bibfnamefont {Xi}~\bibnamefont {Dai}},\
  }\bibfield  {title} {\enquote {\bibinfo {title} {Pseudo landau level
  representation of twisted bilayer graphene: Band topology and implications on
  the correlated insulating phase},}\ }\href {\doibase
  10.1103/PhysRevB.99.155415} {\bibfield  {journal} {\bibinfo  {journal} {Phys.
  Rev. B}\ }\textbf {\bibinfo {volume} {99}},\ \bibinfo {pages} {155415}
  (\bibinfo {year} {2019}{\natexlab{a}})}\BibitemShut {NoStop}%
\bibitem [{\citenamefont {Song}\ \emph {et~al.}(2019)\citenamefont {Song},
  \citenamefont {Wang}, \citenamefont {Shi}, \citenamefont {Li}, \citenamefont
  {Fang},\ and\ \citenamefont {Bernevig}}]{song_2019}%
  \BibitemOpen
  \bibfield  {author} {\bibinfo {author} {\bibfnamefont {Zhida}\ \bibnamefont
  {Song}}, \bibinfo {author} {\bibfnamefont {Zhijun}\ \bibnamefont {Wang}},
  \bibinfo {author} {\bibfnamefont {Wujun}\ \bibnamefont {Shi}}, \bibinfo
  {author} {\bibfnamefont {Gang}\ \bibnamefont {Li}}, \bibinfo {author}
  {\bibfnamefont {Chen}\ \bibnamefont {Fang}}, \ and\ \bibinfo {author}
  {\bibfnamefont {B.~Andrei}\ \bibnamefont {Bernevig}},\ }\bibfield  {title}
  {\enquote {\bibinfo {title} {All magic angles in twisted bilayer graphene are
  topological},}\ }\href {\doibase 10.1103/PhysRevLett.123.036401} {\bibfield
  {journal} {\bibinfo  {journal} {Phys. Rev. Lett.}\ }\textbf {\bibinfo
  {volume} {123}},\ \bibinfo {pages} {036401} (\bibinfo {year}
  {2019})}\BibitemShut {NoStop}%
\bibitem [{\citenamefont {Song}\ \emph {et~al.}(2021)\citenamefont {Song},
  \citenamefont {Lian}, \citenamefont {Regnault},\ and\ \citenamefont
  {Bernevig}}]{song_2020}%
  \BibitemOpen
  \bibfield  {author} {\bibinfo {author} {\bibfnamefont {Zhi-Da}\ \bibnamefont
  {Song}}, \bibinfo {author} {\bibfnamefont {Biao}\ \bibnamefont {Lian}},
  \bibinfo {author} {\bibfnamefont {Nicolas}\ \bibnamefont {Regnault}}, \ and\
  \bibinfo {author} {\bibfnamefont {B.~Andrei}\ \bibnamefont {Bernevig}},\
  }\bibfield  {title} {\enquote {\bibinfo {title} {Twisted bilayer graphene.
  ii. stable symmetry anomaly},}\ }\href {\doibase 10.1103/PhysRevB.103.205412}
  {\bibfield  {journal} {\bibinfo  {journal} {Phys. Rev. B}\ }\textbf {\bibinfo
  {volume} {103}},\ \bibinfo {pages} {205412} (\bibinfo {year}
  {2021})}\BibitemShut {NoStop}%
\bibitem [{\citenamefont {Ahn}\ \emph {et~al.}(2019)\citenamefont {Ahn},
  \citenamefont {Park},\ and\ \citenamefont {Yang}}]{ahn_2019}%
  \BibitemOpen
  \bibfield  {author} {\bibinfo {author} {\bibfnamefont {Junyeong}\
  \bibnamefont {Ahn}}, \bibinfo {author} {\bibfnamefont {Sungjoon}\
  \bibnamefont {Park}}, \ and\ \bibinfo {author} {\bibfnamefont {Bohm-Jung}\
  \bibnamefont {Yang}},\ }\bibfield  {title} {\enquote {\bibinfo {title}
  {Failure of nielsen-ninomiya theorem and fragile topology in two-dimensional
  systems with space-time inversion symmetry: Application to twisted bilayer
  graphene at magic angle},}\ }\href {\doibase 10.1103/PhysRevX.9.021013}
  {\bibfield  {journal} {\bibinfo  {journal} {Phys. Rev. X}\ }\textbf {\bibinfo
  {volume} {9}},\ \bibinfo {pages} {021013} (\bibinfo {year}
  {2019})}\BibitemShut {NoStop}%
\bibitem [{\citenamefont {Po}\ \emph {et~al.}(2019)\citenamefont {Po},
  \citenamefont {Zou}, \citenamefont {Senthil},\ and\ \citenamefont
  {Vishwanath}}]{po_2019}%
  \BibitemOpen
  \bibfield  {author} {\bibinfo {author} {\bibfnamefont {Hoi~Chun}\
  \bibnamefont {Po}}, \bibinfo {author} {\bibfnamefont {Liujun}\ \bibnamefont
  {Zou}}, \bibinfo {author} {\bibfnamefont {T.}~\bibnamefont {Senthil}}, \ and\
  \bibinfo {author} {\bibfnamefont {Ashvin}\ \bibnamefont {Vishwanath}},\
  }\bibfield  {title} {\enquote {\bibinfo {title} {Faithful tight-binding
  models and fragile topology of magic-angle bilayer graphene},}\ }\href
  {\doibase 10.1103/PhysRevB.99.195455} {\bibfield  {journal} {\bibinfo
  {journal} {Phys. Rev. B}\ }\textbf {\bibinfo {volume} {99}},\ \bibinfo
  {pages} {195455} (\bibinfo {year} {2019})}\BibitemShut {NoStop}%
\bibitem [{\citenamefont {Po}\ \emph {et~al.}(2018{\natexlab{a}})\citenamefont
  {Po}, \citenamefont {Zou}, \citenamefont {Vishwanath},\ and\ \citenamefont
  {Senthil}}]{po_2018}%
  \BibitemOpen
  \bibfield  {author} {\bibinfo {author} {\bibfnamefont {Hoi~Chun}\
  \bibnamefont {Po}}, \bibinfo {author} {\bibfnamefont {Liujun}\ \bibnamefont
  {Zou}}, \bibinfo {author} {\bibfnamefont {Ashvin}\ \bibnamefont
  {Vishwanath}}, \ and\ \bibinfo {author} {\bibfnamefont {T.}~\bibnamefont
  {Senthil}},\ }\bibfield  {title} {\enquote {\bibinfo {title} {Origin of mott
  insulating behavior and superconductivity in twisted bilayer graphene},}\
  }\href {\doibase 10.1103/PhysRevX.8.031089} {\bibfield  {journal} {\bibinfo
  {journal} {Phys. Rev. X}\ }\textbf {\bibinfo {volume} {8}},\ \bibinfo {pages}
  {031089} (\bibinfo {year} {2018}{\natexlab{a}})}\BibitemShut {NoStop}%
\bibitem [{\citenamefont {Po}\ \emph {et~al.}(2018{\natexlab{b}})\citenamefont
  {Po}, \citenamefont {Watanabe},\ and\ \citenamefont
  {Vishwanath}}]{po_2018_1}%
  \BibitemOpen
  \bibfield  {author} {\bibinfo {author} {\bibfnamefont {Hoi~Chun}\
  \bibnamefont {Po}}, \bibinfo {author} {\bibfnamefont {Haruki}\ \bibnamefont
  {Watanabe}}, \ and\ \bibinfo {author} {\bibfnamefont {Ashvin}\ \bibnamefont
  {Vishwanath}},\ }\bibfield  {title} {\enquote {\bibinfo {title} {Fragile
  topology and wannier obstructions},}\ }\href {\doibase
  10.1103/PhysRevLett.121.126402} {\bibfield  {journal} {\bibinfo  {journal}
  {Phys. Rev. Lett.}\ }\textbf {\bibinfo {volume} {121}},\ \bibinfo {pages}
  {126402} (\bibinfo {year} {2018}{\natexlab{b}})}\BibitemShut {NoStop}%
\bibitem [{\citenamefont {Kang}\ and\ \citenamefont {Vafek}(2019)}]{kang_2019}%
  \BibitemOpen
  \bibfield  {author} {\bibinfo {author} {\bibfnamefont {Jian}\ \bibnamefont
  {Kang}}\ and\ \bibinfo {author} {\bibfnamefont {Oskar}\ \bibnamefont
  {Vafek}},\ }\bibfield  {title} {\enquote {\bibinfo {title} {Strong coupling
  phases of partially filled twisted bilayer graphene narrow bands},}\ }\href
  {\doibase 10.1103/PhysRevLett.122.246401} {\bibfield  {journal} {\bibinfo
  {journal} {Phys. Rev. Lett.}\ }\textbf {\bibinfo {volume} {122}},\ \bibinfo
  {pages} {246401} (\bibinfo {year} {2019})}\BibitemShut {NoStop}%
\bibitem [{\citenamefont {Kang}\ and\ \citenamefont {Vafek}(2020)}]{kang_2020}%
  \BibitemOpen
  \bibfield  {author} {\bibinfo {author} {\bibfnamefont {Jian}\ \bibnamefont
  {Kang}}\ and\ \bibinfo {author} {\bibfnamefont {Oskar}\ \bibnamefont
  {Vafek}},\ }\bibfield  {title} {\enquote {\bibinfo {title} {Non-abelian dirac
  node braiding and near-degeneracy of correlated phases at odd integer filling
  in magic-angle twisted bilayer graphene},}\ }\href {\doibase
  10.1103/PhysRevB.102.035161} {\bibfield  {journal} {\bibinfo  {journal}
  {Phys. Rev. B}\ }\textbf {\bibinfo {volume} {102}},\ \bibinfo {pages}
  {035161} (\bibinfo {year} {2020})}\BibitemShut {NoStop}%
\bibitem [{\citenamefont {Zhang}\ \emph {et~al.}(2019)\citenamefont {Zhang},
  \citenamefont {Mao},\ and\ \citenamefont {Senthil}}]{zhang_2019}%
  \BibitemOpen
  \bibfield  {author} {\bibinfo {author} {\bibfnamefont {Ya-Hui}\ \bibnamefont
  {Zhang}}, \bibinfo {author} {\bibfnamefont {Dan}\ \bibnamefont {Mao}}, \ and\
  \bibinfo {author} {\bibfnamefont {T.}~\bibnamefont {Senthil}},\ }\bibfield
  {title} {\enquote {\bibinfo {title} {Twisted bilayer graphene aligned with
  hexagonal boron nitride: Anomalous hall effect and a lattice model},}\ }\href
  {\doibase 10.1103/PhysRevResearch.1.033126} {\bibfield  {journal} {\bibinfo
  {journal} {Phys. Rev. Res.}\ }\textbf {\bibinfo {volume} {1}},\ \bibinfo
  {pages} {033126} (\bibinfo {year} {2019})}\BibitemShut {NoStop}%
\bibitem [{\citenamefont {Xu}\ \emph {et~al.}(2018)\citenamefont {Xu},
  \citenamefont {Law},\ and\ \citenamefont {Lee}}]{xu_2018}%
  \BibitemOpen
  \bibfield  {author} {\bibinfo {author} {\bibfnamefont {Xiao~Yan}\
  \bibnamefont {Xu}}, \bibinfo {author} {\bibfnamefont {K.~T.}\ \bibnamefont
  {Law}}, \ and\ \bibinfo {author} {\bibfnamefont {Patrick~A.}\ \bibnamefont
  {Lee}},\ }\bibfield  {title} {\enquote {\bibinfo {title} {Kekul\'e valence
  bond order in an extended hubbard model on the honeycomb lattice with
  possible applications to twisted bilayer graphene},}\ }\href {\doibase
  10.1103/PhysRevB.98.121406} {\bibfield  {journal} {\bibinfo  {journal} {Phys.
  Rev. B}\ }\textbf {\bibinfo {volume} {98}},\ \bibinfo {pages} {121406}
  (\bibinfo {year} {2018})}\BibitemShut {NoStop}%
\bibitem [{\citenamefont {Xie}\ and\ \citenamefont
  {MacDonald}(2020)}]{xie_2020}%
  \BibitemOpen
  \bibfield  {author} {\bibinfo {author} {\bibfnamefont {Ming}\ \bibnamefont
  {Xie}}\ and\ \bibinfo {author} {\bibfnamefont {A.~H.}\ \bibnamefont
  {MacDonald}},\ }\bibfield  {title} {\enquote {\bibinfo {title} {Nature of the
  correlated insulator states in twisted bilayer graphene},}\ }\href {\doibase
  10.1103/PhysRevLett.124.097601} {\bibfield  {journal} {\bibinfo  {journal}
  {Phys. Rev. Lett.}\ }\textbf {\bibinfo {volume} {124}},\ \bibinfo {pages}
  {097601} (\bibinfo {year} {2020})}\BibitemShut {NoStop}%
\bibitem [{\citenamefont {Liu}\ \emph {et~al.}(2018)\citenamefont {Liu},
  \citenamefont {Zhang}, \citenamefont {Chen},\ and\ \citenamefont
  {Yang}}]{liuc_2018}%
  \BibitemOpen
  \bibfield  {author} {\bibinfo {author} {\bibfnamefont {Cheng-Cheng}\
  \bibnamefont {Liu}}, \bibinfo {author} {\bibfnamefont {Li-Da}\ \bibnamefont
  {Zhang}}, \bibinfo {author} {\bibfnamefont {Wei-Qiang}\ \bibnamefont {Chen}},
  \ and\ \bibinfo {author} {\bibfnamefont {Fan}\ \bibnamefont {Yang}},\
  }\bibfield  {title} {\enquote {\bibinfo {title} {Chiral spin density wave and
  $d+id$ superconductivity in the magic-angle-twisted bilayer graphene},}\
  }\href {\doibase 10.1103/PhysRevLett.121.217001} {\bibfield  {journal}
  {\bibinfo  {journal} {Phys. Rev. Lett.}\ }\textbf {\bibinfo {volume} {121}},\
  \bibinfo {pages} {217001} (\bibinfo {year} {2018})}\BibitemShut {NoStop}%
\bibitem [{\citenamefont {Huang}\ \emph {et~al.}(2019)\citenamefont {Huang},
  \citenamefont {Zhang},\ and\ \citenamefont {Ma}}]{huang_2019}%
  \BibitemOpen
  \bibfield  {author} {\bibinfo {author} {\bibfnamefont {Tongyun}\ \bibnamefont
  {Huang}}, \bibinfo {author} {\bibfnamefont {Lufeng}\ \bibnamefont {Zhang}}, \
  and\ \bibinfo {author} {\bibfnamefont {Tianxing}\ \bibnamefont {Ma}},\
  }\bibfield  {title} {\enquote {\bibinfo {title} {Antiferromagnetically
  ordered mott insulator and d+id superconductivity in twisted bilayer
  graphene: a quantum monte carlo study},}\ }\href {\doibase
  https://doi.org/10.1016/j.scib.2019.01.026} {\bibfield  {journal} {\bibinfo
  {journal} {Science Bulletin}\ }\textbf {\bibinfo {volume} {64}},\ \bibinfo
  {pages} {310--314} (\bibinfo {year} {2019})}\BibitemShut {NoStop}%
\bibitem [{\citenamefont {Liu}\ and\ \citenamefont
  {Dai}(2021{\natexlab{a}})}]{liu_2021}%
  \BibitemOpen
  \bibfield  {author} {\bibinfo {author} {\bibfnamefont {Jianpeng}\
  \bibnamefont {Liu}}\ and\ \bibinfo {author} {\bibfnamefont {Xi}~\bibnamefont
  {Dai}},\ }\bibfield  {title} {\enquote {\bibinfo {title} {Theories for the
  correlated insulating states and quantum anomalous hall effect phenomena in
  twisted bilayer graphene},}\ }\href {\doibase 10.1103/PhysRevB.103.035427}
  {\bibfield  {journal} {\bibinfo  {journal} {Phys. Rev. B}\ }\textbf {\bibinfo
  {volume} {103}},\ \bibinfo {pages} {035427} (\bibinfo {year}
  {2021}{\natexlab{a}})}\BibitemShut {NoStop}%
\bibitem [{\citenamefont {Liu}\ \emph {et~al.}(2021)\citenamefont {Liu},
  \citenamefont {Khalaf}, \citenamefont {Lee},\ and\ \citenamefont
  {Vishwanath}}]{lius_2021}%
  \BibitemOpen
  \bibfield  {author} {\bibinfo {author} {\bibfnamefont {Shang}\ \bibnamefont
  {Liu}}, \bibinfo {author} {\bibfnamefont {Eslam}\ \bibnamefont {Khalaf}},
  \bibinfo {author} {\bibfnamefont {Jong~Yeon}\ \bibnamefont {Lee}}, \ and\
  \bibinfo {author} {\bibfnamefont {Ashvin}\ \bibnamefont {Vishwanath}},\
  }\bibfield  {title} {\enquote {\bibinfo {title} {Nematic topological
  semimetal and insulator in magic-angle bilayer graphene at charge
  neutrality},}\ }\href {\doibase 10.1103/PhysRevResearch.3.013033} {\bibfield
  {journal} {\bibinfo  {journal} {Phys. Rev. Res.}\ }\textbf {\bibinfo {volume}
  {3}},\ \bibinfo {pages} {013033} (\bibinfo {year} {2021})}\BibitemShut
  {NoStop}%
\bibitem [{\citenamefont {Bultinck}\ \emph
  {et~al.}(2020{\natexlab{a}})\citenamefont {Bultinck}, \citenamefont
  {Chatterjee},\ and\ \citenamefont {Zaletel}}]{bul1_2020}%
  \BibitemOpen
  \bibfield  {author} {\bibinfo {author} {\bibfnamefont {Nick}\ \bibnamefont
  {Bultinck}}, \bibinfo {author} {\bibfnamefont {Shubhayu}\ \bibnamefont
  {Chatterjee}}, \ and\ \bibinfo {author} {\bibfnamefont {Michael~P.}\
  \bibnamefont {Zaletel}},\ }\bibfield  {title} {\enquote {\bibinfo {title}
  {Mechanism for anomalous hall ferromagnetism in twisted bilayer graphene},}\
  }\href {\doibase 10.1103/PhysRevLett.124.166601} {\bibfield  {journal}
  {\bibinfo  {journal} {Phys. Rev. Lett.}\ }\textbf {\bibinfo {volume} {124}},\
  \bibinfo {pages} {166601} (\bibinfo {year} {2020}{\natexlab{a}})}\BibitemShut
  {NoStop}%
\bibitem [{\citenamefont {Bultinck}\ \emph
  {et~al.}(2020{\natexlab{b}})\citenamefont {Bultinck}, \citenamefont {Khalaf},
  \citenamefont {Liu}, \citenamefont {Chatterjee}, \citenamefont {Vishwanath},\
  and\ \citenamefont {Zaletel}}]{bul_2020}%
  \BibitemOpen
  \bibfield  {author} {\bibinfo {author} {\bibfnamefont {Nick}\ \bibnamefont
  {Bultinck}}, \bibinfo {author} {\bibfnamefont {Eslam}\ \bibnamefont
  {Khalaf}}, \bibinfo {author} {\bibfnamefont {Shang}\ \bibnamefont {Liu}},
  \bibinfo {author} {\bibfnamefont {Shubhayu}\ \bibnamefont {Chatterjee}},
  \bibinfo {author} {\bibfnamefont {Ashvin}\ \bibnamefont {Vishwanath}}, \ and\
  \bibinfo {author} {\bibfnamefont {Michael~P.}\ \bibnamefont {Zaletel}},\
  }\bibfield  {title} {\enquote {\bibinfo {title} {Ground state and hidden
  symmetry of magic-angle graphene at even integer filling},}\ }\href {\doibase
  10.1103/PhysRevX.10.031034} {\bibfield  {journal} {\bibinfo  {journal} {Phys.
  Rev. X}\ }\textbf {\bibinfo {volume} {10}},\ \bibinfo {pages} {031034}
  (\bibinfo {year} {2020}{\natexlab{b}})}\BibitemShut {NoStop}%
\bibitem [{\citenamefont {Wu}\ and\ \citenamefont {Das~Sarma}(2020)}]{wu_2020}%
  \BibitemOpen
  \bibfield  {author} {\bibinfo {author} {\bibfnamefont {Fengcheng}\
  \bibnamefont {Wu}}\ and\ \bibinfo {author} {\bibfnamefont {Sankar}\
  \bibnamefont {Das~Sarma}},\ }\bibfield  {title} {\enquote {\bibinfo {title}
  {Collective excitations of quantum anomalous hall ferromagnets in twisted
  bilayer graphene},}\ }\href {\doibase 10.1103/PhysRevLett.124.046403}
  {\bibfield  {journal} {\bibinfo  {journal} {Phys. Rev. Lett.}\ }\textbf
  {\bibinfo {volume} {124}},\ \bibinfo {pages} {046403} (\bibinfo {year}
  {2020})}\BibitemShut {NoStop}%
\bibitem [{\citenamefont {Chen}\ \emph {et~al.}(2021)\citenamefont {Chen},
  \citenamefont {Liao}, \citenamefont {Chen}, \citenamefont {Vafek},
  \citenamefont {Kang}, \citenamefont {Li},\ and\ \citenamefont
  {Meng}}]{chen_2020}%
  \BibitemOpen
  \bibfield  {author} {\bibinfo {author} {\bibfnamefont {Bin-Bin}\ \bibnamefont
  {Chen}}, \bibinfo {author} {\bibfnamefont {Yuan~Da}\ \bibnamefont {Liao}},
  \bibinfo {author} {\bibfnamefont {Ziyu}\ \bibnamefont {Chen}}, \bibinfo
  {author} {\bibfnamefont {Oskar}\ \bibnamefont {Vafek}}, \bibinfo {author}
  {\bibfnamefont {Jian}\ \bibnamefont {Kang}}, \bibinfo {author} {\bibfnamefont
  {Wei}\ \bibnamefont {Li}}, \ and\ \bibinfo {author} {\bibfnamefont {Zi~Yang}\
  \bibnamefont {Meng}},\ }\bibfield  {title} {\enquote {\bibinfo {title}
  {Realization of topological mott insulator in a twisted bilayer graphene
  lattice model},}\ }\href {\doibase 10.1038/s41467-021-25438-1} {\bibfield
  {journal} {\bibinfo  {journal} {Nature Communications}\ }\textbf {\bibinfo
  {volume} {12}},\ \bibinfo {pages} {5480} (\bibinfo {year}
  {2021})}\BibitemShut {NoStop}%
\bibitem [{\citenamefont {Lu}\ \emph {et~al.}(2022)\citenamefont {Lu},
  \citenamefont {Zhang}, \citenamefont {Zhang}, \citenamefont {Zhang},
  \citenamefont {Liu}, \citenamefont {Wang}, \citenamefont {Gu}, \citenamefont
  {Chen},\ and\ \citenamefont {Yang}}]{lu_2020}%
  \BibitemOpen
  \bibfield  {author} {\bibinfo {author} {\bibfnamefont {Chen}\ \bibnamefont
  {Lu}}, \bibinfo {author} {\bibfnamefont {Yongyou}\ \bibnamefont {Zhang}},
  \bibinfo {author} {\bibfnamefont {Yu}~\bibnamefont {Zhang}}, \bibinfo
  {author} {\bibfnamefont {Ming}\ \bibnamefont {Zhang}}, \bibinfo {author}
  {\bibfnamefont {Cheng-Cheng}\ \bibnamefont {Liu}}, \bibinfo {author}
  {\bibfnamefont {Yu}~\bibnamefont {Wang}}, \bibinfo {author} {\bibfnamefont
  {Zheng-Cheng}\ \bibnamefont {Gu}}, \bibinfo {author} {\bibfnamefont
  {Wei-Qiang}\ \bibnamefont {Chen}}, \ and\ \bibinfo {author} {\bibfnamefont
  {Fan}\ \bibnamefont {Yang}},\ }\bibfield  {title} {\enquote {\bibinfo {title}
  {Chiral so(4) spin-valley density wave and degenerate topological
  superconductivity in magic-angle twisted bilayer graphene},}\ }\href
  {\doibase 10.1103/PhysRevB.106.024518} {\bibfield  {journal} {\bibinfo
  {journal} {Phys. Rev. B}\ }\textbf {\bibinfo {volume} {106}},\ \bibinfo
  {pages} {024518} (\bibinfo {year} {2022})}\BibitemShut {NoStop}%
\bibitem [{\citenamefont {Da~Liao}\ \emph {et~al.}(2021)\citenamefont
  {Da~Liao}, \citenamefont {Kang}, \citenamefont {Brei\o{}}, \citenamefont
  {Xu}, \citenamefont {Wu}, \citenamefont {Andersen}, \citenamefont
  {Fernandes},\ and\ \citenamefont {Meng}}]{liao_2021}%
  \BibitemOpen
  \bibfield  {author} {\bibinfo {author} {\bibfnamefont {Yuan}\ \bibnamefont
  {Da~Liao}}, \bibinfo {author} {\bibfnamefont {Jian}\ \bibnamefont {Kang}},
  \bibinfo {author} {\bibfnamefont {Clara~N.}\ \bibnamefont {Brei\o{}}},
  \bibinfo {author} {\bibfnamefont {Xiao~Yan}\ \bibnamefont {Xu}}, \bibinfo
  {author} {\bibfnamefont {Han-Qing}\ \bibnamefont {Wu}}, \bibinfo {author}
  {\bibfnamefont {Brian~M.}\ \bibnamefont {Andersen}}, \bibinfo {author}
  {\bibfnamefont {Rafael~M.}\ \bibnamefont {Fernandes}}, \ and\ \bibinfo
  {author} {\bibfnamefont {Zi~Yang}\ \bibnamefont {Meng}},\ }\bibfield  {title}
  {\enquote {\bibinfo {title} {Correlation-induced insulating topological
  phases at charge neutrality in twisted bilayer graphene},}\ }\href {\doibase
  10.1103/PhysRevX.11.011014} {\bibfield  {journal} {\bibinfo  {journal} {Phys.
  Rev. X}\ }\textbf {\bibinfo {volume} {11}},\ \bibinfo {pages} {011014}
  (\bibinfo {year} {2021})}\BibitemShut {NoStop}%
\bibitem [{\citenamefont {Lian}\ \emph {et~al.}(2021)\citenamefont {Lian},
  \citenamefont {Song}, \citenamefont {Regnault}, \citenamefont {Efetov},
  \citenamefont {Yazdani},\ and\ \citenamefont {Bernevig}}]{lian_2020}%
  \BibitemOpen
  \bibfield  {author} {\bibinfo {author} {\bibfnamefont {Biao}\ \bibnamefont
  {Lian}}, \bibinfo {author} {\bibfnamefont {Zhi-Da}\ \bibnamefont {Song}},
  \bibinfo {author} {\bibfnamefont {Nicolas}\ \bibnamefont {Regnault}},
  \bibinfo {author} {\bibfnamefont {Dmitri~K.}\ \bibnamefont {Efetov}},
  \bibinfo {author} {\bibfnamefont {Ali}\ \bibnamefont {Yazdani}}, \ and\
  \bibinfo {author} {\bibfnamefont {B.~Andrei}\ \bibnamefont {Bernevig}},\
  }\bibfield  {title} {\enquote {\bibinfo {title} {Twisted bilayer graphene.
  iv. exact insulator ground states and phase diagram},}\ }\href {\doibase
  10.1103/PhysRevB.103.205414} {\bibfield  {journal} {\bibinfo  {journal}
  {Phys. Rev. B}\ }\textbf {\bibinfo {volume} {103}},\ \bibinfo {pages}
  {205414} (\bibinfo {year} {2021})}\BibitemShut {NoStop}%
\bibitem [{\citenamefont {Xie}\ \emph {et~al.}(2021{\natexlab{b}})\citenamefont
  {Xie}, \citenamefont {Cowsik}, \citenamefont {Song}, \citenamefont {Lian},
  \citenamefont {Bernevig},\ and\ \citenamefont {Regnault}}]{xief_2020}%
  \BibitemOpen
  \bibfield  {author} {\bibinfo {author} {\bibfnamefont {Fang}\ \bibnamefont
  {Xie}}, \bibinfo {author} {\bibfnamefont {Aditya}\ \bibnamefont {Cowsik}},
  \bibinfo {author} {\bibfnamefont {Zhi-Da}\ \bibnamefont {Song}}, \bibinfo
  {author} {\bibfnamefont {Biao}\ \bibnamefont {Lian}}, \bibinfo {author}
  {\bibfnamefont {B.~Andrei}\ \bibnamefont {Bernevig}}, \ and\ \bibinfo
  {author} {\bibfnamefont {Nicolas}\ \bibnamefont {Regnault}},\ }\bibfield
  {title} {\enquote {\bibinfo {title} {Twisted bilayer graphene. vi. an exact
  diagonalization study at nonzero integer filling},}\ }\href {\doibase
  10.1103/PhysRevB.103.205416} {\bibfield  {journal} {\bibinfo  {journal}
  {Phys. Rev. B}\ }\textbf {\bibinfo {volume} {103}},\ \bibinfo {pages}
  {205416} (\bibinfo {year} {2021}{\natexlab{b}})}\BibitemShut {NoStop}%
\bibitem [{\citenamefont {Hejazi}\ \emph {et~al.}(2021)\citenamefont {Hejazi},
  \citenamefont {Chen},\ and\ \citenamefont {Balents}}]{hejazi_2020}%
  \BibitemOpen
  \bibfield  {author} {\bibinfo {author} {\bibfnamefont {Kasra}\ \bibnamefont
  {Hejazi}}, \bibinfo {author} {\bibfnamefont {Xiao}\ \bibnamefont {Chen}}, \
  and\ \bibinfo {author} {\bibfnamefont {Leon}\ \bibnamefont {Balents}},\
  }\bibfield  {title} {\enquote {\bibinfo {title} {Hybrid wannier chern bands
  in magic angle twisted bilayer graphene and the quantized anomalous hall
  effect},}\ }\href {\doibase 10.1103/PhysRevResearch.3.013242} {\bibfield
  {journal} {\bibinfo  {journal} {Phys. Rev. Res.}\ }\textbf {\bibinfo {volume}
  {3}},\ \bibinfo {pages} {013242} (\bibinfo {year} {2021})}\BibitemShut
  {NoStop}%
\bibitem [{\citenamefont {Sboychakov}\ \emph {et~al.}(2019)\citenamefont
  {Sboychakov}, \citenamefont {Rozhkov}, \citenamefont {Rakhmanov},\ and\
  \citenamefont {Nori}}]{sb_2019}%
  \BibitemOpen
  \bibfield  {author} {\bibinfo {author} {\bibfnamefont {A.~O.}\ \bibnamefont
  {Sboychakov}}, \bibinfo {author} {\bibfnamefont {A.~V.}\ \bibnamefont
  {Rozhkov}}, \bibinfo {author} {\bibfnamefont {A.~L.}\ \bibnamefont
  {Rakhmanov}}, \ and\ \bibinfo {author} {\bibfnamefont {Franco}\ \bibnamefont
  {Nori}},\ }\bibfield  {title} {\enquote {\bibinfo {title} {Many-body effects
  in twisted bilayer graphene at low twist angles},}\ }\href {\doibase
  10.1103/PhysRevB.100.045111} {\bibfield  {journal} {\bibinfo  {journal}
  {Phys. Rev. B}\ }\textbf {\bibinfo {volume} {100}},\ \bibinfo {pages}
  {045111} (\bibinfo {year} {2019})}\BibitemShut {NoStop}%
\bibitem [{\citenamefont {Rademaker}\ \emph {et~al.}(2019)\citenamefont
  {Rademaker}, \citenamefont {Abanin},\ and\ \citenamefont
  {Mellado}}]{ra_2019}%
  \BibitemOpen
  \bibfield  {author} {\bibinfo {author} {\bibfnamefont {Louk}\ \bibnamefont
  {Rademaker}}, \bibinfo {author} {\bibfnamefont {Dmitry~A.}\ \bibnamefont
  {Abanin}}, \ and\ \bibinfo {author} {\bibfnamefont {Paula}\ \bibnamefont
  {Mellado}},\ }\bibfield  {title} {\enquote {\bibinfo {title} {Charge
  smoothening and band flattening due to hartree corrections in twisted bilayer
  graphene},}\ }\href {\doibase 10.1103/PhysRevB.100.205114} {\bibfield
  {journal} {\bibinfo  {journal} {Phys. Rev. B}\ }\textbf {\bibinfo {volume}
  {100}},\ \bibinfo {pages} {205114} (\bibinfo {year} {2019})}\BibitemShut
  {NoStop}%
\bibitem [{\citenamefont {Guinea}\ and\ \citenamefont {Walet}(2018)}]{gu_2018}%
  \BibitemOpen
  \bibfield  {author} {\bibinfo {author} {\bibfnamefont {Francisco}\
  \bibnamefont {Guinea}}\ and\ \bibinfo {author} {\bibfnamefont {Niels~R.}\
  \bibnamefont {Walet}},\ }\bibfield  {title} {\enquote {\bibinfo {title}
  {Electrostatic effects, band distortions, and superconductivity in twisted
  graphene bilayers},}\ }\href {\doibase 10.1073/pnas.1810947115} {\bibfield
  {journal} {\bibinfo  {journal} {Proceedings of the National Academy of
  Sciences}\ }\textbf {\bibinfo {volume} {115}},\ \bibinfo {pages}
  {13174--13179} (\bibinfo {year} {2018})}\BibitemShut {NoStop}%
\bibitem [{\citenamefont {Seo}\ \emph {et~al.}(2019)\citenamefont {Seo},
  \citenamefont {Kotov},\ and\ \citenamefont {Uchoa}}]{seo_2019}%
  \BibitemOpen
  \bibfield  {author} {\bibinfo {author} {\bibfnamefont {Kangjun}\ \bibnamefont
  {Seo}}, \bibinfo {author} {\bibfnamefont {Valeri~N.}\ \bibnamefont {Kotov}},
  \ and\ \bibinfo {author} {\bibfnamefont {Bruno}\ \bibnamefont {Uchoa}},\
  }\bibfield  {title} {\enquote {\bibinfo {title} {Ferromagnetic mott state in
  twisted graphene bilayers at the magic angle},}\ }\href {\doibase
  10.1103/PhysRevLett.122.246402} {\bibfield  {journal} {\bibinfo  {journal}
  {Phys. Rev. Lett.}\ }\textbf {\bibinfo {volume} {122}},\ \bibinfo {pages}
  {246402} (\bibinfo {year} {2019})}\BibitemShut {NoStop}%
\bibitem [{\citenamefont {Bernevig}\ \emph
  {et~al.}(2021{\natexlab{a}})\citenamefont {Bernevig}, \citenamefont {Song},
  \citenamefont {Regnault},\ and\ \citenamefont {Lian}}]{bernevig_2020}%
  \BibitemOpen
  \bibfield  {author} {\bibinfo {author} {\bibfnamefont {B.~Andrei}\
  \bibnamefont {Bernevig}}, \bibinfo {author} {\bibfnamefont {Zhi-Da}\
  \bibnamefont {Song}}, \bibinfo {author} {\bibfnamefont {Nicolas}\
  \bibnamefont {Regnault}}, \ and\ \bibinfo {author} {\bibfnamefont {Biao}\
  \bibnamefont {Lian}},\ }\bibfield  {title} {\enquote {\bibinfo {title}
  {Twisted bilayer graphene. iii. interacting hamiltonian and exact
  symmetries},}\ }\href {\doibase 10.1103/PhysRevB.103.205413} {\bibfield
  {journal} {\bibinfo  {journal} {Phys. Rev. B}\ }\textbf {\bibinfo {volume}
  {103}},\ \bibinfo {pages} {205413} (\bibinfo {year}
  {2021}{\natexlab{a}})}\BibitemShut {NoStop}%
\bibitem [{\citenamefont {Kang}\ \emph {et~al.}(2021)\citenamefont {Kang},
  \citenamefont {Bernevig},\ and\ \citenamefont {Vafek}}]{kang_2021}%
  \BibitemOpen
  \bibfield  {author} {\bibinfo {author} {\bibfnamefont {Jian}\ \bibnamefont
  {Kang}}, \bibinfo {author} {\bibfnamefont {B.~Andrei}\ \bibnamefont
  {Bernevig}}, \ and\ \bibinfo {author} {\bibfnamefont {Oskar}\ \bibnamefont
  {Vafek}},\ }\bibfield  {title} {\enquote {\bibinfo {title} {Cascades between
  light and heavy fermions in the normal state of magic-angle twisted bilayer
  graphene},}\ }\href {\doibase 10.1103/PhysRevLett.127.266402} {\bibfield
  {journal} {\bibinfo  {journal} {Phys. Rev. Lett.}\ }\textbf {\bibinfo
  {volume} {127}},\ \bibinfo {pages} {266402} (\bibinfo {year}
  {2021})}\BibitemShut {NoStop}%
\bibitem [{\citenamefont {Zhang}\ \emph {et~al.}(2020)\citenamefont {Zhang},
  \citenamefont {Jiang}, \citenamefont {Wang},\ and\ \citenamefont
  {Zhang}}]{zhang_2020}%
  \BibitemOpen
  \bibfield  {author} {\bibinfo {author} {\bibfnamefont {Yi}~\bibnamefont
  {Zhang}}, \bibinfo {author} {\bibfnamefont {Kun}\ \bibnamefont {Jiang}},
  \bibinfo {author} {\bibfnamefont {Ziqiang}\ \bibnamefont {Wang}}, \ and\
  \bibinfo {author} {\bibfnamefont {Fuchun}\ \bibnamefont {Zhang}},\ }\bibfield
   {title} {\enquote {\bibinfo {title} {Correlated insulating phases of twisted
  bilayer graphene at commensurate filling fractions: A hartree-fock study},}\
  }\href {\doibase 10.1103/PhysRevB.102.035136} {\bibfield  {journal} {\bibinfo
   {journal} {Phys. Rev. B}\ }\textbf {\bibinfo {volume} {102}},\ \bibinfo
  {pages} {035136} (\bibinfo {year} {2020})}\BibitemShut {NoStop}%
\bibitem [{\citenamefont {Wang}\ and\ \citenamefont {Vafek}(2024)}]{wang2024}%
  \BibitemOpen
  \bibfield  {author} {\bibinfo {author} {\bibfnamefont {Xiaoyu}\ \bibnamefont
  {Wang}}\ and\ \bibinfo {author} {\bibfnamefont {Oskar}\ \bibnamefont
  {Vafek}},\ }\bibfield  {title} {\enquote {\bibinfo {title} {Theory of
  correlated chern insulators in twisted bilayer graphene},}\ }\href {\doibase
  10.1103/PhysRevX.14.021042} {\bibfield  {journal} {\bibinfo  {journal} {Phys.
  Rev. X}\ }\textbf {\bibinfo {volume} {14}},\ \bibinfo {pages} {021042}
  (\bibinfo {year} {2024})}\BibitemShut {NoStop}%
\bibitem [{\citenamefont {Wang}\ and\ \citenamefont {Vafek}(2022)}]{wang_2022}%
  \BibitemOpen
  \bibfield  {author} {\bibinfo {author} {\bibfnamefont {Xiaoyu}\ \bibnamefont
  {Wang}}\ and\ \bibinfo {author} {\bibfnamefont {Oskar}\ \bibnamefont
  {Vafek}},\ }\bibfield  {title} {\enquote {\bibinfo {title} {Narrow bands in
  magnetic field and strong-coupling hofstadter spectra},}\ }\href {\doibase
  10.1103/PhysRevB.106.L121111} {\bibfield  {journal} {\bibinfo  {journal}
  {Phys. Rev. B}\ }\textbf {\bibinfo {volume} {106}},\ \bibinfo {pages}
  {L121111} (\bibinfo {year} {2022})}\BibitemShut {NoStop}%
\bibitem [{\citenamefont {Chen}\ \emph {et~al.}(2020)\citenamefont {Chen},
  \citenamefont {Sharpe}, \citenamefont {Fox}, \citenamefont {Zhang},
  \citenamefont {Wang}, \citenamefont {Jiang}, \citenamefont {Lyu},
  \citenamefont {Li}, \citenamefont {Watanabe}, \citenamefont {Taniguchi},
  \citenamefont {Shi}, \citenamefont {Senthil}, \citenamefont
  {Goldhaber-Gordon}, \citenamefont {Zhang},\ and\ \citenamefont
  {Wang}}]{chen2020}%
  \BibitemOpen
  \bibfield  {author} {\bibinfo {author} {\bibfnamefont {Guorui}\ \bibnamefont
  {Chen}}, \bibinfo {author} {\bibfnamefont {Aaron~L.}\ \bibnamefont {Sharpe}},
  \bibinfo {author} {\bibfnamefont {Eli~J.}\ \bibnamefont {Fox}}, \bibinfo
  {author} {\bibfnamefont {Ya-Hui}\ \bibnamefont {Zhang}}, \bibinfo {author}
  {\bibfnamefont {Shaoxin}\ \bibnamefont {Wang}}, \bibinfo {author}
  {\bibfnamefont {Lili}\ \bibnamefont {Jiang}}, \bibinfo {author}
  {\bibfnamefont {Bosai}\ \bibnamefont {Lyu}}, \bibinfo {author} {\bibfnamefont
  {Hongyuan}\ \bibnamefont {Li}}, \bibinfo {author} {\bibfnamefont {Kenji}\
  \bibnamefont {Watanabe}}, \bibinfo {author} {\bibfnamefont {Takashi}\
  \bibnamefont {Taniguchi}}, \bibinfo {author} {\bibfnamefont {Zhiwen}\
  \bibnamefont {Shi}}, \bibinfo {author} {\bibfnamefont {T.}~\bibnamefont
  {Senthil}}, \bibinfo {author} {\bibfnamefont {David}\ \bibnamefont
  {Goldhaber-Gordon}}, \bibinfo {author} {\bibfnamefont {Yuanbo}\ \bibnamefont
  {Zhang}}, \ and\ \bibinfo {author} {\bibfnamefont {Feng}\ \bibnamefont
  {Wang}},\ }\bibfield  {title} {\enquote {\bibinfo {title} {Tunable correlated
  chern insulator and ferromagnetism in a moir{\'e}superlattice},}\ }\href
  {\doibase 10.1038/s41586-020-2049-7} {\bibfield  {journal} {\bibinfo
  {journal} {Nature}\ }\textbf {\bibinfo {volume} {579}},\ \bibinfo {pages}
  {56--61} (\bibinfo {year} {2020})}\BibitemShut {NoStop}%
\bibitem [{\citenamefont {Li}\ \emph {et~al.}(2021)\citenamefont {Li},
  \citenamefont {Jiang}, \citenamefont {Shen}, \citenamefont {Zhang},
  \citenamefont {Li}, \citenamefont {Tao}, \citenamefont {Devakul},
  \citenamefont {Watanabe}, \citenamefont {Taniguchi}, \citenamefont {Fu},
  \citenamefont {Shan},\ and\ \citenamefont {Mak}}]{li2021}%
  \BibitemOpen
  \bibfield  {author} {\bibinfo {author} {\bibfnamefont {Tingxin}\ \bibnamefont
  {Li}}, \bibinfo {author} {\bibfnamefont {Shengwei}\ \bibnamefont {Jiang}},
  \bibinfo {author} {\bibfnamefont {Bowen}\ \bibnamefont {Shen}}, \bibinfo
  {author} {\bibfnamefont {Yang}\ \bibnamefont {Zhang}}, \bibinfo {author}
  {\bibfnamefont {Lizhong}\ \bibnamefont {Li}}, \bibinfo {author}
  {\bibfnamefont {Zui}\ \bibnamefont {Tao}}, \bibinfo {author} {\bibfnamefont
  {Trithep}\ \bibnamefont {Devakul}}, \bibinfo {author} {\bibfnamefont {Kenji}\
  \bibnamefont {Watanabe}}, \bibinfo {author} {\bibfnamefont {Takashi}\
  \bibnamefont {Taniguchi}}, \bibinfo {author} {\bibfnamefont {Liang}\
  \bibnamefont {Fu}}, \bibinfo {author} {\bibfnamefont {Jie}\ \bibnamefont
  {Shan}}, \ and\ \bibinfo {author} {\bibfnamefont {Kin~Fai}\ \bibnamefont
  {Mak}},\ }\bibfield  {title} {\enquote {\bibinfo {title} {Quantum anomalous
  hall effect from intertwined moir{\'e}bands},}\ }\href {\doibase
  10.1038/s41586-021-04171-1} {\bibfield  {journal} {\bibinfo  {journal}
  {Nature}\ }\textbf {\bibinfo {volume} {600}},\ \bibinfo {pages} {641--646}
  (\bibinfo {year} {2021})}\BibitemShut {NoStop}%
\bibitem [{\citenamefont {Xu}\ \emph {et~al.}(2023)\citenamefont {Xu},
  \citenamefont {Sun}, \citenamefont {Jia}, \citenamefont {Liu}, \citenamefont
  {Xu}, \citenamefont {Li}, \citenamefont {Gu}, \citenamefont {Watanabe},
  \citenamefont {Taniguchi}, \citenamefont {Tong}, \citenamefont {Jia},
  \citenamefont {Shi}, \citenamefont {Jiang}, \citenamefont {Zhang},
  \citenamefont {Liu},\ and\ \citenamefont {Li}}]{xu2023}%
  \BibitemOpen
  \bibfield  {author} {\bibinfo {author} {\bibfnamefont {Fan}\ \bibnamefont
  {Xu}}, \bibinfo {author} {\bibfnamefont {Zheng}\ \bibnamefont {Sun}},
  \bibinfo {author} {\bibfnamefont {Tongtong}\ \bibnamefont {Jia}}, \bibinfo
  {author} {\bibfnamefont {Chang}\ \bibnamefont {Liu}}, \bibinfo {author}
  {\bibfnamefont {Cheng}\ \bibnamefont {Xu}}, \bibinfo {author} {\bibfnamefont
  {Chushan}\ \bibnamefont {Li}}, \bibinfo {author} {\bibfnamefont
  {Yu}~\bibnamefont {Gu}}, \bibinfo {author} {\bibfnamefont {Kenji}\
  \bibnamefont {Watanabe}}, \bibinfo {author} {\bibfnamefont {Takashi}\
  \bibnamefont {Taniguchi}}, \bibinfo {author} {\bibfnamefont {Bingbing}\
  \bibnamefont {Tong}}, \bibinfo {author} {\bibfnamefont {Jinfeng}\
  \bibnamefont {Jia}}, \bibinfo {author} {\bibfnamefont {Zhiwen}\ \bibnamefont
  {Shi}}, \bibinfo {author} {\bibfnamefont {Shengwei}\ \bibnamefont {Jiang}},
  \bibinfo {author} {\bibfnamefont {Yang}\ \bibnamefont {Zhang}}, \bibinfo
  {author} {\bibfnamefont {Xiaoxue}\ \bibnamefont {Liu}}, \ and\ \bibinfo
  {author} {\bibfnamefont {Tingxin}\ \bibnamefont {Li}},\ }\bibfield  {title}
  {\enquote {\bibinfo {title} {Observation of integer and fractional quantum
  anomalous hall effects in twisted bilayer ${\mathrm{mote}}_{2}$},}\ }\href
  {\doibase 10.1103/PhysRevX.13.031037} {\bibfield  {journal} {\bibinfo
  {journal} {Phys. Rev. X}\ }\textbf {\bibinfo {volume} {13}},\ \bibinfo
  {pages} {031037} (\bibinfo {year} {2023})}\BibitemShut {NoStop}%
\bibitem [{\citenamefont {Park}\ \emph {et~al.}(2023)\citenamefont {Park},
  \citenamefont {Cai}, \citenamefont {Anderson}, \citenamefont {Zhang},
  \citenamefont {Zhu}, \citenamefont {Liu}, \citenamefont {Wang}, \citenamefont
  {Holtzmann}, \citenamefont {Hu}, \citenamefont {Liu}, \citenamefont
  {Taniguchi}, \citenamefont {Watanabe}, \citenamefont {Chu}, \citenamefont
  {Cao}, \citenamefont {Fu}, \citenamefont {Yao}, \citenamefont {Chang},
  \citenamefont {Cobden}, \citenamefont {Xiao},\ and\ \citenamefont
  {Xu}}]{park2023}%
  \BibitemOpen
  \bibfield  {author} {\bibinfo {author} {\bibfnamefont {Heonjoon}\
  \bibnamefont {Park}}, \bibinfo {author} {\bibfnamefont {Jiaqi}\ \bibnamefont
  {Cai}}, \bibinfo {author} {\bibfnamefont {Eric}\ \bibnamefont {Anderson}},
  \bibinfo {author} {\bibfnamefont {Yinong}\ \bibnamefont {Zhang}}, \bibinfo
  {author} {\bibfnamefont {Jiayi}\ \bibnamefont {Zhu}}, \bibinfo {author}
  {\bibfnamefont {Xiaoyu}\ \bibnamefont {Liu}}, \bibinfo {author}
  {\bibfnamefont {Chong}\ \bibnamefont {Wang}}, \bibinfo {author}
  {\bibfnamefont {William}\ \bibnamefont {Holtzmann}}, \bibinfo {author}
  {\bibfnamefont {Chaowei}\ \bibnamefont {Hu}}, \bibinfo {author}
  {\bibfnamefont {Zhaoyu}\ \bibnamefont {Liu}}, \bibinfo {author}
  {\bibfnamefont {Takashi}\ \bibnamefont {Taniguchi}}, \bibinfo {author}
  {\bibfnamefont {Kenji}\ \bibnamefont {Watanabe}}, \bibinfo {author}
  {\bibfnamefont {Jiun-Haw}\ \bibnamefont {Chu}}, \bibinfo {author}
  {\bibfnamefont {Ting}\ \bibnamefont {Cao}}, \bibinfo {author} {\bibfnamefont
  {Liang}\ \bibnamefont {Fu}}, \bibinfo {author} {\bibfnamefont {Wang}\
  \bibnamefont {Yao}}, \bibinfo {author} {\bibfnamefont {Cui-Zu}\ \bibnamefont
  {Chang}}, \bibinfo {author} {\bibfnamefont {David}\ \bibnamefont {Cobden}},
  \bibinfo {author} {\bibfnamefont {Di}~\bibnamefont {Xiao}}, \ and\ \bibinfo
  {author} {\bibfnamefont {Xiaodong}\ \bibnamefont {Xu}},\ }\bibfield  {title}
  {\enquote {\bibinfo {title} {Observation of fractionally quantized anomalous
  hall effect},}\ }\href {\doibase 10.1038/s41586-023-06536-0} {\bibfield
  {journal} {\bibinfo  {journal} {Nature}\ }\textbf {\bibinfo {volume} {622}},\
  \bibinfo {pages} {74--79} (\bibinfo {year} {2023})}\BibitemShut {NoStop}%
\bibitem [{\citenamefont {Lu}\ \emph {et~al.}(2024)\citenamefont {Lu},
  \citenamefont {Han}, \citenamefont {Yao}, \citenamefont {Reddy},
  \citenamefont {Yang}, \citenamefont {Seo}, \citenamefont {Watanabe},
  \citenamefont {Taniguchi}, \citenamefont {Fu},\ and\ \citenamefont
  {Ju}}]{lu2024}%
  \BibitemOpen
  \bibfield  {author} {\bibinfo {author} {\bibfnamefont {Zhengguang}\
  \bibnamefont {Lu}}, \bibinfo {author} {\bibfnamefont {Tonghang}\ \bibnamefont
  {Han}}, \bibinfo {author} {\bibfnamefont {Yuxuan}\ \bibnamefont {Yao}},
  \bibinfo {author} {\bibfnamefont {Aidan~P.}\ \bibnamefont {Reddy}}, \bibinfo
  {author} {\bibfnamefont {Jixiang}\ \bibnamefont {Yang}}, \bibinfo {author}
  {\bibfnamefont {Junseok}\ \bibnamefont {Seo}}, \bibinfo {author}
  {\bibfnamefont {Kenji}\ \bibnamefont {Watanabe}}, \bibinfo {author}
  {\bibfnamefont {Takashi}\ \bibnamefont {Taniguchi}}, \bibinfo {author}
  {\bibfnamefont {Liang}\ \bibnamefont {Fu}}, \ and\ \bibinfo {author}
  {\bibfnamefont {Long}\ \bibnamefont {Ju}},\ }\bibfield  {title} {\enquote
  {\bibinfo {title} {Fractional quantum anomalous hall effect in multilayer
  graphene},}\ }\href {\doibase 10.1038/s41586-023-07010-7} {\bibfield
  {journal} {\bibinfo  {journal} {Nature}\ }\textbf {\bibinfo {volume} {626}},\
  \bibinfo {pages} {759--764} (\bibinfo {year} {2024})}\BibitemShut {NoStop}%
\bibitem [{\citenamefont {He}\ \emph {et~al.}(2020)\citenamefont {He},
  \citenamefont {Goldhaber-Gordon},\ and\ \citenamefont {Law}}]{he_2020}%
  \BibitemOpen
  \bibfield  {author} {\bibinfo {author} {\bibfnamefont {Wen-Yu}\ \bibnamefont
  {He}}, \bibinfo {author} {\bibfnamefont {David}\ \bibnamefont
  {Goldhaber-Gordon}}, \ and\ \bibinfo {author} {\bibfnamefont {K.~T.}\
  \bibnamefont {Law}},\ }\bibfield  {title} {\enquote {\bibinfo {title} {Giant
  orbital magnetoelectric effect and current-induced magnetization switching in
  twisted bilayer graphene},}\ }\href {\doibase 10.1038/s41467-020-15473-9}
  {\bibfield  {journal} {\bibinfo  {journal} {Nature Communications}\ }\textbf
  {\bibinfo {volume} {11}},\ \bibinfo {pages} {1650} (\bibinfo {year}
  {2020})}\BibitemShut {NoStop}%
\bibitem [{\citenamefont {Su}\ and\ \citenamefont {Lin}(2020)}]{su_2020}%
  \BibitemOpen
  \bibfield  {author} {\bibinfo {author} {\bibfnamefont {Ying}\ \bibnamefont
  {Su}}\ and\ \bibinfo {author} {\bibfnamefont {Shi-Zeng}\ \bibnamefont
  {Lin}},\ }\bibfield  {title} {\enquote {\bibinfo {title} {Current-induced
  reversal of anomalous hall conductance in twisted bilayer graphene},}\ }\href
  {\doibase 10.1103/PhysRevLett.125.226401} {\bibfield  {journal} {\bibinfo
  {journal} {Phys. Rev. Lett.}\ }\textbf {\bibinfo {volume} {125}},\ \bibinfo
  {pages} {226401} (\bibinfo {year} {2020})}\BibitemShut {NoStop}%
\bibitem [{\citenamefont {Huang}\ \emph {et~al.}(2021)\citenamefont {Huang},
  \citenamefont {Wei},\ and\ \citenamefont {MacDonald}}]{huang_2021}%
  \BibitemOpen
  \bibfield  {author} {\bibinfo {author} {\bibfnamefont {Chunli}\ \bibnamefont
  {Huang}}, \bibinfo {author} {\bibfnamefont {Nemin}\ \bibnamefont {Wei}}, \
  and\ \bibinfo {author} {\bibfnamefont {Allan~H.}\ \bibnamefont {MacDonald}},\
  }\bibfield  {title} {\enquote {\bibinfo {title} {Current-driven magnetization
  reversal in orbital chern insulators},}\ }\href {\doibase
  10.1103/PhysRevLett.126.056801} {\bibfield  {journal} {\bibinfo  {journal}
  {Phys. Rev. Lett.}\ }\textbf {\bibinfo {volume} {126}},\ \bibinfo {pages}
  {056801} (\bibinfo {year} {2021})}\BibitemShut {NoStop}%
\bibitem [{\citenamefont {Lee}\ \emph {et~al.}(2019)\citenamefont {Lee},
  \citenamefont {Khalaf}, \citenamefont {Liu}, \citenamefont {Liu},
  \citenamefont {Hao}, \citenamefont {Kim},\ and\ \citenamefont
  {Vishwanath}}]{lee_2020}%
  \BibitemOpen
  \bibfield  {author} {\bibinfo {author} {\bibfnamefont {Jong~Yeon}\
  \bibnamefont {Lee}}, \bibinfo {author} {\bibfnamefont {Eslam}\ \bibnamefont
  {Khalaf}}, \bibinfo {author} {\bibfnamefont {Shang}\ \bibnamefont {Liu}},
  \bibinfo {author} {\bibfnamefont {Xiaomeng}\ \bibnamefont {Liu}}, \bibinfo
  {author} {\bibfnamefont {Zeyu}\ \bibnamefont {Hao}}, \bibinfo {author}
  {\bibfnamefont {Philip}\ \bibnamefont {Kim}}, \ and\ \bibinfo {author}
  {\bibfnamefont {Ashvin}\ \bibnamefont {Vishwanath}},\ }\bibfield  {title}
  {\enquote {\bibinfo {title} {Theory of correlated insulating behaviour and
  spin-triplet superconductivity in twisted double bilayer graphene},}\ }\href
  {\doibase 10.1038/s41467-019-12981-1} {\bibfield  {journal} {\bibinfo
  {journal} {Nature Communications}\ }\textbf {\bibinfo {volume} {10}},\
  \bibinfo {pages} {5333} (\bibinfo {year} {2019})}\BibitemShut {NoStop}%
\bibitem [{\citenamefont {Zhang}\ \emph {et~al.}(2022)\citenamefont {Zhang},
  \citenamefont {Dai},\ and\ \citenamefont {Liu}}]{zhangs_2020}%
  \BibitemOpen
  \bibfield  {author} {\bibinfo {author} {\bibfnamefont {Shihao}\ \bibnamefont
  {Zhang}}, \bibinfo {author} {\bibfnamefont {Xi}~\bibnamefont {Dai}}, \ and\
  \bibinfo {author} {\bibfnamefont {Jianpeng}\ \bibnamefont {Liu}},\ }\bibfield
   {title} {\enquote {\bibinfo {title} {Spin-polarized nematic order, quantum
  valley hall states, and field-tunable topological transitions in twisted
  multilayer graphene systems},}\ }\href {\doibase
  10.1103/PhysRevLett.128.026403} {\bibfield  {journal} {\bibinfo  {journal}
  {Phys. Rev. Lett.}\ }\textbf {\bibinfo {volume} {128}},\ \bibinfo {pages}
  {026403} (\bibinfo {year} {2022})}\BibitemShut {NoStop}%
\bibitem [{\citenamefont {Tschirhart}\ \emph {et~al.}(2021)\citenamefont
  {Tschirhart}, \citenamefont {Serlin}, \citenamefont {Polshyn}, \citenamefont
  {Shragai}, \citenamefont {Xia}, \citenamefont {Zhu}, \citenamefont {Zhang},
  \citenamefont {Watanabe}, \citenamefont {Taniguchi}, \citenamefont {Huber},\
  and\ \citenamefont {Young}}]{tsch2021}%
  \BibitemOpen
  \bibfield  {author} {\bibinfo {author} {\bibfnamefont {C.~L.}\ \bibnamefont
  {Tschirhart}}, \bibinfo {author} {\bibfnamefont {M.}~\bibnamefont {Serlin}},
  \bibinfo {author} {\bibfnamefont {H.}~\bibnamefont {Polshyn}}, \bibinfo
  {author} {\bibfnamefont {A.}~\bibnamefont {Shragai}}, \bibinfo {author}
  {\bibfnamefont {Z.}~\bibnamefont {Xia}}, \bibinfo {author} {\bibfnamefont
  {J.}~\bibnamefont {Zhu}}, \bibinfo {author} {\bibfnamefont {Y.}~\bibnamefont
  {Zhang}}, \bibinfo {author} {\bibfnamefont {K.}~\bibnamefont {Watanabe}},
  \bibinfo {author} {\bibfnamefont {T.}~\bibnamefont {Taniguchi}}, \bibinfo
  {author} {\bibfnamefont {M.~E.}\ \bibnamefont {Huber}}, \ and\ \bibinfo
  {author} {\bibfnamefont {A.~F.}\ \bibnamefont {Young}},\ }\bibfield  {title}
  {\enquote {\bibinfo {title} {Imaging orbital ferromagnetism in a moiré chern
  insulator},}\ }\href {\doibase 10.1126/science.abd3190} {\bibfield  {journal}
  {\bibinfo  {journal} {Science}\ }\textbf {\bibinfo {volume} {372}},\ \bibinfo
  {pages} {1323--1327} (\bibinfo {year} {2021})}\BibitemShut {NoStop}%
\bibitem [{\citenamefont {Grover}\ \emph {et~al.}(2022)\citenamefont {Grover},
  \citenamefont {Bocarsly}, \citenamefont {Uri}, \citenamefont {Stepanov},
  \citenamefont {Di~Battista}, \citenamefont {Roy}, \citenamefont {Xiao},
  \citenamefont {Meltzer}, \citenamefont {Myasoedov}, \citenamefont {Pareek},
  \citenamefont {Watanabe}, \citenamefont {Taniguchi}, \citenamefont {Yan},
  \citenamefont {Stern}, \citenamefont {Berg}, \citenamefont {Efetov},\ and\
  \citenamefont {Zeldov}}]{grover2022}%
  \BibitemOpen
  \bibfield  {author} {\bibinfo {author} {\bibfnamefont {Sameer}\ \bibnamefont
  {Grover}}, \bibinfo {author} {\bibfnamefont {Matan}\ \bibnamefont
  {Bocarsly}}, \bibinfo {author} {\bibfnamefont {Aviram}\ \bibnamefont {Uri}},
  \bibinfo {author} {\bibfnamefont {Petr}\ \bibnamefont {Stepanov}}, \bibinfo
  {author} {\bibfnamefont {Giorgio}\ \bibnamefont {Di~Battista}}, \bibinfo
  {author} {\bibfnamefont {Indranil}\ \bibnamefont {Roy}}, \bibinfo {author}
  {\bibfnamefont {Jiewen}\ \bibnamefont {Xiao}}, \bibinfo {author}
  {\bibfnamefont {Alexander~Y.}\ \bibnamefont {Meltzer}}, \bibinfo {author}
  {\bibfnamefont {Yuri}\ \bibnamefont {Myasoedov}}, \bibinfo {author}
  {\bibfnamefont {Keshav}\ \bibnamefont {Pareek}}, \bibinfo {author}
  {\bibfnamefont {Kenji}\ \bibnamefont {Watanabe}}, \bibinfo {author}
  {\bibfnamefont {Takashi}\ \bibnamefont {Taniguchi}}, \bibinfo {author}
  {\bibfnamefont {Binghai}\ \bibnamefont {Yan}}, \bibinfo {author}
  {\bibfnamefont {Ady}\ \bibnamefont {Stern}}, \bibinfo {author} {\bibfnamefont
  {Erez}\ \bibnamefont {Berg}}, \bibinfo {author} {\bibfnamefont {Dmitri~K.}\
  \bibnamefont {Efetov}}, \ and\ \bibinfo {author} {\bibfnamefont {Eli}\
  \bibnamefont {Zeldov}},\ }\bibfield  {title} {\enquote {\bibinfo {title}
  {Chern mosaic and berry-curvature magnetism in magic-angle graphene},}\
  }\href {\doibase 10.1038/s41567-022-01635-7} {\bibfield  {journal} {\bibinfo
  {journal} {Nature Physics}\ }\textbf {\bibinfo {volume} {18}},\ \bibinfo
  {pages} {885--892} (\bibinfo {year} {2022})}\BibitemShut {NoStop}%
\bibitem [{\citenamefont {Liu}\ \emph {et~al.}(2019{\natexlab{b}})\citenamefont
  {Liu}, \citenamefont {Ma}, \citenamefont {Gao},\ and\ \citenamefont
  {Dai}}]{liu_2019_1}%
  \BibitemOpen
  \bibfield  {author} {\bibinfo {author} {\bibfnamefont {Jianpeng}\
  \bibnamefont {Liu}}, \bibinfo {author} {\bibfnamefont {Zhen}\ \bibnamefont
  {Ma}}, \bibinfo {author} {\bibfnamefont {Jinhua}\ \bibnamefont {Gao}}, \ and\
  \bibinfo {author} {\bibfnamefont {Xi}~\bibnamefont {Dai}},\ }\bibfield
  {title} {\enquote {\bibinfo {title} {Quantum valley hall effect, orbital
  magnetism, and anomalous hall effect in twisted multilayer graphene
  systems},}\ }\href {\doibase 10.1103/PhysRevX.9.031021} {\bibfield  {journal}
  {\bibinfo  {journal} {Phys. Rev. X}\ }\textbf {\bibinfo {volume} {9}},\
  \bibinfo {pages} {031021} (\bibinfo {year} {2019}{\natexlab{b}})}\BibitemShut
  {NoStop}%
\bibitem [{\citenamefont {Liu}\ and\ \citenamefont
  {Dai}(2021{\natexlab{b}})}]{review_tbg}%
  \BibitemOpen
  \bibfield  {author} {\bibinfo {author} {\bibfnamefont {Jianpeng}\
  \bibnamefont {Liu}}\ and\ \bibinfo {author} {\bibfnamefont {Xi}~\bibnamefont
  {Dai}},\ }\bibfield  {title} {\enquote {\bibinfo {title} {Orbital magnetic
  states in moir{\'e}graphene systems},}\ }\href {\doibase
  10.1038/s42254-021-00297-3} {\bibfield  {journal} {\bibinfo  {journal}
  {Nature Reviews Physics}\ }\textbf {\bibinfo {volume} {3}},\ \bibinfo {pages}
  {367--382} (\bibinfo {year} {2021}{\natexlab{b}})}\BibitemShut {NoStop}%
\bibitem [{\citenamefont {Zhu}\ \emph {et~al.}(2020)\citenamefont {Zhu},
  \citenamefont {Su},\ and\ \citenamefont {MacDonald}}]{zhu_2020}%
  \BibitemOpen
  \bibfield  {author} {\bibinfo {author} {\bibfnamefont {Jihang}\ \bibnamefont
  {Zhu}}, \bibinfo {author} {\bibfnamefont {Jung-Jung}\ \bibnamefont {Su}}, \
  and\ \bibinfo {author} {\bibfnamefont {A.~H.}\ \bibnamefont {MacDonald}},\
  }\bibfield  {title} {\enquote {\bibinfo {title} {Voltage-controlled magnetic
  reversal in orbital chern insulators},}\ }\href {\doibase
  10.1103/PhysRevLett.125.227702} {\bibfield  {journal} {\bibinfo  {journal}
  {Phys. Rev. Lett.}\ }\textbf {\bibinfo {volume} {125}},\ \bibinfo {pages}
  {227702} (\bibinfo {year} {2020})}\BibitemShut {NoStop}%
\bibitem [{\citenamefont {Kang}\ \emph {et~al.}(2026)\citenamefont {Kang},
  \citenamefont {Wang},\ and\ \citenamefont {Vafek}}]{kang_2025}%
  \BibitemOpen
  \bibfield  {author} {\bibinfo {author} {\bibfnamefont {Jian}\ \bibnamefont
  {Kang}}, \bibinfo {author} {\bibfnamefont {Minxuan}\ \bibnamefont {Wang}}, \
  and\ \bibinfo {author} {\bibfnamefont {Oskar}\ \bibnamefont {Vafek}},\
  }\bibfield  {title} {\enquote {\bibinfo {title} {Orbital magnetization and
  magnetic susceptibility of interacting electrons},}\ }\href {\doibase
  10.1103/9l9c-mp3t} {\bibfield  {journal} {\bibinfo  {journal} {Phys. Rev.
  Lett.}\ }\textbf {\bibinfo {volume} {137}},\ \bibinfo {pages} {106703}
  (\bibinfo {year} {2026})}\BibitemShut {NoStop}%
\bibitem [{\citenamefont {Zhu}\ and\ \citenamefont {Huang}(2026)}]{zhu_2025}%
  \BibitemOpen
  \bibfield  {author} {\bibinfo {author} {\bibfnamefont {Jihang}\ \bibnamefont
  {Zhu}}\ and\ \bibinfo {author} {\bibfnamefont {Chunli}\ \bibnamefont
  {Huang}},\ }\bibfield  {title} {\enquote {\bibinfo {title} {Orbital
  magnetization and streda formula of interacting electrons in the mean-field
  approximation},}\ }\href {\doibase 10.1103/w49b-25pd} {\bibfield  {journal}
  {\bibinfo  {journal} {Phys. Rev. B}\ }\textbf {\bibinfo {volume} {113}},\
  \bibinfo {pages} {205114} (\bibinfo {year} {2026})}\BibitemShut {NoStop}%
\bibitem [{\citenamefont {Liu}\ \emph {et~al.}(2026)\citenamefont {Liu},
  \citenamefont {Wang}, \citenamefont {Chen}, \citenamefont {Zhang},
  \citenamefont {Cao},\ and\ \citenamefont {Xiao}}]{liu_2025}%
  \BibitemOpen
  \bibfield  {author} {\bibinfo {author} {\bibfnamefont {Xiaoyu}\ \bibnamefont
  {Liu}}, \bibinfo {author} {\bibfnamefont {Chong}\ \bibnamefont {Wang}},
  \bibinfo {author} {\bibfnamefont {Haoran}\ \bibnamefont {Chen}}, \bibinfo
  {author} {\bibfnamefont {Xiao-Wei}\ \bibnamefont {Zhang}}, \bibinfo {author}
  {\bibfnamefont {Ting}\ \bibnamefont {Cao}}, \ and\ \bibinfo {author}
  {\bibfnamefont {Di}~\bibnamefont {Xiao}},\ }\bibfield  {title} {\enquote
  {\bibinfo {title} {Orbital magnetization of correlated states in twisted
  bilayer transition metal dichalcogenides},}\ }\href {\doibase
  10.1103/k46f-8m8t} {\bibfield  {journal} {\bibinfo  {journal} {Phys. Rev.
  Lett.}\ }\textbf {\bibinfo {volume} {136}},\ \bibinfo {pages} {166606}
  (\bibinfo {year} {2026})}\BibitemShut {NoStop}%
\bibitem [{\citenamefont {{Ye}}(2026)}]{ye_2026}%
  \BibitemOpen
  \bibfield  {author} {\bibinfo {author} {\bibfnamefont {Mengxing}\
  \bibnamefont {{Ye}}},\ }\bibfield  {title} {\enquote {\bibinfo {title} {{A
  Quantum Many-Body Approach for Orbital Magnetism in Correlated Multiband
  Electron Systems}},}\ }\href {\doibase 10.48550/arXiv.2601.14372} {\bibfield
  {journal} {\bibinfo  {journal} {arXiv e-prints}\ ,\ \bibinfo {pages}
  {arXiv:2601.14372}} (\bibinfo {year} {2026})}\BibitemShut {NoStop}%
\bibitem [{\citenamefont {{Chen}}\ and\ \citenamefont
  {{Song}}(2026)}]{song_2026}%
  \BibitemOpen
  \bibfield  {author} {\bibinfo {author} {\bibfnamefont {Xi}~\bibnamefont
  {{Chen}}}\ and\ \bibinfo {author} {\bibfnamefont {Zhi-Da}\ \bibnamefont
  {{Song}}},\ }\bibfield  {title} {\enquote {\bibinfo {title} {{Orbital
  Magnetization of Interacting Electrons}},}\ }\href {\doibase
  10.48550/arXiv.2602.02478} {\bibfield  {journal} {\bibinfo  {journal} {arXiv
  e-prints}\ ,\ \bibinfo {eid} {arXiv:2602.02478}} (\bibinfo {year}
  {2026})}\BibitemShut {NoStop}%
\bibitem [{\citenamefont {Ceresoli}\ \emph {et~al.}(2006)\citenamefont
  {Ceresoli}, \citenamefont {Thonhauser}, \citenamefont {Vanderbilt},\ and\
  \citenamefont {Resta}}]{cere_2006}%
  \BibitemOpen
  \bibfield  {author} {\bibinfo {author} {\bibfnamefont {Davide}\ \bibnamefont
  {Ceresoli}}, \bibinfo {author} {\bibfnamefont {T.}~\bibnamefont
  {Thonhauser}}, \bibinfo {author} {\bibfnamefont {David}\ \bibnamefont
  {Vanderbilt}}, \ and\ \bibinfo {author} {\bibfnamefont {R.}~\bibnamefont
  {Resta}},\ }\bibfield  {title} {\enquote {\bibinfo {title} {Orbital
  magnetization in crystalline solids: Multi-band insulators, chern insulators,
  and metals},}\ }\href {\doibase 10.1103/PhysRevB.74.024408} {\bibfield
  {journal} {\bibinfo  {journal} {Phys. Rev. B}\ }\textbf {\bibinfo {volume}
  {74}},\ \bibinfo {pages} {024408} (\bibinfo {year} {2006})}\BibitemShut
  {NoStop}%
\bibitem [{\citenamefont {Xia}\ \emph {et~al.}(2026)\citenamefont {Xia},
  \citenamefont {Han}, \citenamefont {Zhu}, \citenamefont {Zhang},
  \citenamefont {Kn{\"u}ppel}, \citenamefont {Watanabe}, \citenamefont
  {Taniguchi}, \citenamefont {Mak},\ and\ \citenamefont {Shan}}]{xia_2026}%
  \BibitemOpen
  \bibfield  {author} {\bibinfo {author} {\bibfnamefont {Yiyu}\ \bibnamefont
  {Xia}}, \bibinfo {author} {\bibfnamefont {Zhongdong}\ \bibnamefont {Han}},
  \bibinfo {author} {\bibfnamefont {Jiacheng}\ \bibnamefont {Zhu}}, \bibinfo
  {author} {\bibfnamefont {Yichi}\ \bibnamefont {Zhang}}, \bibinfo {author}
  {\bibfnamefont {Patrick}\ \bibnamefont {Kn{\"u}ppel}}, \bibinfo {author}
  {\bibfnamefont {Kenji}\ \bibnamefont {Watanabe}}, \bibinfo {author}
  {\bibfnamefont {Takashi}\ \bibnamefont {Taniguchi}}, \bibinfo {author}
  {\bibfnamefont {Kin~Fai}\ \bibnamefont {Mak}}, \ and\ \bibinfo {author}
  {\bibfnamefont {Jie}\ \bibnamefont {Shan}},\ }\bibfield  {title} {\enquote
  {\bibinfo {title} {Bandwidth-tuned mott transition and superconductivity in
  moir{\'e}wse2},}\ }\href {\doibase 10.1038/s41586-025-10049-3} {\bibfield
  {journal} {\bibinfo  {journal} {Nature}\ }\textbf {\bibinfo {volume} {650}},\
  \bibinfo {pages} {585--591} (\bibinfo {year} {2026})}\BibitemShut {NoStop}%
\bibitem [{\citenamefont {Lopes~dos Santos}\ \emph {et~al.}(2012)\citenamefont
  {Lopes~dos Santos}, \citenamefont {Peres},\ and\ \citenamefont
  {Castro~Neto}}]{lop_2012}%
  \BibitemOpen
  \bibfield  {author} {\bibinfo {author} {\bibfnamefont {J.~M.~B.}\
  \bibnamefont {Lopes~dos Santos}}, \bibinfo {author} {\bibfnamefont
  {N.~M.~R.}\ \bibnamefont {Peres}}, \ and\ \bibinfo {author} {\bibfnamefont
  {A.~H.}\ \bibnamefont {Castro~Neto}},\ }\bibfield  {title} {\enquote
  {\bibinfo {title} {Continuum model of the twisted graphene bilayer},}\ }\href
  {\doibase 10.1103/PhysRevB.86.155449} {\bibfield  {journal} {\bibinfo
  {journal} {Phys. Rev. B}\ }\textbf {\bibinfo {volume} {86}},\ \bibinfo
  {pages} {155449} (\bibinfo {year} {2012})}\BibitemShut {NoStop}%
\bibitem [{\citenamefont {Nam}\ and\ \citenamefont {Koshino}(2017)}]{nam_2017}%
  \BibitemOpen
  \bibfield  {author} {\bibinfo {author} {\bibfnamefont {Nguyen N.~T.}\
  \bibnamefont {Nam}}\ and\ \bibinfo {author} {\bibfnamefont {Mikito}\
  \bibnamefont {Koshino}},\ }\bibfield  {title} {\enquote {\bibinfo {title}
  {Lattice relaxation and energy band modulation in twisted bilayer
  graphene},}\ }\href {\doibase 10.1103/PhysRevB.96.075311} {\bibfield
  {journal} {\bibinfo  {journal} {Phys. Rev. B}\ }\textbf {\bibinfo {volume}
  {96}},\ \bibinfo {pages} {075311} (\bibinfo {year} {2017})}\BibitemShut
  {NoStop}%
\bibitem [{\citenamefont {Kang}\ and\ \citenamefont {Vafek}(2023)}]{kang_2023}%
  \BibitemOpen
  \bibfield  {author} {\bibinfo {author} {\bibfnamefont {Jian}\ \bibnamefont
  {Kang}}\ and\ \bibinfo {author} {\bibfnamefont {Oskar}\ \bibnamefont
  {Vafek}},\ }\bibfield  {title} {\enquote {\bibinfo {title} {Pseudomagnetic
  fields, particle-hole asymmetry, and microscopic effective continuum
  hamiltonians of twisted bilayer graphene},}\ }\href {\doibase
  10.1103/PhysRevB.107.075408} {\bibfield  {journal} {\bibinfo  {journal}
  {Phys. Rev. B}\ }\textbf {\bibinfo {volume} {107}},\ \bibinfo {pages}
  {075408} (\bibinfo {year} {2023})}\BibitemShut {NoStop}%
\bibitem [{\citenamefont {Kang}\ and\ \citenamefont
  {Vafek}(2025)}]{kang_2025_1}%
  \BibitemOpen
  \bibfield  {author} {\bibinfo {author} {\bibfnamefont {Jian}\ \bibnamefont
  {Kang}}\ and\ \bibinfo {author} {\bibfnamefont {Oskar}\ \bibnamefont
  {Vafek}},\ }\bibfield  {title} {\enquote {\bibinfo {title} {Analytical
  solution for the relaxed atomic configuration of twisted bilayer graphene
  including heterostrain},}\ }\href {\doibase 10.1103/s3s7-513d} {\bibfield
  {journal} {\bibinfo  {journal} {Phys. Rev. B}\ }\textbf {\bibinfo {volume}
  {112}},\ \bibinfo {pages} {125138} (\bibinfo {year} {2025})}\BibitemShut
  {NoStop}%
\bibitem [{\citenamefont {Vafek}\ and\ \citenamefont
  {Kang}(2020)}]{kang_2020_1}%
  \BibitemOpen
  \bibfield  {author} {\bibinfo {author} {\bibfnamefont {Oskar}\ \bibnamefont
  {Vafek}}\ and\ \bibinfo {author} {\bibfnamefont {Jian}\ \bibnamefont
  {Kang}},\ }\bibfield  {title} {\enquote {\bibinfo {title} {Renormalization
  group study of hidden symmetry in twisted bilayer graphene with coulomb
  interactions},}\ }\href {\doibase 10.1103/PhysRevLett.125.257602} {\bibfield
  {journal} {\bibinfo  {journal} {Phys. Rev. Lett.}\ }\textbf {\bibinfo
  {volume} {125}},\ \bibinfo {pages} {257602} (\bibinfo {year}
  {2020})}\BibitemShut {NoStop}%
\bibitem [{sm()}]{sm}%
  \BibitemOpen
  \href@noop {} {}\bibinfo {note} {See Supplemental Material for more detailed
  discussions.}\BibitemShut {Stop}%
\bibitem [{\citenamefont {Bernevig}\ \emph
  {et~al.}(2021{\natexlab{b}})\citenamefont {Bernevig}, \citenamefont {Song},
  \citenamefont {Regnault},\ and\ \citenamefont {Lian}}]{bern_2021}%
  \BibitemOpen
  \bibfield  {author} {\bibinfo {author} {\bibfnamefont {B.~Andrei}\
  \bibnamefont {Bernevig}}, \bibinfo {author} {\bibfnamefont {Zhi-Da}\
  \bibnamefont {Song}}, \bibinfo {author} {\bibfnamefont {Nicolas}\
  \bibnamefont {Regnault}}, \ and\ \bibinfo {author} {\bibfnamefont {Biao}\
  \bibnamefont {Lian}},\ }\bibfield  {title} {\enquote {\bibinfo {title}
  {Twisted bilayer graphene. i. matrix elements, approximations, perturbation
  theory, and a $k\ifmmode\cdot\else\textperiodcentered\fi{}p$ two-band
  model},}\ }\href {\doibase 10.1103/PhysRevB.103.205411} {\bibfield  {journal}
  {\bibinfo  {journal} {Phys. Rev. B}\ }\textbf {\bibinfo {volume} {103}},\
  \bibinfo {pages} {205411} (\bibinfo {year} {2021}{\natexlab{b}})}\BibitemShut
  {NoStop}%
\bibitem [{\citenamefont {Fukui}\ \emph {et~al.}(2005)\citenamefont {Fukui},
  \citenamefont {Hatsugai},\ and\ \citenamefont {Suzuki}}]{fukui_2005}%
  \BibitemOpen
  \bibfield  {author} {\bibinfo {author} {\bibfnamefont {Takahiro}\
  \bibnamefont {Fukui}}, \bibinfo {author} {\bibfnamefont {Yasuhiro}\
  \bibnamefont {Hatsugai}}, \ and\ \bibinfo {author} {\bibfnamefont {Hiroshi}\
  \bibnamefont {Suzuki}},\ }\bibfield  {title} {\enquote {\bibinfo {title}
  {Chern numbers in discretized brillouin zone: Efficient method of computing
  (spin) hall conductances},}\ }\href {\doibase 10.1143/JPSJ.74.1674}
  {\bibfield  {journal} {\bibinfo  {journal} {Journal of the Physical Society
  of Japan}\ }\textbf {\bibinfo {volume} {74}},\ \bibinfo {pages} {1674--1677}
  (\bibinfo {year} {2005})}\BibitemShut {NoStop}%
\bibitem [{\citenamefont {Souza}\ and\ \citenamefont
  {Vanderbilt}(2008)}]{souza_2008}%
  \BibitemOpen
  \bibfield  {author} {\bibinfo {author} {\bibfnamefont {Ivo}\ \bibnamefont
  {Souza}}\ and\ \bibinfo {author} {\bibfnamefont {David}\ \bibnamefont
  {Vanderbilt}},\ }\bibfield  {title} {\enquote {\bibinfo {title} {Dichroic
  $f$-sum rule and the orbital magnetization of crystals},}\ }\href {\doibase
  10.1103/PhysRevB.77.054438} {\bibfield  {journal} {\bibinfo  {journal} {Phys.
  Rev. B}\ }\textbf {\bibinfo {volume} {77}},\ \bibinfo {pages} {054438}
  (\bibinfo {year} {2008})}\BibitemShut {NoStop}%
\bibitem [{\citenamefont {Resta}(2020)}]{resta_2020}%
  \BibitemOpen
  \bibfield  {author} {\bibinfo {author} {\bibfnamefont {Raffaele}\
  \bibnamefont {Resta}},\ }\bibfield  {title} {\enquote {\bibinfo {title}
  {Magnetic circular dichroism versus orbital magnetization},}\ }\href
  {\doibase 10.1103/PhysRevResearch.2.023139} {\bibfield  {journal} {\bibinfo
  {journal} {Phys. Rev. Res.}\ }\textbf {\bibinfo {volume} {2}},\ \bibinfo
  {pages} {023139} (\bibinfo {year} {2020})}\BibitemShut {NoStop}%
\bibitem [{\citenamefont {{Ji}}\ \emph {et~al.}(2025)\citenamefont {{Ji}},
  \citenamefont {{Palomino}}, \citenamefont {{Goldman}}, \citenamefont
  {{Ozawa}}, \citenamefont {{Riseborough}}, \citenamefont {{Wang}},\ and\
  \citenamefont {{Mera}}}]{wang_2025}%
  \BibitemOpen
  \bibfield  {author} {\bibinfo {author} {\bibfnamefont {Guangyue}\
  \bibnamefont {{Ji}}}, \bibinfo {author} {\bibfnamefont {David~E.}\
  \bibnamefont {{Palomino}}}, \bibinfo {author} {\bibfnamefont {Nathan}\
  \bibnamefont {{Goldman}}}, \bibinfo {author} {\bibfnamefont {Tomoki}\
  \bibnamefont {{Ozawa}}}, \bibinfo {author} {\bibfnamefont {Peter}\
  \bibnamefont {{Riseborough}}}, \bibinfo {author} {\bibfnamefont {Jie}\
  \bibnamefont {{Wang}}}, \ and\ \bibinfo {author} {\bibfnamefont {Bruno}\
  \bibnamefont {{Mera}}},\ }\bibfield  {title} {\enquote {\bibinfo {title}
  {{Density Matrix Geometry and Sum Rules}},}\ }\href {\doibase
  10.48550/arXiv.2507.14028} {\bibfield  {journal} {\bibinfo  {journal} {arXiv
  e-prints}\ ,\ \bibinfo {pages} {arXiv:2507.14028}} (\bibinfo {year}
  {2025})}\BibitemShut {NoStop}%
\bibitem [{\citenamefont {Xiao}\ \emph {et~al.}(2010)\citenamefont {Xiao},
  \citenamefont {Chang},\ and\ \citenamefont {Niu}}]{xiao_2010}%
  \BibitemOpen
  \bibfield  {author} {\bibinfo {author} {\bibfnamefont {Di}~\bibnamefont
  {Xiao}}, \bibinfo {author} {\bibfnamefont {Ming-Che}\ \bibnamefont {Chang}},
  \ and\ \bibinfo {author} {\bibfnamefont {Qian}\ \bibnamefont {Niu}},\
  }\bibfield  {title} {\enquote {\bibinfo {title} {Berry phase effects on
  electronic properties},}\ }\href {\doibase 10.1103/RevModPhys.82.1959}
  {\bibfield  {journal} {\bibinfo  {journal} {Rev. Mod. Phys.}\ }\textbf
  {\bibinfo {volume} {82}},\ \bibinfo {pages} {1959--2007} (\bibinfo {year}
  {2010})}\BibitemShut {NoStop}%
\end{thebibliography}%


\begin{thebibliography}{5}%

\providecommand \@ifxundefined [1]{%
 \@ifx{#1\undefined}
}%
\providecommand \@ifnum [1]{%
 \ifnum #1\expandafter \@firstoftwo
 \else \expandafter \@secondoftwo
 \fi
}%
\providecommand \@ifx [1]{%
 \ifx #1\expandafter \@firstoftwo
 \else \expandafter \@secondoftwo
 \fi
}%
\providecommand \natexlab [1]{#1}%
\providecommand \enquote  [1]{``#1''}%
\providecommand \bibnamefont  [1]{#1}%
\providecommand \bibfnamefont [1]{#1}%
\providecommand \citenamefont [1]{#1}%
\providecommand \href@noop [0]{\@secondoftwo}%
\providecommand \href [0]{\begingroup \@sanitize@url \@href}%
\providecommand \@href[1]{\@@startlink{#1}\@@href}%
\providecommand \@@href[1]{\endgroup#1\@@endlink}%
\providecommand \@sanitize@url [0]{\catcode `\\12\catcode `\$12\catcode `\&12\catcode `\#12\catcode `\^12\catcode `\_12\catcode `\%12\relax}%
\providecommand \@@startlink[1]{}%
\providecommand \@@endlink[0]{}%
\providecommand \url  [0]{\begingroup\@sanitize@url \@url }%
\providecommand \@url [1]{\endgroup\@href {#1}{\urlprefix }}%
\providecommand \urlprefix  [0]{URL }%
\providecommand \Eprint [0]{\href }%
\providecommand \doibase [0]{https://doi.org/}%
\providecommand \selectlanguage [0]{\@gobble}%
\providecommand \bibinfo  [0]{\@secondoftwo}%
\providecommand \bibfield  [0]{\@secondoftwo}%
\providecommand \translation [1]{[#1]}%
\providecommand \BibitemOpen [0]{}%
\providecommand \bibitemStop [0]{}%
\providecommand \bibitemNoStop [0]{.\EOS\space}%
\providecommand \EOS [0]{\spacefactor3000\relax}%
\providecommand \BibitemShut  [1]{\csname bibitem#1\endcsname}%
\let\auto@bib@innerbib\@empty
\bibitem [{\citenamefont {Bernevig}\ \emph {et~al.}(2021)\citenamefont {Bernevig}, \citenamefont {Song}, \citenamefont {Regnault},\ and\ \citenamefont {Lian}}]{bernevig_2020s}%
  \BibitemOpen
  \bibfield  {author} {\bibinfo {author} {\bibfnamefont {B.~A.}\ \bibnamefont {Bernevig}}, \bibinfo {author} {\bibfnamefont {Z.-D.}\ \bibnamefont {Song}}, \bibinfo {author} {\bibfnamefont {N.}~\bibnamefont {Regnault}},\ and\ \bibinfo {author} {\bibfnamefont {B.}~\bibnamefont {Lian}},\ }\bibfield  {title} {\bibinfo {title} {Twisted bilayer graphene. iii. interacting hamiltonian and exact symmetries},\ }\href {https://doi.org/10.1103/PhysRevB.103.205413} {\bibfield  {journal} {\bibinfo  {journal} {Phys. Rev. B}\ }\textbf {\bibinfo {volume} {103}},\ \bibinfo {pages} {205413} (\bibinfo {year} {2021})}\BibitemShut {NoStop}%
\bibitem [{\citenamefont {Kang}\ \emph {et~al.}(2021)\citenamefont {Kang}, \citenamefont {Bernevig},\ and\ \citenamefont {Vafek}}]{kang_2021s}%
  \BibitemOpen
  \bibfield  {author} {\bibinfo {author} {\bibfnamefont {J.}~\bibnamefont {Kang}}, \bibinfo {author} {\bibfnamefont {B.~A.}\ \bibnamefont {Bernevig}},\ and\ \bibinfo {author} {\bibfnamefont {O.}~\bibnamefont {Vafek}},\ }\bibfield  {title} {\bibinfo {title} {Cascades between light and heavy fermions in the normal state of magic-angle twisted bilayer graphene},\ }\href {https://doi.org/10.1103/PhysRevLett.127.266402} {\bibfield  {journal} {\bibinfo  {journal} {Phys. Rev. Lett.}\ }\textbf {\bibinfo {volume} {127}},\ \bibinfo {pages} {266402} (\bibinfo {year} {2021})}\BibitemShut {NoStop}%
\bibitem [{\citenamefont {Zhang}\ \emph {et~al.}(2020)\citenamefont {Zhang}, \citenamefont {Jiang}, \citenamefont {Wang},\ and\ \citenamefont {Zhang}}]{zhang_2020s}%
  \BibitemOpen
  \bibfield  {author} {\bibinfo {author} {\bibfnamefont {Y.}~\bibnamefont {Zhang}}, \bibinfo {author} {\bibfnamefont {K.}~\bibnamefont {Jiang}}, \bibinfo {author} {\bibfnamefont {Z.}~\bibnamefont {Wang}},\ and\ \bibinfo {author} {\bibfnamefont {F.}~\bibnamefont {Zhang}},\ }\bibfield  {title} {\bibinfo {title} {Correlated insulating phases of twisted bilayer graphene at commensurate filling fractions: A hartree-fock study},\ }\href {https://doi.org/10.1103/PhysRevB.102.035136} {\bibfield  {journal} {\bibinfo  {journal} {Phys. Rev. B}\ }\textbf {\bibinfo {volume} {102}},\ \bibinfo {pages} {035136} (\bibinfo {year} {2020})}\BibitemShut {NoStop}%
\bibitem [{\citenamefont {Bultinck}\ \emph {et~al.}(2020)\citenamefont {Bultinck}, \citenamefont {Khalaf}, \citenamefont {Liu}, \citenamefont {Chatterjee}, \citenamefont {Vishwanath},\ and\ \citenamefont {Zaletel}}]{bul_2020s}%
  \BibitemOpen
  \bibfield  {author} {\bibinfo {author} {\bibfnamefont {N.}~\bibnamefont {Bultinck}}, \bibinfo {author} {\bibfnamefont {E.}~\bibnamefont {Khalaf}}, \bibinfo {author} {\bibfnamefont {S.}~\bibnamefont {Liu}}, \bibinfo {author} {\bibfnamefont {S.}~\bibnamefont {Chatterjee}}, \bibinfo {author} {\bibfnamefont {A.}~\bibnamefont {Vishwanath}},\ and\ \bibinfo {author} {\bibfnamefont {M.~P.}\ \bibnamefont {Zaletel}},\ }\bibfield  {title} {\bibinfo {title} {Ground state and hidden symmetry of magic-angle graphene at even integer filling},\ }\href {https://doi.org/10.1103/PhysRevX.10.031034} {\bibfield  {journal} {\bibinfo  {journal} {Phys. Rev. X}\ }\textbf {\bibinfo {volume} {10}},\ \bibinfo {pages} {031034} (\bibinfo {year} {2020})}\BibitemShut {NoStop}%
\bibitem [{\citenamefont {Liu}\ and\ \citenamefont {Dai}(2021)}]{liu_2021s}%
  \BibitemOpen
  \bibfield  {author} {\bibinfo {author} {\bibfnamefont {J.}~\bibnamefont {Liu}}\ and\ \bibinfo {author} {\bibfnamefont {X.}~\bibnamefont {Dai}},\ }\bibfield  {title} {\bibinfo {title} {Theories for the correlated insulating states and quantum anomalous hall effect phenomena in twisted bilayer graphene},\ }\href {https://doi.org/10.1103/PhysRevB.103.035427} {\bibfield  {journal} {\bibinfo  {journal} {Phys. Rev. B}\ }\textbf {\bibinfo {volume} {103}},\ \bibinfo {pages} {035427} (\bibinfo {year} {2021})}\BibitemShut {NoStop}%
  \bibitem [{\citenamefont {Ceresoli}\ \emph {et~al.}(2006)\citenamefont {Ceresoli}, \citenamefont {Thonhauser}, \citenamefont {Vanderbilt},\ and\ \citenamefont {Resta}}]{cere_2006s}%
  \BibitemOpen
  \bibfield  {author} {\bibinfo {author} {\bibfnamefont {D.}~\bibnamefont {Ceresoli}}, \bibinfo {author} {\bibfnamefont {T.}~\bibnamefont {Thonhauser}}, \bibinfo {author} {\bibfnamefont {D.}~\bibnamefont {Vanderbilt}},\ and\ \bibinfo {author} {\bibfnamefont {R.}~\bibnamefont {Resta}},\ }\bibfield  {title} {\bibinfo {title} {Orbital magnetization in crystalline solids: Multi-band insulators, chern insulators, and metals},\ }\href {https://doi.org/10.1103/PhysRevB.74.024408} {\bibfield  {journal} {\bibinfo  {journal} {Phys. Rev. B}\ }\textbf {\bibinfo {volume} {74}},\ \bibinfo {pages} {024408} (\bibinfo {year} {2006})}\BibitemShut {NoStop}%
\end{thebibliography}

\clearpage
\onecolumngrid
\begin{center}
\textbf{\large Supplemental Material: Contribution of remote bands to orbital magnetization in twisted bilayer graphene}
\end{center}

\setcounter{equation}{0}
\setcounter{figure}{0}
\setcounter{table}{0}
\setcounter{page}{1}
\setcounter{section}{0}
\makeatletter
\renewcommand{\theequation}{S\arabic{equation}}
\renewcommand{\thefigure}{S\arabic{figure}}
\renewcommand{\bibnumfmt}[1]{[S#1]}
\renewcommand{\citenumfont}[1]{S#1}

\section{HF approximation in the projected basis}
In this section, we summarize the standard projected-basis Hartree--Fock formulation used in the main text. We project the Coulomb interaction onto the two lowest-energy flat bands per spin and per valley and work directly in momentum space, following Refs.~\cite{bernevig_2020s,kang_2021s,zhang_2020s}. The interacting part of the Hamiltonian is written as
\begin{equation}
	H_{int}=\frac{1}{2 S}\sum_{\bq\neq0}V(\bq)\delta\rho_{\bq}\delta\rho_{-\bq} \ ,
	\label{eq:coulomb}
\end{equation}
where $S$ is the total area and $\delta\rho_{\bq}=\rho_{\bq}-\bar{\rho}_{\bq}$
is the density operator relative to charge neutrality of the decoupled bilayer graphene in momentum space. Compared with the normal-ordered interaction supplemented by double-counting subtraction schemes used in some previous HF studies~\cite{bul_2020s,liu_2021s,zhang_2020s}, the interaction in Eq.~\ref{eq:coulomb} partially incorporates the renormalization and screening effects of the remote bands on the active flat bands~\cite{kang_2021s,bernevig_2020s}. The density operator can be expanded in the eigen-basis of $H_0$ as
\begin{equation}
	\rho_{\bq} =\sum_{s,\tau}\sum_{\bk\in \text{mBZ},m_1,m_2} \lambda_{m_1,m_2;\tau}(\bk,\bk+\bq) d^{\dagger}_{\bk,m_1,s,\tau}d_{\bk+\bq,m_2,s,\tau} \,
\end{equation}
with the quantum-averaged background density
\begin{equation}
	\bar{\rho}_{\bq} =\sum_{\bG,\tau,m}\delta_{\bq,\bG}\sum_{\bk\in \text{mBZ}} \lambda_{m,m;\tau}(\bk,\bk+\bG) \ ,
\end{equation}
where the form factor is defined by
\begin{equation}
\begin{split}
	\lambda_{m_1 m_2;\tau}(\bk,\bk+\bq)&=
	\left<u_{m_1,\tau}(\bk)|u_{m_2,\tau}(\bk+\bq)\right>=\sum_{\bG^{'},X}u^{*}_{m_1,\tau;\bG^{'},X}(\bk)u_{m_2,\tau;\bG^{'},X}(\bk+\bq).
\end{split}
\end{equation}
Here, $\bk$ is restricted to the mBZ while $\bq$ can extend to the original Brillouin zone of monolayer graphene. The operator $d^{\dagger}$ obeys the periodic condition $d^{\dagger}_{\bk,m,s,\tau}=d^{\dagger}_{\bk+\bG,m,s,\tau}$, which is equivalent to the Bloch-wavefunction relation $u_{m,\tau;\bG,X}(\bk+\bG_0)=u_{m,\tau;\bG+\bG_0,X}(\bk)$. We take $V(\bq)$ to be the single-gate-screened Coulomb potential~\cite{bul_2020s},
\begin{equation}
	V(\bq)=\frac{e^2}{2\varepsilon\varepsilon_0 q}(1-e^{-2q d_s}) \ ,
	\label{eq:Vg}
\end{equation}
with $\varepsilon=7$ and $d_s=40$ nm.

Within this projected basis, the quartic interaction is decoupled in the standard Hartree and Fock channels in terms of the single-particle density matrix
\begin{equation}
	\rho(\bk)_{m_1,s,\tau;m_2,s^{\prime}\tau^{\prime}}=\langle d^{\dagger}_{\bk,m_1,s,\tau}d_{\bk,m_2,s^{\prime},\tau^{\prime}}\rangle .
\end{equation}
We then solve the resulting HF Hamiltonian self-consistently at each integer filling. The converged HF eigenvalues and eigenvectors are used in Eqs.~(6)--(8) and (12)--(15) of the main text to evaluate the Berry curvature, orbital magnetization, and self-rotation contribution.

\section{Orbital Magnetization for a Gapped Dirac Cone}
In this section, we consider a noninteracting two-band toy model describing a gapped Dirac cone, with the Hamiltonian
\begin{align}
    H(\fvec k) = \hbar v_F \fvec \sigma \cdot \fvec k + \Delta \sigma_z \ .
\end{align}
The spectrum consists of two states at each $\fvec k$,
\begin{align}
    E_{\pm} = \pm \sqrt{\Delta^2 + (\hbar v_F k)^2} \ ,
\end{align}
with corresponding eigenstates
\begin{align}
    | u_+ \rangle &= \left( \cos\frac{\theta}2, \ \sin\frac{\theta}2 e^{i \phi} \right)^T , \\
    | u_- \rangle &= \left( -\sin\frac{\theta}2, \ \cos\frac{\theta}2 e^{i \phi} \right)^T ,
\end{align}
where
\begin{align}
    \theta = \tan^{-1}\frac{\hbar v_F k}{\Delta} \ , \qquad
    \phi = \tan^{-1}\frac{k_y}{k_x} \ .
\end{align}
Using the identity
\begin{align}
    \langle \partial_{\mu} u_- | u_+ \rangle = \frac{\langle u_- | \partial_{\mu} H | u_+ \rangle}{E_- - E_+}  =\hbar v_F  \frac{\langle u_- | \sigma_{\mu}  | u_+ \rangle}{E_- - E_+} \ ,
\end{align}
we obtain
\begin{align}
    \langle \pd_x u_- | u_+ \rangle & = \hbar v_F \frac{\langle u_- | \sigma_x | u_+ \rangle}{E_- - E_+} = - \frac{\hbar v_F}{2 \sqrt{\Delta^2 + (\hbar v_F k)^2}} \left( \cos^2\frac{\theta}2 e^{-i \phi} - \sin^2\frac{\theta}2 e^{i \phi} \right) \ , \\
    \langle \pd_y u_- | u_+ \rangle & = \hbar v_F \frac{\langle u_- | \sigma_y | u_+ \rangle}{E_- - E_+} = -i \frac{\hbar v_F}{2 \sqrt{\Delta^2 + (\hbar v_F k)^2}} \left( \cos^2\frac{\theta}2 e^{-i \phi} + \sin^2\frac{\theta}2 e^{i \phi} \right) \  .
\end{align}
Substituting these matrix elements into the general expression for the orbital magnetization gives
\begin{align}
    M_{\text{orb}} & = \frac{e}{2i \hbar} \sum_{\fvec k} \Theta(\mu - E_-(\fvec k)) \Theta(E_+(\fvec k) - \mu) \left( \langle \pd_x u_- | u_+ \rangle \langle u_+ | \pd_y u_- \rangle (E_+ + E_- - 2\mu) - \mathrm{c.c.}
    \right) \nonumber \\
    & =  \frac{e \mu}{\hbar} \sum_{\fvec k} \Theta(\mu - E_-(\fvec k)) \Theta(E_+(\fvec k) - \mu) \frac{(\hbar v_F)^2}{\Delta^2 + (\hbar v_F k)^2 }
    \left( \cos^4\frac{\theta}2 - \sin^4 \frac{\theta}2 \right) \ . \\
  \Longrightarrow \quad   \frac{M_{\text{orb}}}S  & =  \frac{e \mu}{\hbar} \int_0^{\infty} \frac{k \rmd k}{2 \pi} \frac{(\hbar v_F)^2}{\Delta^2 + (\hbar v_F k)^2 } \cos\theta \nonumber \\
    & =  \frac{e \mu}{2 h} \int_0^{\infty} k \rmd k\ \Theta\left( \mu + \sqrt{\Delta^2 + (\hbar v_F k)^2} \right) \Theta\left( \sqrt{\Delta^2 + (\hbar v_F k)^2} - \mu \right) \frac{ (\hbar v_F)^2 \Delta}{( \Delta^2 + (\hbar v_F k)^2 )^{3/2}} \nonumber \\
    & = \left\{ \begin{array}{ll}
         \dfrac{e \mu}{2 h} \ , & |\mu| \leq \Delta \  , \\[6pt]
         \dfrac{e \Delta}{2h} \mathrm{sgn}(\mu) \  , & |\mu| > \Delta \  .
    \end{array}  \right.
\end{align}

It is also instructive to evaluate the self-rotation contribution $m_{\text{SR}}$. This yields
\begin{align}
    \frac{m_{\text{SR}}}{S}   & = - \frac{e}{2h} \int_0^{\infty} k \rmd k\ \frac{(\hbar v_F)^2 \Delta}{\Delta^2 + (\hbar v_F k)^2 } \ .
\end{align}
This integral diverges logarithmically in the ultraviolet limit $k \rightarrow \infty$. Within the present continuum description, this indicates that the valley self-rotation contribution $m_{\text{SR}}$ is ultraviolet divergent, and therefore would also diverge in hexagonal boron nitride (hBN)-aligned Dirac systems unless a suitable high-energy cutoff is introduced.

\section{Projection matrix formalism for orbital magnetization}
In this section, we derive the expression for $M_{\text{orb}}$ and $m_{\text{SR}}$ in terms of projection operators. We begin from Eq.~\ref{eq:Morb} and \ref{eq:msr}, which give the general expression for the orbital magnetization. However, these formulas are not convenient for numerical implementation, since the derivative of the Bloch states is gauge dependent, and the summation over the occupied bands involves a large number of bands for moir\'e systems. 

We first note that the expression of $M_{orb}$ in Eq.~\ref{eq:Morb} can be rewritten as
\begin{align}
    M_{\text{orb}} & = \frac{e}{2i \hbar} \sum_{n, \fvec k} n_F\left(E_n^{\text{HF}}(\fvec k) - \mu \right)    \epsilon_{\mu\nu} \sum_{\alpha} \langle \pd_{\mu} u_n(\fvec k) | u_{\alpha}(\fvec k) \rangle  \big( E_{\alpha}^{\text{HF}}(\fvec k) + E_n^{\text{HF}}(\fvec k) - 2\mu  \big) \langle u_{\alpha}(\fvec k) | \pd_{\nu} u_n(\fvec k) \rangle
\end{align}
The summation over $\alpha$ can be decomposed into contributions from occupied and empty states. Accordingly, $M_{\text{orb}}$ can be expressed as
\begin{align}
  M_{\text{orb}} & =  M_{\text{orb}}^{(1)} + M_{\text{orb}}^{(2)} \\
  M_{\text{orb}}^{(1)} & = \frac{e}{2i \hbar} \sum_{n, \alpha, \fvec k} n_F\left(E_n^{\text{HF}}(\fvec k) - \mu \right) n_F\left(E_{\alpha}^{\text{HF}}(\fvec k) - \mu \right) \epsilon_{\mu\nu} \langle \pd_{\mu} u_n(\fvec k) | u_{\alpha}(\fvec k) \rangle  \big( E_{\alpha}^{\text{HF}}(\fvec k) + E_n^{\text{HF}}(\fvec k) - 2\mu  \big) \langle u_{\alpha}(\fvec k) | \pd_{\nu} u_n(\fvec k) \rangle \\
  M_{\text{orb}}^{(2)} & = \frac{e}{2i \hbar} \sum_{n, \alpha, \fvec k} n_F\left(E_n^{\text{HF}}(\fvec k) - \mu \right) \left( 1 -  n_F\left(E_{\alpha}^{\text{HF}}(\fvec k) - \mu \right) \right) \epsilon_{\mu\nu} \langle \pd_{\mu} u_n(\fvec k) | u_{\alpha}(\fvec k) \rangle  \big( E_{\alpha}^{\text{HF}}(\fvec k) + E_n^{\text{HF}}(\fvec k) - 2\mu  \big) \langle u_{\alpha}(\fvec k) | \pd_{\nu} u_n(\fvec k) \rangle \label{Eqn:Morb2}
\end{align}
Using the orthogonality relation,
\[  \langle u_n(\fvec k) | u_{\alpha}(\fvec k) \rangle = \delta_{\alpha n} \ , \]
we obtain
\[  \langle \pd_{\mu} u_n(\fvec k) | u_{\alpha}(\fvec k) \rangle = - \langle u_n(\fvec k) | \pd_{\mu} u_{\alpha}(\fvec k) \rangle \ . \]
It then follows that $M_{\text{orb}}^{(1)}$ can be rewritten as
\begin{align}
    M_{\text{orb}}^{(1)} & = \frac{e}{2i \hbar} \sum_{n, \alpha, \fvec k} n_F\left(E_n^{\text{HF}}(\fvec k) - \mu \right) n_F\left(E_{\alpha}^{\text{HF}}(\fvec k) - \mu \right) \epsilon_{\mu\nu} \langle u_n(\fvec k) | \pd_{\mu}  u_{\alpha}(\fvec k) \rangle  \big( E_{\alpha}^{\text{HF}}(\fvec k) + E_n^{\text{HF}}(\fvec k) - 2\mu  \big) \langle \pd_{\nu} u_{\alpha}(\fvec k) |  u_n(\fvec k) \rangle \nonumber \\
    & = \frac{e}{2i \hbar} \sum_{n, \alpha, \fvec k} n_F\left(E_n^{\text{HF}}(\fvec k) - \mu \right) n_F\left(E_{\alpha}^{\text{HF}}(\fvec k) - \mu \right) \epsilon_{\mu\nu} \langle \pd_{\nu} u_{\alpha}(\fvec k) |  u_n(\fvec k) \rangle  \big( E_{\alpha}^{\text{HF}}(\fvec k) + E_n^{\text{HF}}(\fvec k) - 2\mu  \big) \langle u_n(\fvec k) | \pd_{\mu}  u_{\alpha}(\fvec k) \rangle \nonumber \\
    & = - M_{\text{orb}}^{(1)} \ .
\end{align}
Therefore, 
\begin{equation}
    M_{\text{orb}}^{(1)} = 0 \ ,
\end{equation}
so that this term does not contribute to the orbital magnetization.

We next rewrite $M_{\mathrm{orb}}^{(2)}$ in terms of projectors. For the full HF Hamiltonian $H^{\mathrm{HF}}(\fvec k)$, we define the projectors onto the occupied and empty subspaces as
\begin{align}
    P(\fvec k) & = \sum_{n \in \mathrm{occ}} |u_n(\fvec k)\rangle \langle u_n(\fvec k)| \  , \\
    Q(\fvec k) & = \sum_{\alpha \in \mathrm{emp}} |u_{\alpha}(\fvec k)\rangle \langle u_{\alpha}(\fvec k)| \  .  
\end{align}
We then introduce the following gauge invariant quantities,
\begin{align}
    W_{\mu\nu}(\fvec k)
    & =
    \mtr\left[ P(\fvec k)\big(\pd_{\mu} P(\fvec k)\big)\big(\pd_{\nu} Q(\fvec k)\big)\big(H^{\mathrm{HF}}(\fvec k)-\mu\big) \right] \   , \label{Eqn:Wmunu} \\
    N_{\mu\nu}(\fvec k)
    &=
    \mtr\left[ Q(\fvec k)\big(\pd_{\mu} P(\fvec k)\big)\big(\pd_{\nu} Q(\fvec k)\big)\big(H^{\mathrm{HF}}(\fvec k)-\mu\big) \right] \  .  \label{Eqn:Nmunu}
\end{align}
Since $P Q = 0$, one has
\[  \big( \pd_{\mu} P \big) Q = - P \big( \pd_{\mu} Q \big) \ . \]
Moreover, both $P$ and $Q$ commute with $H^{\text{HF}}$. Consequently, 
\begin{align}
    W_{\mu\nu}(\fvec k) & = \mtr\left[ P(\fvec k)\big(\pd_{\mu} P(\fvec k)\big)\big(\pd_{\nu} Q(\fvec k)\big)\big(H^{\mathrm{HF}}(\fvec k)-\mu\big) \right] = \mtr\left[ \big(\pd_{\mu} P(\fvec k)\big) \big(\pd_{\nu} Q(\fvec k)\big) P(\fvec k) \big(H^{\mathrm{HF}}(\fvec k)-\mu\big) \right]  \nonumber \\
    & = - \mtr\left[ \big(\pd_{\mu} P(\fvec k)\big) Q(\fvec k) \big(\pd_{\nu} P(\fvec k)\big)  \big(H^{\mathrm{HF}}(\fvec k)-\mu\big) \right] \nonumber \\
    & = - \sum_{n, \alpha} n_F(E_n(\fvec k) - \mu) \big( 1- n_F(E_{\alpha}(\fvec k) - \mu) \big) \langle \pd_{\mu} u_n(\fvec k) | u_{\alpha}(\fvec k) \rangle \langle u_{\alpha}(\fvec k) | \pd_{\nu} u_n(\fvec k)  \rangle  \left( E_n^{\text{HF}}(\fvec k) - \mu \right) 
\end{align}
which implies
\begin{equation}
    \big( W_{\mu\nu}(\fvec k)  \big)^*  = - \sum_{n, \alpha} n_F(E_n(\fvec k) - \mu) \big( 1- n_F(E_{\alpha}(\fvec k) - \mu) \big) \langle u_{\alpha}(\fvec k)  | \pd_{\mu} u_n(\fvec k)  \rangle \langle \pd_{\nu} u_n(\fvec k) | u_{\alpha}(\fvec k)  \rangle  \left( E_n^{\text{HF}}(\fvec k) - \mu \right) = W_{\nu\mu}(\fvec k) \ .
\end{equation}
Similarly,
\begin{align}
    N_{\mu\nu}(\fvec k) & = \mtr\left[ Q(\fvec k)\big(\pd_{\mu} P(\fvec k)\big)\big(\pd_{\nu} Q(\fvec k)\big)\big(H^{\mathrm{HF}}(\fvec k)-\mu\big) \right] = - \mtr\left[ \big(\pd_{\mu} Q(\fvec k)\big) P(\fvec k) \big(\pd_{\nu} Q(\fvec k)\big)  \big(H^{\mathrm{HF}}(\fvec k)-\mu\big) \right]  \nonumber \\
    & = - \sum_{n, \alpha} n_F(E^{\mathrm{HF}}_n(\fvec k) - \mu) \big( 1- n_F(E^{\mathrm{HF}}_{\alpha}(\fvec k) - \mu) \big) \langle \pd_{\mu} u_{\alpha}(\fvec k) | u_n(\fvec k) \rangle \langle u_n(\fvec k) | \pd_{\nu} u_{\alpha}(\fvec k)  \rangle  \left( E_{\alpha}^{\text{HF}}(\fvec k) - \mu \right) \nonumber \\
    & = - \sum_{n, \alpha} n_F(E^{\mathrm{HF}}_n(\fvec k) - \mu) \big( 1- n_F(E^{\mathrm{HF}}_{\alpha}(\fvec k) - \mu) \big) \langle \pd_{\nu} u_n(\fvec k) |  u_{\alpha}(\fvec k)  \rangle       \langle  u_{\alpha}(\fvec k) | \pd_{\mu} u_n(\fvec k) \rangle  \left( E_{\alpha}^{\text{HF}}(\fvec k) - \mu \right) 
\end{align}
and therefore
\begin{equation}
    \big( N_{\mu\nu}(\fvec k)  \big)^* = - \sum_{n, \alpha} n_F(E^{\mathrm{HF}}_n(\fvec k) - \mu) \big( 1- n_F(E^{\mathrm{HF}}_{\alpha}(\fvec k) - \mu) \big) \langle  u_{\alpha}(\fvec k) | \pd_{\nu} u_n(\fvec k)   \rangle       \langle  \pd_{\mu} u_n(\fvec k) | u_{\alpha}(\fvec k)  \rangle  \left( E_{\alpha}^{\text{HF}}(\fvec k) - \mu \right) = N_{\nu\mu}(\fvec k) \ . 
\end{equation}
Substituting these results into Eq.~\ref{Eqn:Morb2}, we obtain
\begin{equation}
    M_{\mathrm{orb}} = M_{\mathrm{orb}}^{(2)} = - \frac{e}{\hbar} \sum_{\fvec k} \mathrm{Im}\left[ W_{xy}(\fvec k) - N_{xy}(\fvec k) \right] \ .
\end{equation}
Similarly,  the self-rotation $m_{\text{SR}}$ is given by
\begin{equation}
    m_{\mathrm{SR}}  =  \frac{e}{\hbar} \sum_{\fvec k} \mathrm{Im}\left[ W_{xy}(\fvec k) + N_{xy}(\fvec k) \right] \  .
\end{equation}
These expressions are manifestly gauge invariant since they depend only on the projection operators, and are therefore numerically more stable than formulas involving explicit derivatives of Bloch wave functions.

We further emphasize that the derivation above does not explicitly rely on the identity $P + Q = I$. This observation suggests that $P$ and $Q$ may be replaced by their counterparts defined in the truncated occupied and empty space. More specifically, we therefore introduce the truncated projectors defined in the main text,
\begin{align}
    P_{n_{\rm cut}}(\mathbf{k}) & = \sum_{n \in \mathrm{occupied}} |u^{\rm A}_n(\mathbf{k})\rangle\langle u^{\rm A}_n(\mathbf{k})| + \sum_{n \le n_{\rm cut}} |u^{R-}_n(\mathbf{k})\rangle \langle u^{R-}_n(\mathbf{k})| , \\
    Q_{n_{\rm cut}}(\mathbf{k}) & = \sum_{n \in \mathrm{empty}} |u^{\rm A}_n(\mathbf{k})\rangle\langle u^{\rm A}_n(\mathbf{k})| + \sum_{n \le n_{\rm cut}} |u^{R+}_n(\mathbf{k})\rangle \langle u^{R+}_n(\mathbf{k})| .
\end{align}
Here, $A$, $R-$, and $R+$ denote the active bands, the occupied remote bands, and the empty remote bands, respectively. Replacing $P(\fvec k)$ and $Q(\fvec k)$ by $P_{n_{\rm cut}}(\fvec k)$ and $Q_{n_{\rm cut}}(\fvec k)$ in the definitions of $W_{\mu\nu}$ and $N_{\mu\nu}$ yields the truncated quantities $W_{\mu\nu}^{\rm trun}$ and $N_{\mu\nu}^{\rm trun}$, and hence Eqs.~\ref{eq:Morb1} and~\ref{eq:msr1} of the main text used in the numerical calculations. The full result is recovered in the limit $n_{\mathrm{cut}}\to\infty$, while in practice we find rapid convergence with increasing $n_{\mathrm{cut}}$, as benchmarked in sections below.

\section{Numerical stability of the symmetric truncation scheme}
The numerical stability of the present construction follows from the way in which the Hilbert space is partitioned. Specifically, we decompose the exact projectors as
\begin{equation}
    P(\fvec k)=P_{n_{\rm cut}}(\fvec k) + P_1(\fvec k)\ , \qquad
    Q(\fvec k)=Q_{n_{\rm cut}}(\fvec k) + Q_1(\fvec k)   \ ,
\end{equation}
where $P_1(\fvec k)$ and $Q_1(\fvec k)$ denote the omitted occupied and empty remote-band sectors, respectively.

As discussed in the previous section, the orbital magnetization is expressed in terms of the two gauge-invariant quantities $W_{\mu\nu}$ and $N_{\mu\nu}$ defined in Eqs.~\ref{Eqn:Wmunu} and \ref{Eqn:Nmunu}. In moir\'e systems, however, the large number of occupied and empty bands leads to a substantial computational cost for calculations in the full Hilbert space. One therefore replaces these projectors by their truncated counterparts, and correspondingly replaces $W_{\mu\nu}$ and $N_{\mu\nu}$ by the truncated quantities $W_{\mu\nu}^{\text{trun}}$ and $N_{\mu\nu}^{\text{trun}}$ defined in Eqs.~\ref{Eqn:WmunuTrunc} and \ref{Eqn:NmunuTrunc}. 

Following the formalism introduced in the previous section, one readily obtains
\begin{align}
    W^{\rm trun}_{\mu\nu}(\fvec k) & = - \sum_{\substack{n \leq n_{\rm cut} \\ n \in \rm occ}} \sum_{\substack{\alpha \leq n_{\rm cut} \\ \alpha \in \rm emp}} \langle \pd_{\mu} u_n | u_{\alpha} \rangle \langle u_{\alpha} | \pd_{\nu} u_n  \rangle  \left( E_n^{\rm HF} - \mu \right)   =  - \sum_{\substack{n \leq n_{\rm cut} \\ n \in \rm occ}} \sum_{\substack{\alpha \leq n_{\rm cut} \\ \alpha \in \rm emp}} \frac{\langle u_n | \pd_{\mu} H^{\rm HF} | u_{\alpha} \rangle}{ E^{\rm HF}_n - E^{\rm HF}_{\alpha} } \frac{\langle u_{\alpha} | \pd_{\nu} H^{\rm HF} | u_n \rangle}{ E^{\rm HF}_n - E^{\rm HF}_{\alpha} } \left(  E^{\rm HF}_n - \mu  \right) \\
    N^{\rm trun}_{\mu\nu}(\fvec k) & = - \sum_{\substack{n \leq n_{\rm cut} \\ n \in \rm occ}} \sum_{\substack{\alpha \leq n_{\rm cut} \\ \alpha \in \rm emp}} \langle \pd_{\nu} u_n | u_{\alpha} \rangle \langle u_{\alpha} | \pd_{\mu} u_n  \rangle  \left( E^{\rm HF}_{\alpha} - \mu \right)  =  - \sum_{\substack{n \leq n_{\rm cut} \\ n \in \rm occ}} \sum_{\substack{\alpha \leq n_{\rm cut} \\ \alpha \in \rm emp}} \frac{\langle u_n | \pd_{\nu} H^{\rm HF} | u_{\alpha} \rangle}{ E^{\rm HF}_n - E^{\rm HF}_{\alpha} } \frac{\langle u_{\alpha} | \pd_{\mu} H^{\rm HF} | u_n \rangle}{ E^{\rm HF}_n - E^{\rm HF}_{\alpha} } \left(  E^{\rm HF}_{\alpha} - \mu  \right) \ ,
\end{align}
where the explicit $\fvec k$ dependence of the Hamiltonian, the Bloch states, and the corresponding energies has been omitted for notational convenience. In deriving these expressions, we have also used
\begin{equation}
    \left\langle u_{\alpha} \middle| \pd_{\mu} u_n \right\rangle  =  \frac{\left\langle u_{\alpha} \middle| \pd_{\mu} H^{\mathrm{HF}} \middle| u_n \right\rangle}{E^{\rm HF}_n - E^{\rm HF}_{\alpha}} \ . 
\end{equation}

\begin{figure}[htb]
	\begin{center}
		\fig{6.8in}{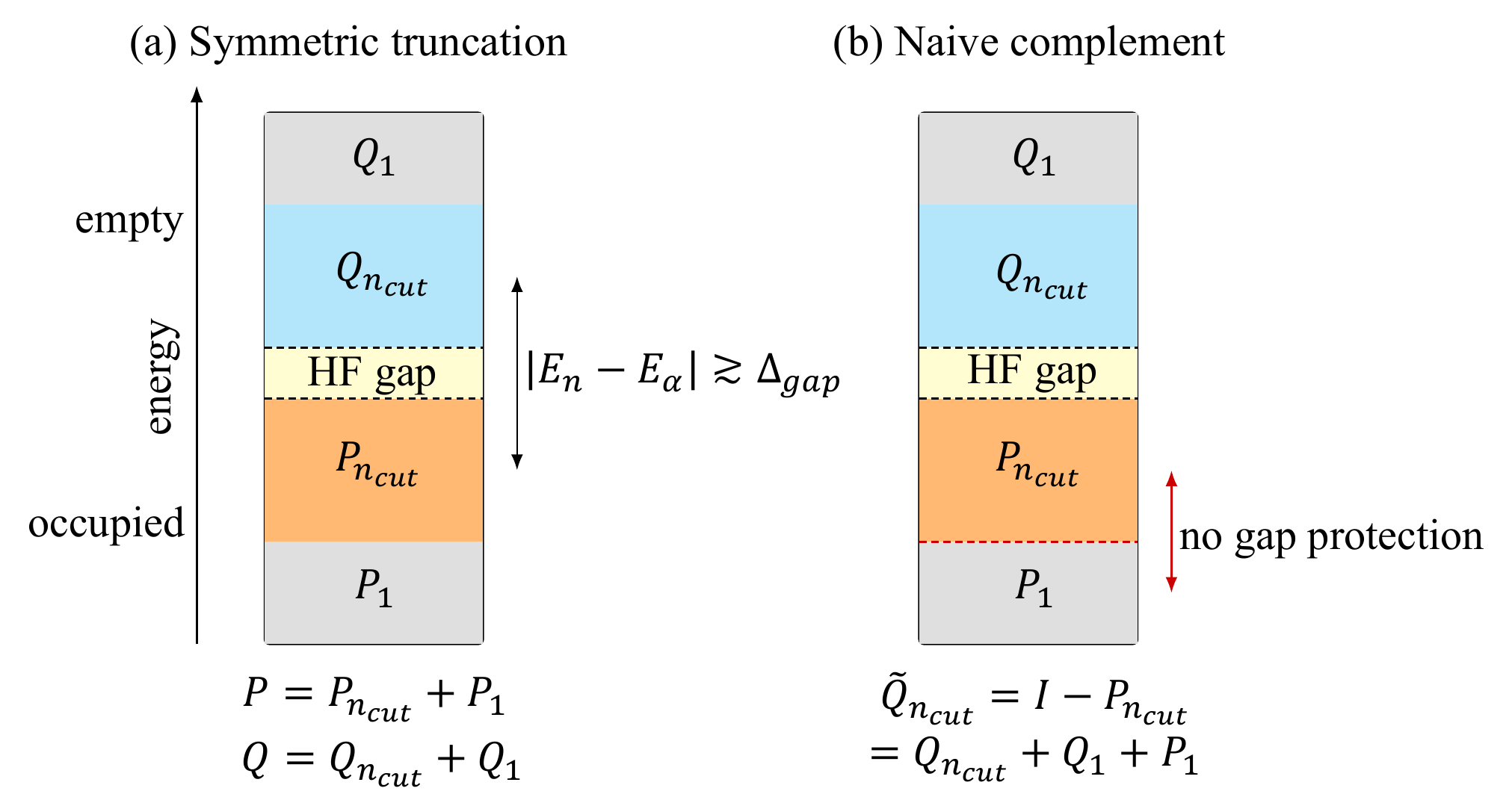}\caption{Schematic illustration of the Hilbert-space decomposition responsible for the numerical stability of the symmetric truncation scheme. In panel (a), the retained occupied and empty subspaces, $P_{n_{\rm cut}}$ and $Q_{n_{\rm cut}}$, are separated by the insulating gap, so that the interband energy denominators entering $W_{\mu\nu}$ and $N_{\mu\nu}$ remain finite. In panel (b), the alternative choice $\widetilde Q_{n_{\rm cut}} = I - P_{n_{\rm cut}}$ additionally includes the omitted occupied sector $P_1$, therefore generating matrix elements across an artificial cutoff boundary within the occupied subspace. These terms are not gap-protected and therefore lead to unstable, oscillatory convergence.
			\label{fig:proj_stability}}
	\end{center}
	\vskip-0.5cm
\end{figure}

In the symmetric truncation scheme adopted in this work, the matrix elements entering $W_{\mu\nu}^{\rm trun}$ and $N_{\mu\nu}^{\rm trun}$ always connect a retained occupied state $| u_n \rangle$ belonging to $P_{n_{\rm cut}}$ to a retained empty state $| u_{\alpha} \rangle$ belonging to $Q_{n_{\rm cut}}$. These two states lie on opposite sides of the insulating gap, as illustrated schematically in Fig.~\ref{fig:proj_stability}. The denominator $E_n^{HF} - E_{\alpha}^{HF}$ is therefore bounded below by the direct gap, which in turn guarantees smooth convergence of the truncated sums despite the formally infinite remote-band sector.

We now examine the difference between the truncated expressions and those defined in the full Hilbert space. Using the relations $P Q = 0$ and the fact that both projectors commute with the Hamiltonian, we obtain
\begin{align} 
    W_{\mu\nu} - W_{\mu\nu}^{\text{trun}} & = - \mtr\left\{ \left[ \big(\pd_{\mu} P_{n_{\text{cut}}} + \pd_{\mu} P_1 \big) \big( Q_{n_{\text{cut}}} + Q_1 \big) \big( \pd_{\nu} P_{n_{\text{cut}}} + \pd_{\nu} P_1 \big) - \big(\pd_{\mu} P_{n_{\text{cut}}} \big) Q_{n_{\text{cut}}} \big(\pd_{\nu} P_{n_{\text{cut}}} \big) \right] \big( H^{\mathrm{HF}} - \mu \big) \right\} 
\end{align}
where the explicit $\fvec k$ dependence has been omitted for notational convenience. Furthermore, one finds that
\begin{equation} 
    \mtr\left[ (\pd_{\mu} P_{n_{\text{cut}}}) Q (\pd_{\nu} P_1 \big) \big(H^{\mathrm{HF}} - \mu \big) \right] = \mtr\left[ (\pd_{\mu} P_1) Q (\pd_{\nu} P_{n_{\text{cut}}} \big) \big(H^{\mathrm{HF}} - \mu \big) \right] = 0 \ . 
\end{equation}
It then follows that
\begin{align} 
    W_{\mu\nu} - W_{\mu\nu}^{\text{trun}} & = - \mtr\left\{ \left[ \big( \pd_{\mu} P_1 \big) Q \big( \pd_{\nu} P_1 \big) + \big(\pd_{\mu} P_{n_{\text{cut}}} \big) Q_1 \big(\pd_{\nu} P_{n_{\text{cut}}} \big) \right] \big(H^{\mathrm{HF}} - \mu \big) \right\} \nonumber \\
    & = - \left( \sum_{ \substack{n > n_{{\text{cut}}} \\ n \in \rm occ}} \sum_{\alpha \in \text{emp}} + \sum_{ \substack{n \leq n_{{\text{cut}}} \\ n \in \rm occ}} \sum_{\substack{\alpha > n_{\text{cut}} \\ \alpha \in \text{emp}}} \right)   \frac{ \left\langle u_n \middle| \pd_{\mu} H^{\mathrm{HF}} \middle| u_{\alpha} \right\rangle }{ E^{\rm HF}_n - E^{\rm HF}_{\alpha} }  \frac{ \left\langle u_{\alpha} \middle| \pd_{\nu} H^{\mathrm{HF}} \middle| u_n \right\rangle }{ E^{\rm HF}_n - E^{\rm HF}_{\alpha} }  \left(  E^{\rm HF}_n - \mu \right) \   .
\end{align}

Similarly, for $N_{\mu\nu}$ we obtain
\begin{align}
    N_{\mu\nu} - N_{\mu\nu}^{\text{trun}} & = - \mtr\left\{ \left[ \big( \pd_{\nu} Q_1 \big) P \big( \pd_{\mu} Q_1 \big) + \big(\pd_{\nu} Q_{n_{\text{cut}}} \big) P_1 \big(\pd_{\mu} Q_{n_{\text{cut}}} \big) \right] \big(H^{\mathrm{HF}} - \mu \big) \right\}  \nonumber \\
    & = - \left( \sum_{ \substack{\alpha > n_{{\text{cut}}} \\ \alpha \in \rm emp}} \sum_{n \in \text{occ}} + \sum_{ \substack{\alpha \leq n_{{\text{cut}}} \\ \alpha \in \rm emp}} \sum_{\substack{n > n_{\text{cut}} \\ n \in \text{occ}}} \right)    \frac{ \left\langle u_{\alpha} \middle| \pd_{\nu} H^{\mathrm{HF}} \middle| u_n \right\rangle }{ E^{\rm HF}_{\alpha} - E^{\rm HF}_n }  \frac{ \left\langle u_n \middle| \pd_{\mu} H^{\mathrm{HF}} \middle| u_{\alpha} \right\rangle }{ E^{\rm HF}_{\alpha} - E^{\rm HF}_n  }  \left(  E^{\rm HF}_{\alpha} - \mu \right) \ .
\end{align}
In the above expressions for differences, the summation involves a pair of states: either a state $| u_n\rangle$ belonging to $P_1$ together with a state $| u_{\alpha} \rangle$ belonging to $Q$, or a state $| u_n \rangle$ belonging to $P$ together with a state $| u_{\alpha} \rangle$ belonging to $Q_1$. As illustrated schematically in Fig.~\ref{fig:proj_stability}, the two states in each pair are always separated by a continuously retained truncated subspace: $P_{n_{\rm cut}}$ in the former case, and $Q_{n_{\rm cut}}$ in the latter. As the cutoff $n_{\rm cut}$ is increased, both truncated subspaces expand, so that the denominator $E^{\rm HF}_n - E^{\rm HF}_{\alpha}$ correspondingly increases in magnitude. As a result, both differences $W_{\mu\nu} - W_{\mu\nu}^{\text{trun}}$ and $N_{\mu\nu} - N_{\mu\nu}^{\text{trun}}$ converge to zero. This guarantees the numerical stability of the present construction based on two truncated projectors.

It may be tempting instead to truncate only the occupied projector, denoted $P_{n_{\rm cut}}$, and to naively identify its complement with the projector onto the empty subspace, namely
\begin{equation}
    \widetilde Q_{n_{\rm cut}}(\fvec k) \equiv I - P_{n_{\rm cut}}(\fvec k) = Q_{n_{\rm cut}}(\fvec k) + Q_1(\fvec k) + P_1(\fvec k) \ .
\end{equation}
Within this prescription, one immediately finds
\begin{align}
    \widetilde{W}^{\rm trun}_{\mu\nu}(\fvec k) & = - \mtr\left[ \big( \pd_{\mu} P_{n_{\rm cut}} \big) \big( Q_{n_{\rm cut}} + Q_1 + P_1 \big)  \big( \pd_{\nu} P_{n_{\rm cut}} \big) \big( H^{\rm HF} - \mu \big) \right] \nonumber \\
    & = W_{\mu\nu}^{\rm trun} +  \mtr\left[ \big( \pd_{\mu} P_{n_{\rm cut}} \big)  Q_1  \big( \pd_{\nu} P_{n_{\rm cut}} \big) \big( H^{\rm HF} - \mu \big) \right] +  \mtr\left[ \big( \pd_{\mu} P_{n_{\rm cut}} \big)  P_1  \big( \pd_{\nu} P_{n_{\rm cut}} \big) \big( H^{\rm HF} - \mu \big) \right] \  .
\end{align}
As discussed above, $W_{\mu\nu}^{\rm trun}$ is numerically stable because the occupied and empty subspaces are separated by the insulating gap. The second correction term,
\[  \mtr\left[ \big( \pd_{\mu} P_{n_{\rm cut}} \big)  Q_1  \big( \pd_{\nu} P_{n_{\rm cut}} \big) \big( H^{\rm HF} - \mu \big) \right]  \]
is also stable, since the subspace $Q_{n_{\rm cut}}$ separates $P_{n_{\rm cut}}$ from $Q_1$, so that the energy interval between them increases as $n_{\rm cut}$ becomes larger. By contrast, no such interval exists between $P_{n_{\rm cut}}$ and $P_1$. More explicitly,
\begin{align}
    \mtr\left[ \big( \pd_{\mu} P_{n_{\rm cut}} \big)  P_1  \big( \pd_{\nu} P_{n_{\rm cut}} \big) \big( H^{\rm HF} - \mu \big) \right]  & = \sum_{\substack{n \leq n_{\rm cut} \\ n \in \rm occ}} \sum_{\substack{m > n_{\rm cut} \\ m \in \rm occ}} \langle \pd_{\mu} u_n | u_m \rangle \langle u_m | \pd_{\nu} u_n  \rangle  \left( E^{\rm HF}_n - \mu \right)   \nonumber \\
    & =  \sum_{\substack{n \leq n_{\rm cut} \\ n \in \rm occ}} \sum_{\substack{m > n_{\rm cut} \\ m \in \rm occ}}  \frac{\langle u_n | \pd_{\mu} H^{\rm HF} | u_m \rangle}{ E^{\rm HF}_n - E^{\rm HF}_m } \frac{\langle u_m | \pd_{\nu} H^{\rm HF} | u_n \rangle}{ E^{\rm HF}_n - E^{\rm HF}_m } \left(  E^{\rm HF}_n - \mu  \right)  \  .
\end{align}
As illustrated schematically in Fig.~\ref{fig:proj_stability}, the two subspaces $P_{n_{\rm cut}}$ and $P_1$ are directly connected, so that the denominator $E^{\rm HF}_n - E^{\rm HF}_m$ can, in principle, be arbitrarily small. Moreover, in the high-energy limit of a Dirac system, the density of states scales as $|E|$, so that the spacing between consecutive bands decreases approximately as $1/|E|$ as $n_{\rm cut}$ increases. As a result, the denominator $E^{\rm HF}_n - E^{\rm HF}_m$ becomes increasingly small, leading to large oscillatory contributions and consequently poor numerical convergence, as observed in scheme m$_1$ shown in Fig.~\ref{fig:om_bm}(b) of the main text.   

The same issue also arises in approaches based on the covariant derivative, which is defined as~\cite{cere_2006s}  
\begin{equation}
    |\widetilde{\pd}_{\mu} u_l \rangle
    =
    \sum_{\alpha \neq l} | u_{\alpha} \rangle \langle u_{\alpha} | \pd_{\mu} u_l \rangle \   ,
\end{equation}
if the complementary subspace is implemented through an incomplete projector of the form $I - P_{n_{\rm cut}}$. In that case, the projected derivative mixes the retained band with states lying on the same side of the gap, and the corresponding matrix elements are therefore no longer protected by the insulating gap.

\section{Derivation of the particle-hole symmetry of $M_{\mathrm{orb}}$ and $m_{\mathrm{SR}}$  }
The same projector formalism also makes the particle-hole relation between the HF states at fillings $\nu$ and $-\nu$ transparent. To establish the corresponding symmetry of the orbital magnetization, it is convenient to label the projectors and the self-consistent HF Hamiltonian by the filling, namely $P^{(\nu)}(\bk)$, $Q^{(\nu)}(\bk)$, and $H^{(\nu)}(\bk)$. Let $\mathcal{P}$ denote the unitary particle-hole transformation of the BM model~\cite{bernevig_2020s}. For a pair of particle-hole related HF solutions, $\mathcal{P}$ exchanges the occupied and empty subspaces according to
\begin{align}
    \mathcal{P} P^{(\nu)}(\bk) \mathcal{P}^{-1} &= Q^{(-\nu)}(-\bk) , \\
    \mathcal{P} Q^{(\nu)}(\bk) \mathcal{P}^{-1} &= P^{(-\nu)}(-\bk) ,
\end{align}
while the HF Hamiltonian, measured relative to the corresponding chemical potential, transforms as
\begin{equation}
    \mathcal{P}\big(H_{\nu}^{\mathrm{HF}}(\bk)-\mu_{\nu}\big)\mathcal{P}^{-1}
    =
    -\big(H_{-\nu}^{\mathrm{HF}}(-\bk)-\mu_{-\nu}\big).
\end{equation}

Substituting these relations into the operator definitions of $W_{\mu\nu}$ and $N_{\mu\nu}$ and using the invariance of the trace under a unitary transformation, we find
\begin{align}
    W_{\mu\nu}^{(\nu)}(\bk) & = \mtr\left[ P^{(\nu)}(\bk) \big( \pd_{\mu} P^{(\nu)}(\bk) \big) \big( \pd_{\nu} Q^{(\nu)}(\bk)\big)\big(H_{\nu}^{\mathrm{HF}}(\bk)-\mu_{\nu}\big) \right] \nonumber \\
    & = - \mtr\left[ Q^{(-\nu)}(-\bk) \big(\pd_{\mu} Q^{(-\nu)}(-\bk) \big) \big(\pd_{\nu} P^{(-\nu)}(-\bk)\big)\big(H_{-\nu}^{\mathrm{HF}}(-\bk)-\mu_{-\nu}\big) \right] \nonumber \\
    & = - \mtr\left[ \big(\pd_{\mu} Q^{(-\nu)}(-\bk) \big) \big(\pd_{\nu} P^{(-\nu)}(-\bk)\big) Q^{(-\nu)}(-\bk)   \big(H_{-\nu}^{\mathrm{HF}}(-\bk)-\mu_{-\nu}\big) \right] \nonumber \\
    & = - \mtr\left[  Q^{(-\nu)}(-\bk) \big( \pd_{\mu} P^{(-\nu)}(-\bk) \big) \big(\pd_{\nu} Q^{(-\nu)}(-\bk) \big)   \big(H_{-\nu}^{\mathrm{HF}}(-\bk)-\mu_{-\nu}\big) \right] \nonumber \\
    & = - N_{\mu\nu}^{(-\nu)}(-\bk) \    ,
\end{align}
and similarly
\begin{equation}
    N_{\mu\nu}^{(\nu)}(\bk)    =    - W_{\mu\nu}^{(-\nu)}(-\bk) \  .
\end{equation}
Substituting these identities into Eqs.~\ref{eq:Morb} and~\ref{eq:msr} of the main text, and then summing over the mBZ, yields the general particle-hole relations
\begin{equation}
    M_{\mathrm{orb}}(\nu)=M_{\mathrm{orb}}(-\nu)\ , \qquad
    m_{\mathrm{SR}}(\nu)=-m_{\mathrm{SR}}(-\nu) \ .  \label{Eqn:PHOrbM}
\end{equation}
Importantly, the derivation above does not rely on the identity $P + Q = I$. The same symmetry relations therefore remain valid for the truncated quantities, provided that the truncation preserves particle-hole-related remote-band pairs symmetrically:
\begin{equation}
    M^{\rm trun}_{\mathrm{orb}}(\nu) = M^{\rm trun}_{\mathrm{orb}}(-\nu)\ , \qquad    m^{\rm trun}_{\mathrm{SR}}(\nu) = - m^{\rm trun}_{\mathrm{SR}}(-\nu) \ .  
\end{equation}
As shown in Fig.~\ref{fig:om_bm} and \ref{fig:fig3}, the numerical results rapidly converge to the exact symmetry above as $n_{\mathrm{cut}}$ increases. In particular, for $\nu=\pm3$ this gives the symmetry relations quoted in the main text and observed numerically in Fig.~\ref{fig:fig3}(c,d) of the main text.

\section{Additional numerical results for benchmarking the formalism}
\begin{figure}[htb]
	\begin{center}
		\fig{6.8in}{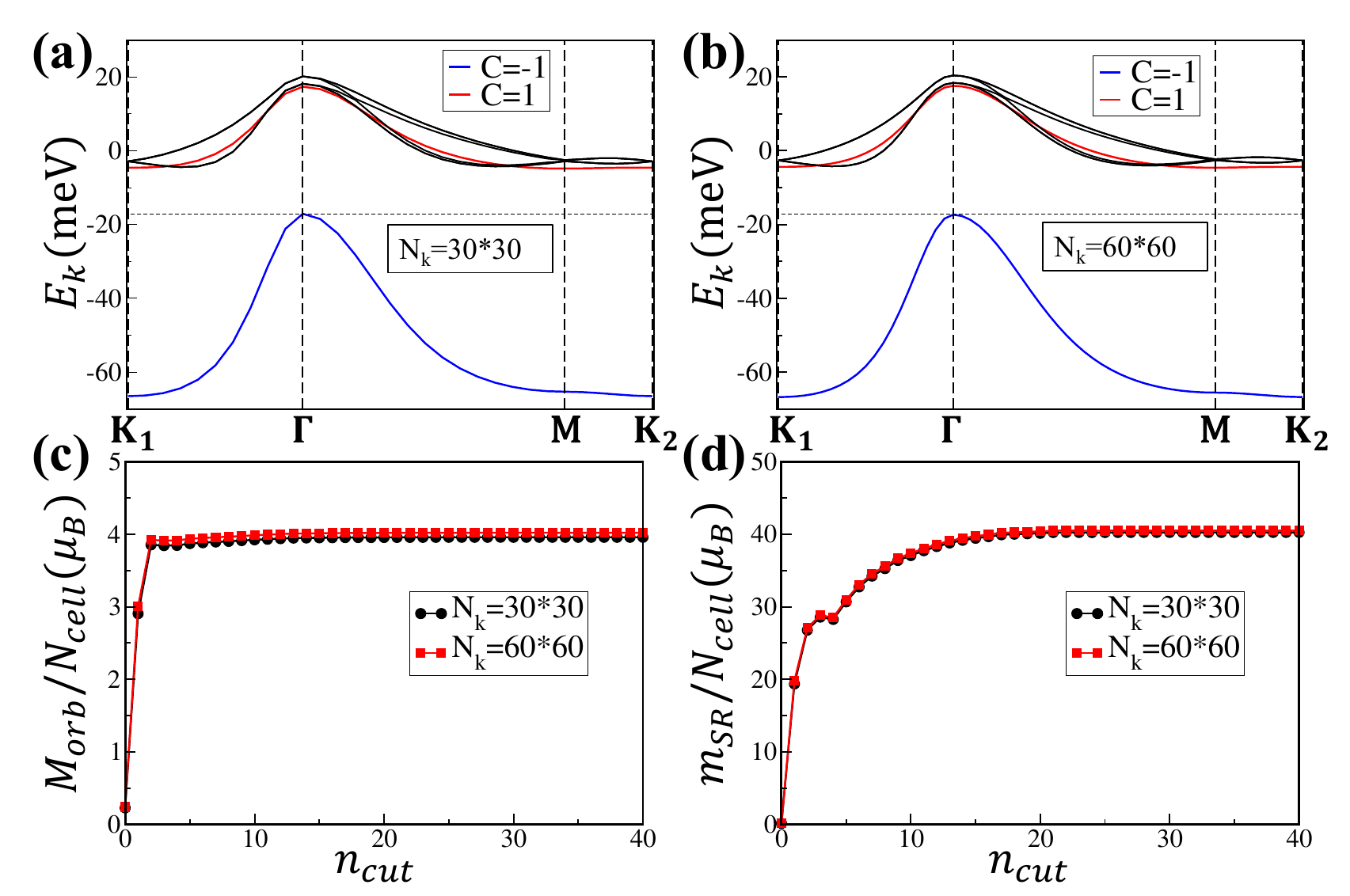}\caption{(a,b) HF band structures of the Chern insulating state at filling $\nu = -3$, obtained using $900$ (a) and $3600$ (b) $k$ points in the mBZ. Bands carrying nontrivial Chern numbers are highlighted in red ($C = 1$) and blue ($C = -1$). 
	    (c,d) Total orbital magnetization $M_{\mathrm{orb}}$ and self-rotation contribution $m_{\mathrm{SR}}$ per moir\'e cell for the states shown in (a) and (b), plotted as functions of the number of remote-band pairs $n_{\mathrm{cut}}$ included in the projection-matrix formalism. Both quantities are well converged with respect to the $k$-mesh density. When evaluating $M_{\mathrm{orb}}$, the chemical potential $\mu$ is fixed at the top of the valence band, as indicated by the horizontal dashed lines in panels (a) and (b).  
			\label{fig:om_nk}}
	\end{center}
	\vskip-0.5cm
\end{figure}

In this section, we present additional numerical benchmarks for the projection-matrix formalism. We first repeat the HF and orbital-magnetization calculations for $\nu=-3$ on a denser $k$-point mesh with $N_k=60\times60=3600$ points in the mBZ. The corresponding HF band structures are shown in Fig.~\ref{fig:om_nk}(a,b), while the resulting $M_{\mathrm{orb}}$ and $m_{\mathrm{SR}}$ are displayed in Fig.~\ref{fig:om_nk}(c,d). The results obtained from the two $k$-point meshes agree within 1\% error, indicating that the calculation is well converged with respect to the $k$-mesh density in the mBZ.

\begin{figure}[htb]
	\begin{center}
		\fig{6.8in}{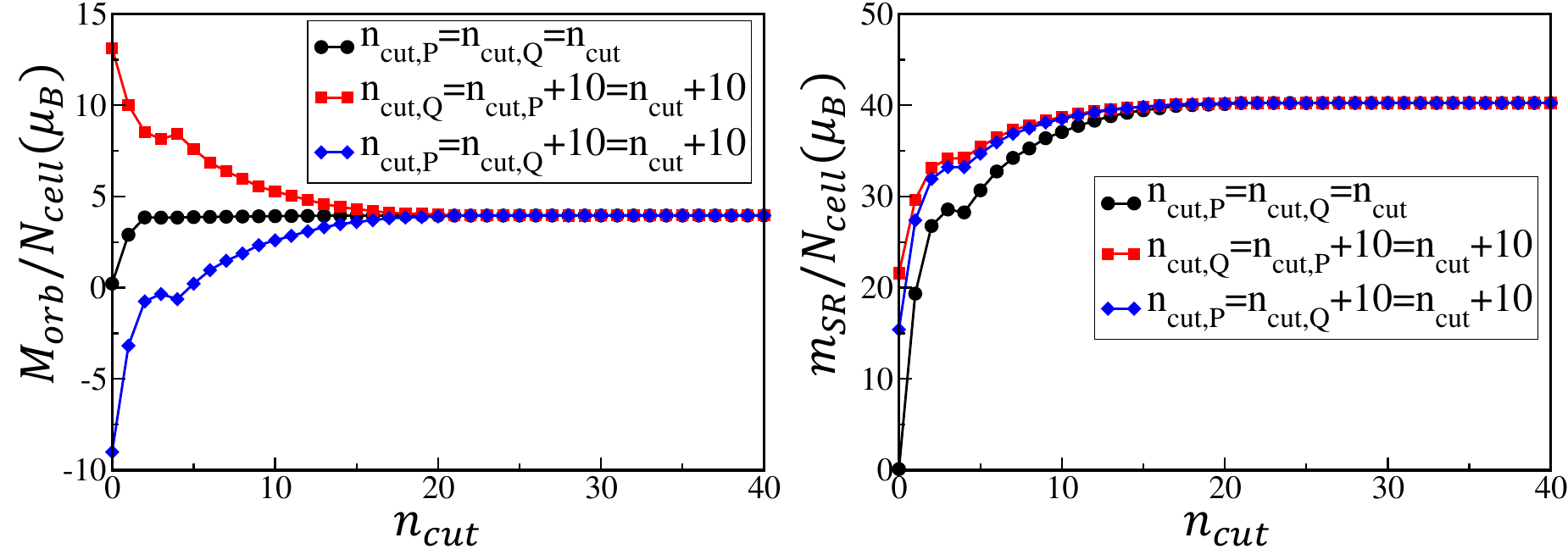}\caption{Total orbital magnetization $M_{\mathrm{orb}}$ (a) and self-rotation contribution $m_{\mathrm{SR}}$ (b) per moir\'e cell for the Chern insulating state at $\nu=-3$, plotted as functions of the independently chosen remote-band cutoffs $n_{{\rm cut},P}$ and $n_{{\rm cut},Q}$ in the projection-matrix formalism. Both quantities show clear convergence around $n_{\mathrm{cut}} \approx 20$, even when $n_{{\rm cut},P}\neq n_{{\rm cut},Q}$. When evaluating $M_{\mathrm{orb}}$, the chemical potential $\mu$ is fixed at the top of the valence band.
			\label{fig:om_PQ}}
	\end{center}
	\vskip-0.5cm
\end{figure}

In the main text, for simplicity, we take the same cutoff $n_{\mathrm{cut}}$ for both projection operators $P_{n_{\rm cut}}(\mathbf{k})$ and $Q_{n_{\rm cut}}(\mathbf{k})$. More generally, the two cutoffs can be chosen independently. To illustrate this point, we repeat the calculation for $\nu=-3$ using $n_{{\rm cut},P}=n_{{\rm cut},Q}+10$ and $n_{{\rm cut},Q}=n_{{\rm cut},P}+10$. As shown in Fig.~\ref{fig:om_PQ}, both $M_{\mathrm{orb}}$ and $m_{\mathrm{SR}}$ exhibit convergence behavior very similar to the case with $n_{{\rm cut},P}=n_{{\rm cut},Q}$.     

We additionally compare the symmetric truncation scheme with the $m_1$ scheme using the HF band structure obtained at $\nu=-3$. As demonstrated by the valley orbital magnetization of the noninteracting, hBN-aligned system considered in Sec.~IV(A) of the main text, the $m_1$ scheme generally exhibits poor numerical stability and does not converge. In the present case, however, it converges at approximately the same cutoff, $n_{\rm cut}\approx20$, as the symmetric truncation scheme. A more detailed examination shows that, although individual terms in the $m_1$ scheme still contain the small energy denominators discussed in the preceding sections, the potentially singular contribution from the remote band sector vanishes because the frozen remote bands preserve $C_2\mathcal T$ symmetry. The apparent stability of the $m_1$ scheme in this particular case is therefore symmetry dependent and should not be interpreted as evidence of its general convergence.

\begin{figure}[htb]
	\begin{center}
		\fig{6.2in}{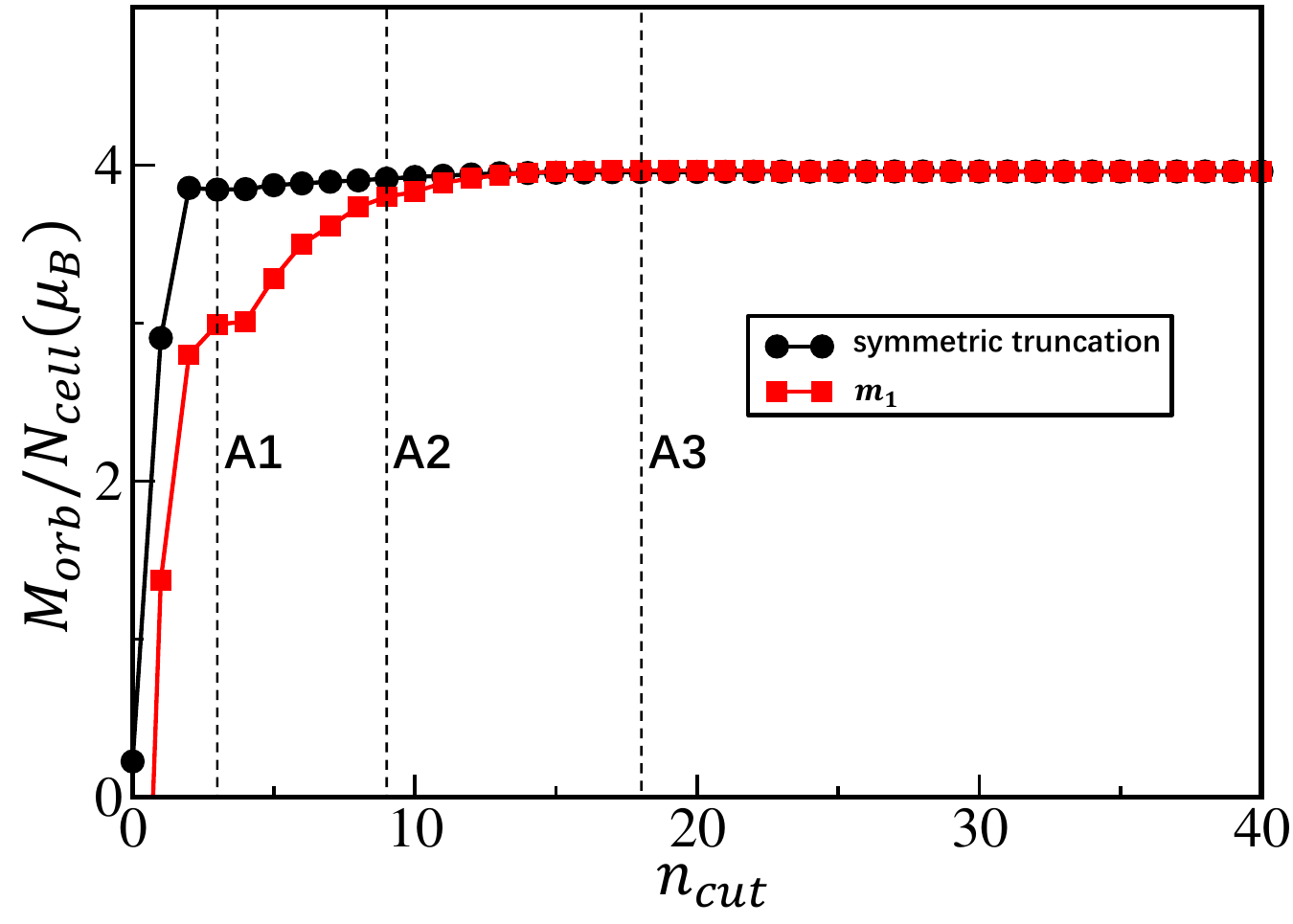}\caption{Calculated total orbital magnetization $M_{\rm orb}/N_{\rm cell}$ for the Chern-insulating HF state at $\nu=-3$ shown in Fig.~\ref{fig:fig3}(a), obtained using the symmetric truncation scheme (black circles) and the $m_1$ scheme (red squares), and plotted as a function of the remote-band cutoff $n_{\rm cut}$. The vertical dashed lines at $n_{\rm cut}=3$, $9$, and $18$ mark the cumulative cutoffs obtained upon completing the momentum-space shells $A1$, $A2$, and $A3$, respectively. Although the $m_1$ result converges more slowly at small cutoffs, both schemes approach the same converged value for $n_{\rm cut}\gtrsim20$. The chemical potential is fixed at the top of the valence band, $\mu=-17.08\,\mathrm{meV}$.
			\label{fig:hf_m1_comp}}
	\end{center}
	\vskip-0.5cm
\end{figure}

\section{Orbital magnetization for $\nu=\pm2$}
\begin{figure}[htb]
	\begin{center}
		\fig{6.8in}{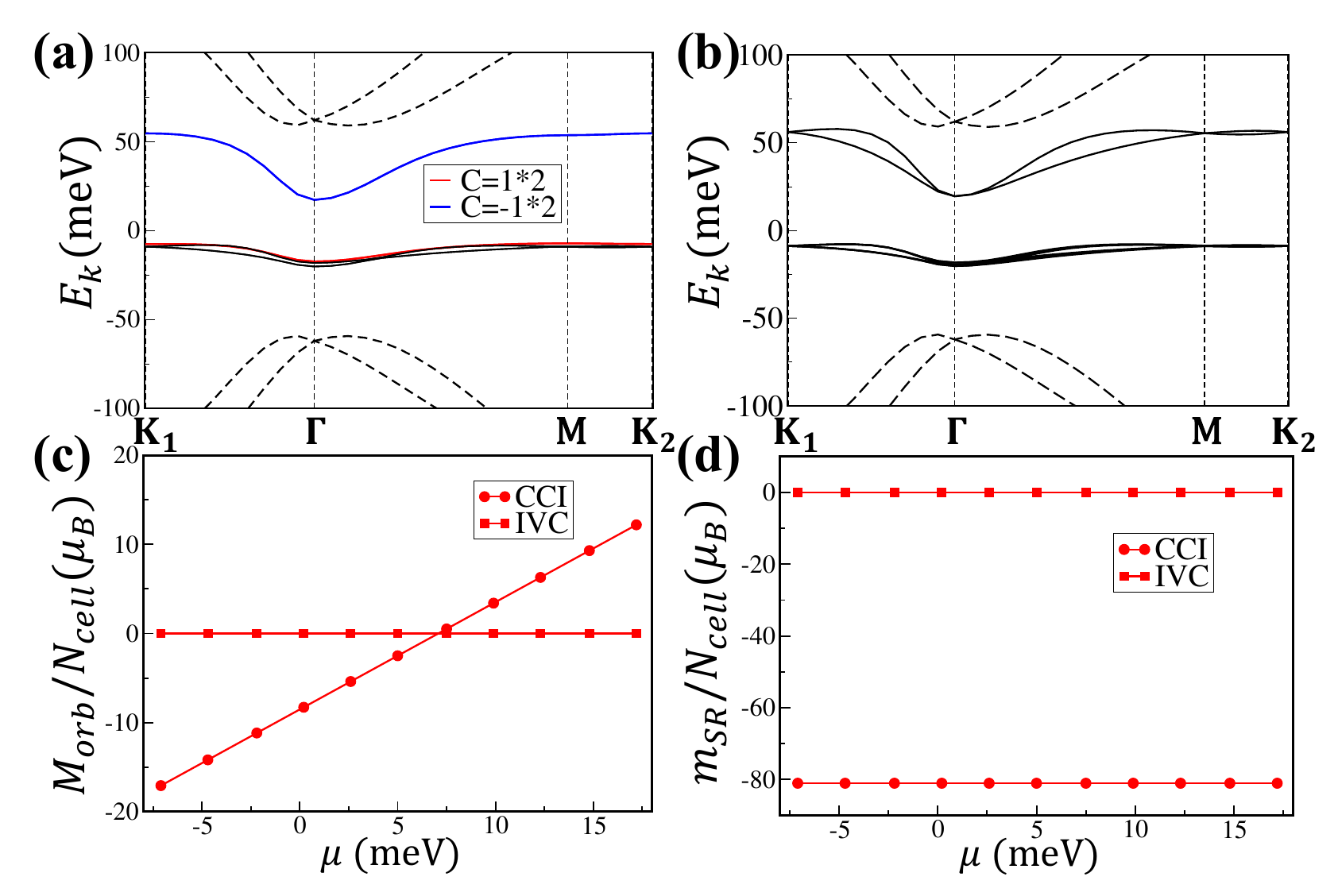}\caption{(a,b) HF band dispersions of the correlated Chern insulating state (CCI) (a) and the intervalley-coherent (IVC) state (b) at filling $\nu=2$. The solid curves represent the eight interaction-renormalized HF bands, while the dashed curves denote the remote bands excluded from the self-consistent HF calculation. Bands carrying nontrivial Chern numbers are highlighted in red ($C=1$) and blue ($C=-1$). For the CCI, the Chern bands are doubly degenerate, yielding a total Chern number $C=2$.
        (c,d) Total orbital magnetization $M_{\mathrm{orb}}$ (c) and self-rotation contribution $m_{\mathrm{SR}}$ (d) per moir\'e cell as functions of the chemical potential $\mu$ inside the insulating gap for the two states shown in (a) and (b). Solid circles correspond to the CCI, and solid squares to the IVC state. For the CCI, $M_{\mathrm{orb}}$ varies linearly with $\mu$ and changes sign across the gap, while $m_{\mathrm{SR}}$ remains constant. For the IVC state, both $M_{\mathrm{orb}}$ and $m_{\mathrm{SR}}$ vanish because all bands are topologically trivial. The number of remote-band pairs is fixed at $n_{\mathrm{cut}}=30$ in the calculation.
			\label{fig:om_nv2}}
	\end{center}
	\vskip-0.5cm
\end{figure}

We next present HF results for fillings $\nu=\pm2$. Owing to the particle-hole symmetry relating $\nu$ and $-\nu$, it is sufficient to focus on $\nu=2$. The Hartree--Fock calculation yields two nearly degenerate correlated insulating states: a correlated Chern insulator (CCI) and an intervalley-coherent (IVC) insulator.
The lower-energy solution is the IVC state, whose bands are topologically trivial, as shown in Fig.~\ref{fig:om_nv2}(b). It is lower in energy by approximately 0.4 meV per moir\'e cell than the CCI state, in which two valence bands break $C_2\mathcal{T}$ symmetry and each carry a finite Chern number. Among the degenerate CCI solutions, we show the state with total Chern number $C=2$, whose band dispersion is plotted in Fig.~\ref{fig:om_nv2}(a).

We then evaluate the orbital magnetization for both states. As shown in Fig.~\ref{fig:om_nv2}(c), the total orbital magnetization $M_{\mathrm{orb}}$ of the CCI depends linearly on the chemical potential within the insulating gap and changes sign across the gap, whereas $M_{\mathrm{orb}}$ vanishes for the IVC state because its bands are topologically trivial. This indicates that the CCI is energetically favored over the IVC state in the presence of a finite magnetic field. The self-rotation contribution $m_{\mathrm{SR}}$, shown in Fig.~\ref{fig:om_nv2}(d), is likewise sizable for the CCI but vanishes for the IVC state.
Using the magnetization of CCI in the center of the gap and the energy difference between the two insulating states, we also estimate a critical magnetic field for the transition from the IVC state to the CCI of $B_c \sim \delta E/\delta M_{\mathrm{orb}} \approx 9.2$ T.

\section{Cutoff on Remote bands}
In this section, we provide a physical interpretation of the remote band cutoff $n_{\rm cut}$. For the parameters considered in the main text, we find numerically that convergence is achieved for $n_{\rm cut} \sim 20$. This value can be understood from the localization properties of the BM model eigenstates in momentum space.

As an illustrative example, we consider eigenstates at the corner of the mBZ, denoted by $\fvec K_t$. As shown in Eq.~\eqref{eq:bloch}, the corresponding Bloch wavefunctions can be expanded as
\begin{align}
    \psi^X_{m, \tau, \fvec k}(\fvec r) = \sum_{\fvec G} u_{m, \tau, \fvec G, X} e^{i (\fvec k + \fvec G) \cdot \fvec r} \ ,
\end{align}
where $X=(l,s)$ labels the layer $l$ and sublattice $s$, $\fvec G$ is a moir\'e reciprocal lattice vector, and $m$ and $\tau$ denote the band and valley indices, respectively.

\begin{figure}[htb]
	\begin{center}
		\fig{6.8in}{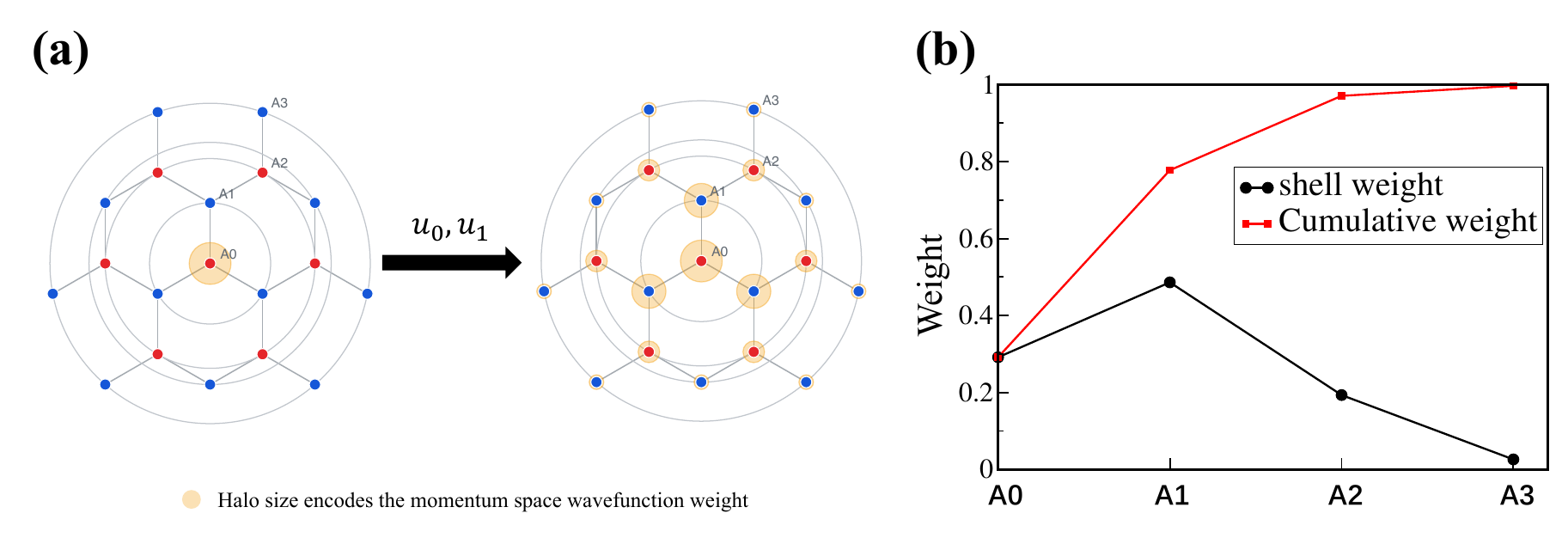}\caption{Momentum space localization of an active-band wavefunction in the BM model. (a) Schematic momentum space honeycomb lattice centered at the mBZ corner $\fvec K_t$. In the decoupled limit $u_0=u_1=0$, the momentum sites are disconnected and the active band wavefunctions at $\fvec K_t$  are localized at the central site in shell $A0$. Finite intra- and inter-sublattice interlayer tunneling amplitudes $u_0$ and $u_1$ hybridize neighboring momentum sites and extend the wavefunction over successive graph-distance shells $An$. The area of each orange halo schematically represents the wavefunction weight on the corresponding momentum site. (b) Calculated shell-resolved weight (black circles) and cumulative weight (red squares) of the active-band state at $\fvec K_t$. The calculation uses the same BM-model parameters as in the main text. The rapid saturation of the cumulative weight demonstrates that the active-band wavefunction is strongly localized within the first few momentum shells.
			\label{fig:shell}}
	\end{center}
\end{figure}

The structure of these Bloch wavefunctions, particularly those in the active bands, can be understood by viewing the BM model as an effective tight-binding model in the momentum space. In this representation, each vector $\fvec G + \fvec K_1$ and $\fvec G + \fvec K_2$ corresponds to a momentum space site whose on-site kinetic energy is $\hbar v_F|\fvec G|$ and $\hbar v_F|\fvec G + \fvec K_2 - \fvec K_1|$. The interlayer tunneling amplitudes $u_0$ and $u_1$ describe intra-sublattice and inter-sublattice hopping between neighboring momentum space sites, respectively. In the limit $u_0 = u_1 = 0$, as shown in Fig.~\ref{fig:shell}(a), these sites are decoupled, and each eigenstate is completely localized on an individual momentum space site. States associated with sites in the same momentum shell are then degenerate.

When the interlayer tunnelings are switched on, electrons can hop between neighboring momentum space sites, leading the eigenstates to extend over several momentum shells. The degree of this momentum space delocalization is controlled primarily by two dimensionless parameters
\[ \frac{u_0}{\hbar v_F K_{\theta}} \quad \mbox{and} \quad \frac{u_1}{\hbar v_F K_{\theta}} \quad \mbox{with} \ K_{\theta} = 2 |\fvec K| \sin\frac{\theta}2 = \frac{8 \pi}{3a_0} \sin\frac{\theta}2  \  .  \]
Nevertheless, the momentum shell structure established in the decoupled limit remains qualitatively valid: each eigenstate retains most of its spectral weight within a characteristic momentum shell, with additional weight distributed over neighboring shells. In particular, the active band wavefunctions are predominantly localized near the central momentum space sites highlighted in Fig.~\ref{fig:shell}(a), while acquiring finite weight on nearby shells through interlayer tunneling.

For the BM model parameters used in the main text, Fig.~\ref{fig:shell}(b) shows that the active band wavefunctions are concentrated primarily within the three innermost momentum shells. Consequently, remote band states with substantial weight in these shells have appreciable overlap with the active band states and can therefore contribute significantly to $M_{\rm orb}$. This overlap, rather than the proximity of the remote band energies to the charge neutrality point, determines the importance of a given remote band.

The three innermost shells contain $18$ momentum space sites. Each site contributes one positive energy and one negative energy state, corresponding to $18$ remote bands on each side of charge neutrality. This estimate is consistent with the numerically determined cutoff $n_{\rm cut} \sim 20$ and provides a physical interpretation of the observed convergence scale.

This shell correspondence is also visible in Fig.~\ref{fig:hf_m1_comp}, where the vertical dashed lines at $n_{\rm cut}=3$, $9$, and $18$ mark completion of the $A1$, $A2$, and $A3$ shells, respectively. Most of the slower cutoff dependence of the $m_1$ result occurs while these shells are being completed, and only small changes remain beyond $n_{\rm cut}=18$, further supporting the momentum-space-shell interpretation of the convergence scale.

\section{Further decomposition of $M_{\rm orb}$}
In the main text and the preceding sections of the Supplemental Material, we examined the stability and convergence of the two truncation schemes and explained why the $m_1$ scheme is numerically unstable and fails to converge. In this section, we further decompose $M_{\rm orb}$ into four contributions according to whether the occupied and empty states belong to the active or remote bands.
\begin{align}
    P^{A-}(\fvec k) & =   \sum_{n \in \mathrm{occupied}} |u^{\rm A}_n(\mathbf{k})\rangle\langle u^{\rm A}_n(\mathbf{k})| &
    P^{R-}_{\rm ncut}(\fvec k) & = \sum_{n \le n_{\rm cut}} |u^{R-}_n(\mathbf{k})\rangle \langle u^{R-}_n(\mathbf{k})|  \\
    Q^{A+}(\fvec k) & = \sum_{n \in \mathrm{empty}} |u^{\rm A}_n(\mathbf{k})\rangle\langle u^{\rm A}_n(\mathbf{k})| &
    Q^{R+}_{\rm ncut}(\fvec k) & = \sum_{n \le n_{\rm cut}} |u^{R+}_n(\mathbf{k})\rangle \langle u^{R+}_n(\mathbf{k})| \  .
\end{align}
Here, $P^{A-}$ and $P_{n_{\rm cut}}^{R-}$ project onto the occupied active and remote bands, respectively, whereas $Q^{A+}$ and $Q_{n_{\rm cut}}^{R+}$ project onto the corresponding empty band sectors.

The quantities $W_{\mu\nu}$ and $N_{\mu\nu}$ defined in Eqs.~\eqref{Eqn:Wmunu} and \eqref{Eqn:Nmunu} can then be decomposed into contributions from the four combinations:
\[ \big(  A- ; A+ \big) \qquad  \big(  A- ; R+ \big) \qquad \big(  R- ; A+ \big) \quad \mbox{and} \quad \big(  R- ; R+ \big)  \  . \]
More explicitly, the four contributions to $W_{\mu\nu}$ are
\begin{align}
    W_{\mu\nu}^{(A-;A+)}(\fvec k) & = \mtr\left[ P^{A-}(\fvec k) \big(\pd_{\mu} P^{A-}(\fvec k) \big) \big(\pd_{\nu} Q^{A+}(\fvec k)\big) \big( H^{\mathrm{HF}}(\fvec k)-\mu\big) \right] \   , \label{Eqn:WmunuAA} \\
    W_{\mu\nu}^{(A-; R+)}(\fvec k) & = \mtr\left[ P^{A-}(\fvec k) \big( \pd_{\mu} P^{A-}(\fvec k) \big) \big(\pd_{\nu} Q^{R+}(\fvec k) \big) \big( H^{\mathrm{HF}}(\fvec k)-\mu\big) \right] \   , \label{Eqn:WmunuAR} \\
    W_{\mu\nu}^{(R-; A+)}(\fvec k) & = \mtr\left[ P^{R-}(\fvec k) \big(\pd_{\mu} P^{R-}(\fvec k) \big) \big(\pd_{\nu} Q^{A+}(\fvec k) \big) \big(H^{\mathrm{HF}}(\fvec k)-\mu\big) \right] \   , \label{Eqn:WmunuRA} \\
    W_{\mu\nu}^{(R-; R+)}(\fvec k) & = \mtr\left[ P^{R-}(\fvec k) \big(\pd_{\mu} P^{R-}(\fvec k) \big) \big(\pd_{\nu} Q^{R+}(\fvec k) \big) \big(H^{\mathrm{HF}}(\fvec k)-\mu\big) \right] \   , \label{Eqn:WmunuRR} 
\end{align}
where for simplicity, we omit the subscript $\rm ncut$ in $P^{R-}$ and $Q^{R+}$.  The tensor $N_{\mu\nu}$ can be decomposed analogously into the same four sectors.

\begin{figure}[htb]
	\begin{center}
		\fig{6.8in}{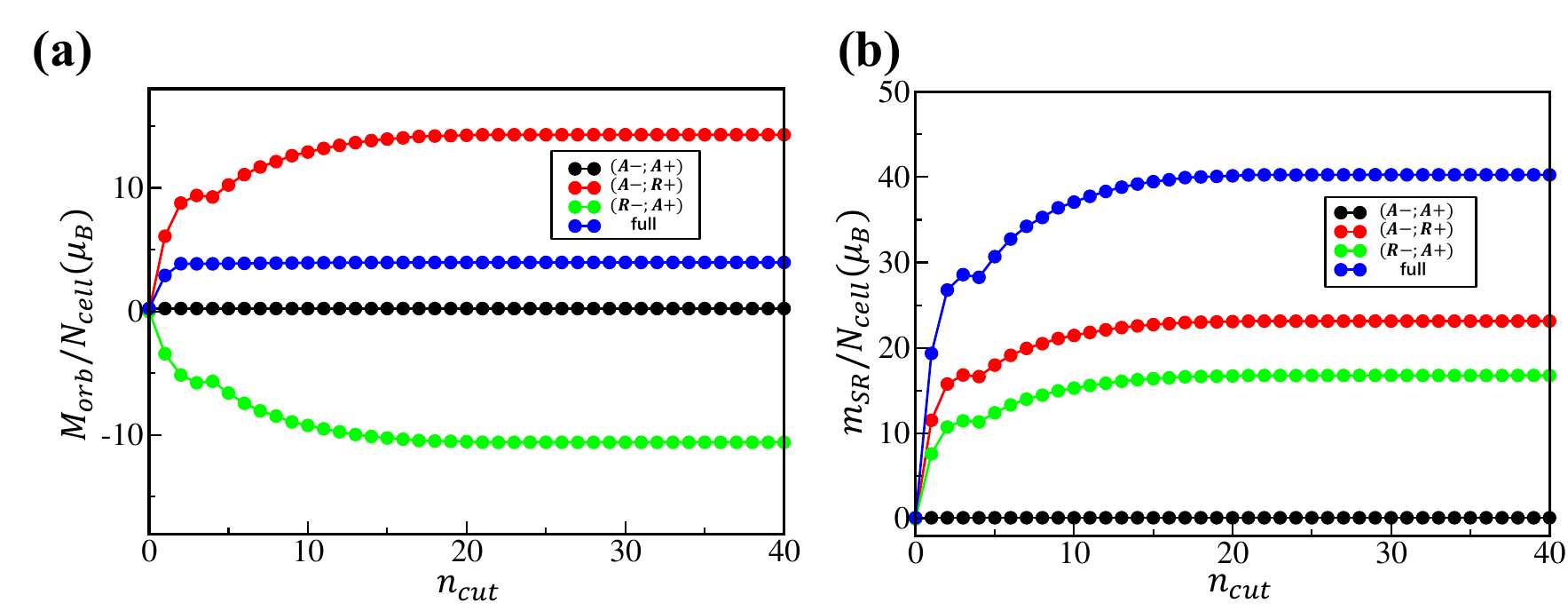}\caption{Sector resolved contributions to the orbital magnetization of the Chern-insulating state at filling $\nu=-3$, with the chemical potential fixed at the valence band maximum, $\mu=-17.08\,\mathrm{meV}$, as functions of the remote band cutoff $n_{\rm cut}$ used in the symmetric-truncation scheme. (a) Total orbital magnetization $M_{\rm orb}/N_{\rm cell}$ and (b) self-rotation contribution $m_{\rm SR}/N_{\rm cell}$, both in units of $\mu_B$. The black, red, and green curves denote the $(A-; A+)$, $(A-;R+)$, and $(R-;A+)$ sectors, respectively, whereas the blue curve shows the directly evaluated full result. The $(R-; R+)$ contribution vanishes within numerical precision and is therefore not shown. The mixed active-remote sectors provide the dominant contributions, and the results converge for $n_{\rm cut}\gtrsim 20$.
			\label{fig:om_sector}}
	\end{center}
	\vskip-0.5cm
\end{figure}

\begin{table}[htb]
\caption{Converged sector-resolved contributions to the orbital magnetization and its self-rotation part. All values are per moir\'e unit cell and are given in units of $\mu_B$. The full quantities are evaluated directly rather than obtained by summing the four separately evaluated sectors. The small differences between the directly evaluated full quantities and the sums of their sector contributions originate from the numerical momentum derivatives on the finite $k$ mesh.}
\label{tab:om_sector}
\begin{ruledtabular}
\begin{tabular}{lcc}
Sector & $M_{\rm orb}/N_{\rm cell}$  & $m_{\rm SR}/N_{\rm cell}$  \\
\hline
$(A-;A+)$ & $0.225$ & $0.109$ \\
$(A-;R+)$ & $14.29$ & $23.14$ \\
$(R-;A+)$ & $-10.59$ & $16.77$ \\
$(R-;R+)$ & $0.0$ & $0.0$ \\
Full & $3.96$ & $40.26$ \\
\end{tabular}
\end{ruledtabular}
\end{table}

Within the frozen remote band approximation employed in our self-consistent HF calculations, $C_2 \mathcal T$ is broken only in the active band sector. Consequently, the $(R-;R+)$ sector does not contribute to the orbital magnetization. The $(A-; A+)$ sector contains only active band contributions and therefore coincides with the result obtained for $n_{\rm cut} = 0$. By contrast, the mixed sectors $(R-; A+)$ and $(A-; R+)$ provide sizable contributions. These contributions depend appreciably on $n_{\rm cut}$ but converge once $n_{\rm cut}\gtrsim 20$.
The cutoff dependence is shown in Fig.~\ref{fig:om_sector}, and the corresponding converged values are summarized in Table~\ref{tab:om_sector}. For the parameters considered in our manuscript, both $M_{\rm orb}$ and $m_{\rm SR}$ are dominated by these mixed active-remote channels, whereas the purely active $(A-; A+)$ sector contributes only weakly. Additionally, we note that the two mixed contributions partially cancel in $M_{\rm orb}$ but add constructively in $m_{\rm SR}$.

\section{Orbital Magnetization at $\nu = -3$ in the Chiral Limit}

In the chiral limit, each valley and spin hosts two exactly flat bands with Chern numbers $C=\pm1$, which are fully polarized on opposite sublattices. This property allows the interacting ground states at integer fillings to be obtained analytically. Up to symmetry transformations, the ground state at $\nu=-3$ consists of a single occupied active band that is fully polarized in spin, valley, and sublattice, while the remaining seven active bands are empty and degenerate. We denote the occupied active band by $A-$ and all the empty active bands with the same spin and valley but the opposite sublattice polarization by $A_+$. The contribution of this pair to the orbital magnetization is
\begin{align}
    M_{\rm orb}^{(A-; A+)} = \frac{e}{2i\hbar}\epsilon_{\mu\nu} \sum_{\fvec k} \left\langle \pd_\mu u^{A-} \middle| u^{A+} \right\rangle \left( E_{A-}^{\rm HF} + E_{A+}^{\rm HF} - 2\mu \right) \left\langle u^{A+}\middle| \pd_\nu u^{A-} \right\rangle \ .
\end{align}
Every empty active band differs from the occupied band in at least one of its spin, valley, or sublattice quantum numbers. Because these internal quantum numbers are independent of momentum, momentum differentiation does not mix the corresponding sectors. Consequently,
\begin{align}
    \left\langle \pd_\mu u^{A-}\middle|u^{A+} \right\rangle = 0
\end{align}
for all empty active bands. It follows that all active-active interband contributions to both $M_{\rm orb}$ and $m_{\rm SR}$ vanish when only the active bands are retained, corresponding to $n_{\rm cut}=0$.

We next consider the contributions to the orbital magnetization arising from transitions between the active and remote bands, namely, the $(A-;R+)$ and $(R-;A+)$ processes. As shown above, the active--active contribution vanishes in the chiral limit. Moreover, the purely remote contribution vanishes because the remote-band sector remains $C_2\mathcal T$ symmetric. The remaining orbital magnetization can therefore be decomposed as follows. For simplicity, we set $\mu=0$:
\begin{align}
    M_{\rm orb} & =  M_{\rm orb}^{(A-; R+)} + M_{\rm orb}^{(R-; A+)}  \nonumber \\
    M_{\rm orb}^{(A-; R+)}  & = \frac{e}{2i \hbar} \epsilon_{\mu\nu} \sum_{\fvec k}   \mtr\left[  \big( \pd_{\mu} P_A  \big) \hat Q_R  \hat H_{\rm BM} \big( \pd_{\nu} \hat P_A \big) - \hat Q_R \big( \pd_{\mu} \hat P_A \big) \hat H_{\rm HF} \big( \pd_{\nu} \hat P_A \big) \right] \label{Eqn:MA-R+} \\
    M_{\rm orb}^{(R-; A+)}  & = \frac{e}{2i \hbar} \epsilon_{\mu\nu} \sum_{\fvec k}   \mtr\left[  \big( \pd_{\mu} P_R  \big) \hat Q_A  \hat H_{\rm HF} \big( \pd_{\nu} \hat P_R \big) - \hat Q_A \big( \pd_{\mu} \hat P_R \big) \hat H_{\rm BM} \big( \pd_{\nu} \hat P_R \big)  \right]  \label{Eqn:MR-A+}
\end{align}
Here, $\hat P_A$ and $\hat Q_A$ denote the projectors onto the occupied and unoccupied active band subspaces, respectively, while $\hat P_R$ and $\hat Q_R$ denote the corresponding projectors onto the occupied and unoccupied remote band subspaces.

Because the active bands are fully sublattice polarized in the chiral limit, both active-band projectors commute with the sublattice operator $\sigma_z$:
\begin{align}
    \sigma_z \hat P_A \sigma_z = \hat P_A \ ,\qquad \sigma_z \hat Q_A \sigma_z = \hat Q_A \ .
\end{align}
The restriction of the HF Hamiltonian to the active-band subspace also commutes with $\sigma_z$. In addition, the chiral symmetry generated by $\sigma_z$\footnote{The unitary chiral symmetry considered here is distinct from the particle--hole symmetry discussed in the main text and is exact only in the chiral limit of the BM model.} implies
\begin{align}
    \sigma_z \hat H_{\rm BM}\sigma_z & = -\hat H_{\rm BM} \ ,  &  \sigma_z \hat P_R \sigma_z & = \hat Q_R \ , &  \sigma_z \hat Q_R \sigma_z & = \hat P_R \ . \label{eq:chiralPH}
\end{align}
Applying this transformation to the first term of Eq.~\eqref{Eqn:MA-R+} gives
\begin{align}
    \mtr\left[ \big( \pd_{\mu} P_A  \big) \hat Q_R  \hat H_{\rm BM} \big( \pd_{\nu} \hat P_A \big) \right] = - \mtr\left[ \big( \pd_{\mu} P_A  \big) \hat P_R  \hat H_{\rm BM} \big( \pd_{\nu} \hat P_A \big) \right] = - \mtr\left[ \hat P_A \big( \pd_{\mu} \hat P_R   \big)  \hat H_{\rm BM} \big( \pd_{\nu} \hat P_R \big) \right] \label{Eqn:PH1}
\end{align}
where, in the second equality, we used the orthogonality of the active and remote projectors and the identities obtained by differentiating $\hat P_A \hat P_R = 0$.

Combining Eq.~\eqref{Eqn:PH1} with the second term of Eq.~\eqref{Eqn:MR-A+}, the terms involving $\hat H_{\rm BM}$ reduce to
\begin{align}
    -\frac{e}{2 i \hbar} \epsilon_{\mu\nu}  \sum_{\fvec k} \mtr\left[ \bigl( \hat P_A + \hat Q_A \bigr) \bigl( \pd_\mu \hat P_R \bigr) \hat H_{\rm BM} \bigl( \pd_\nu \hat P_R \bigr) \right]   \   . \label{Eqn:PH2}
\end{align}
The combination $\hat P_A+\hat Q_A$ is the projector onto the complete active band subspace and is therefore invariant under $C_2\mathcal T$. The remaining operators in Eq.~\eqref{Eqn:PH2} are also $C_2\mathcal T$ symmetric. Because the orbital magnetization is odd under $C_2\mathcal T$, the antisymmetric contraction in Eq.~\eqref{Eqn:PH2} vanishes. Consequently, the terms proportional to $\hat H_{\rm BM}$ do not contribute to $M_{\rm orb}$.

Similarly, applying the chiral transformation to the second term in $M_{\rm orb}^{(A-;R+)}$ gives
\begin{align}
   - \mtr\left[ \hat Q_R \big( \pd_{\mu} \hat P_A \big) \hat H_{\rm HF} \big( \pd_{\nu} \hat P_A \big) \right] = - \mtr\left[ \hat P_R \big( \pd_{\mu} \hat P_A \big) \hat H_{\rm HF} \big( \pd_{\nu} \hat P_A \big) \right] = - \mtr\left[ \big( \pd_{\mu} \hat P_R \big) \hat P_A \hat H_{\rm HF} \big( \pd_{\nu} \hat P_R \big) \right] \label{eq:PH3}
\end{align}
where we have used the invariance of the active-space HF Hamiltonian under $\sigma_z$, together with the orthogonality relation $\hat P_A \hat P_R = 0$. Combining Eq.~\eqref{eq:PH3} with the first term in $M_{\rm orb}^{(R-;A+)}$ yields
\begin{align}
   M_{\rm orb}(\mu = 0) & = \frac{e}{2i \hbar} \epsilon_{\mu\nu} \sum_{\fvec k} \mtr\left[ \big( \pd_{\mu} \hat P_R \big) \big( \hat Q_A - \hat P_A \big) \hat H_{\rm HF} \big( \pd_{\nu} \hat P_R \big) \right] \label{eq:PH5}
\end{align}

Suppose that the occupied active band belongs to the spin valley sector $(s, \tau)$. The $C_2\mathcal T$ symmetry is broken only in this sector, whereas all other spin valley sectors remain $C_2\mathcal T$ symmetric and therefore do not contribute to the orbital magnetization. We may thus restrict the following discussion to the $(s, \tau)$ sector. Within this sector, $\hat P_A$ and $\hat Q_A$ denote the rank one projectors onto the occupied and unoccupied active bands, respectively. The restriction of the HF Hamiltonian to the active band subspace can be written as
\begin{align}
    \hat P_{\rm act} \hat H_{\rm HF} \hat P_{\rm act} = a(\fvec k) \bigl( \hat P_A + \hat Q_A \bigr) + b(\fvec k) \bigl( \hat Q_A - \hat P_A \bigr) \ , \qquad \hat P_{\rm act} = \hat P_A + \hat Q_A \ ,   \label{eq:Parameter}
\end{align}
where
\begin{align}
    a(\fvec k) & = \frac{ E_{A-}^{\rm HF}(\fvec k) + E_{A+}^{\rm HF}(\fvec k) }2 \ , & b(\fvec k) & = \frac{ E_{A+}^{\rm HF}(\fvec k) - E_{A-}^{\rm HF}(\fvec k) }2  \ .
\end{align}
Thus, $a(\fvec k)$ specifies the midpoint of the local HF gap, while $2b(\fvec k)$ is the corresponding gap size.

Substituting Eq.~\eqref{eq:Parameter} into Eq.~\eqref{eq:PH5}, and using
\begin{align}
    \bigl( \hat Q_A - \hat P_A \bigr) \bigl( \hat P_A + \hat Q_A \bigr) & = \hat Q_A - \hat P_A \ , & \bigl( \hat Q_A - \hat P_A \bigr)^2 & = \hat P_A + \hat Q_A  \ ,
\end{align}
we obtain
\begin{align}
        M_{\rm orb}(\mu = 0) & = \frac{e}{2i \hbar} \epsilon_{\mu\nu} \sum_{\fvec k} \mtr\left[ \big( \pd_{\mu} \hat P_R \big) \left( a(\fvec k) \big( \hat Q_A - \hat P_A  \big) + b(\fvec k) \left( \hat Q_A + \hat P_A  \right)  \right) \big( \pd_{\nu} \hat P_R \big) \right] \nonumber \\
    & = \frac{e}{2i \hbar} \epsilon_{\mu\nu} \sum_{\fvec k} a(\fvec k) \mtr\left[ \big( \pd_{\mu} \hat P_R \big) \big( \hat Q_A - \hat P_A  \big) \big( \pd_{\nu} \hat P_R \big)   \right] \   .  \label{eq:PH4}
\end{align}
In the second equality, we used the fact that the term proportional to $b(\fvec k)$ is $C_2\mathcal T$ symmetric and therefore vanishes upon contraction with $\epsilon_{\mu\nu}$.

Using the orthogonality of the active and remote projectors, Eq.~\eqref{eq:PH4} can be recast as
\begin{align}
    M_{\rm orb}(\mu=0) = \frac{e}{2 i \hbar} \epsilon_{\mu\nu} \sum_{\fvec k} a(\fvec k) \mtr\left[ \hat P_R \left\{ \bigl( \pd_\mu \hat Q_A \bigr) \bigl( \pd_\nu \hat Q_A \bigr) - \bigl( \pd_\mu \hat P_A \bigr) \bigl( \pd_\nu \hat P_A \bigr) \right\} \right]  \ . \label{eq:PH6}
\end{align}
The chiral symmetry exchanges the occupied and unoccupied remote-band projectors while leaving the active-band projectors invariant. Consequently, Eq.~\eqref{eq:PH6} is unchanged upon replacing $\hat P_R$ by $\hat Q_R$. Averaging the two equivalent expressions gives
\begin{align}
    M_{\rm orb}(\mu=0) = \frac{e}{4 i \hbar} \epsilon_{\mu\nu} \sum_{\fvec k} a(\fvec k) \mtr\left[ \bigl( \hat P_R + \hat Q_R \bigr) \left\{ \bigl( \pd_\mu \hat Q_A \bigr) \bigl( \pd_\nu \hat Q_A \bigr) - \bigl( \pd_\mu \hat P_A \bigr) \bigl( \pd_\nu \hat P_A \bigr) \right\} \right] \ . \label{eq:PH7}
\end{align}

Because $\hat P_A$ and $\hat Q_A$ are polarized on opposite sublattices, they satisfy
\begin{align}
    \hat P_A \bigl( \pd_\mu \hat Q_A \bigr) = 0 \ , \qquad \hat Q_A \bigl( \pd_\mu \hat P_A \bigr) = 0 \ .
\end{align}
We may therefore add $\hat P_A$ to the first projector factor in Eq.~\eqref{eq:PH7} and $\hat Q_A$ to the second without changing the result. Using the completeness relation
\begin{align}
    \hat P_R + \hat Q_R + \hat P_A + \hat Q_A = \hat I \ ,
\end{align}
we obtain
\begin{align}
    M_{\rm orb}(\mu=0) = \frac{e}{4i\hbar} \epsilon_{\mu\nu} \sum_{\fvec k} a(\fvec k) \mtr\left[ \bigl( \hat I - \hat Q_A \bigr) \bigl( \pd_\mu \hat Q_A \bigr) \bigl( \pd_\nu \hat Q_A \bigr) - \bigl(\hat I - \hat P_A \bigr) \bigl( \pd_\mu \hat P_A \bigr) \bigl( \pd_\nu \hat P_A \bigr) \right] \ . \label{eq:PH8}
\end{align}

The Berry curvatures of the two active bands can be expressed as
\begin{align}
    \epsilon_{\mu\nu} \mtr\left[ \bigl( \hat I - \hat Q_A \bigr) \bigl( \pd_\mu \hat Q_A \bigr) \bigl( \pd_\nu \hat Q_A \bigr) \right] & = i \mathcal F^{A+}(\fvec k) \ , & \epsilon_{\mu\nu} \mtr\left[ \bigl( \hat I - \hat P_A \bigr) \bigl( \pd_\mu \hat P_A \bigr) \bigl( \pd_\nu \hat P_A \bigr) \right] & = i \mathcal F^{A-}(\fvec k) \ .
\end{align}
Since the two active bands are related by $C_2\mathcal T$ symmetry, their Berry curvatures satisfy
\begin{align}
    \mathcal F^{A+}(\fvec k) = - \mathcal F^{A-}(\fvec k) \  .
\end{align}
Equation~\eqref{eq:PH8} therefore reduces to
\begin{align}
    M_{\rm orb}(\mu=0) = -\frac{e}{2\hbar} \left\langle a(\fvec k) \mathcal F^{A-}(\fvec k) \right\rangle_{\rm mBZ} \ . \label{eq:PH9}
\end{align}
If $a(\fvec k)$ varies only weakly across the moir\'e Brillouin zone, it may be replaced by its mBZ average. With the momentum space normalization chosen such that
\begin{align}
    \left\langle \mathcal F^{A-}(\fvec k) \right\rangle_{\rm mBZ} = C \ ,
\end{align}
we finally obtain
\begin{align}
    M_{\rm orb}(\mu=0) \approx -\frac{e}{2\hbar} C \left\langle a(\fvec k)\right\rangle_{\rm mBZ}  \ ,
    \label{eq:PH10}
\end{align}
where $a(\fvec k)$ is the midpoint energy of the HF gap. 
For the nonzero $\mu$ inside the gap,  Eq.~\ref{eq:PH9} and Eq.~\ref{eq:PH10} become 
\begin{align}
	M_{\rm orb}(\mu) = -\frac{e}{2\hbar} \left\langle \left[a(\fvec k)-2\mu\right] \mathcal F^{A-}(\fvec k)  \right\rangle_{\rm mBZ} 
	 = -\frac{e}{2\hbar} \left\langle a(\fvec k) \mathcal F^{A-}(\fvec k) \right\rangle_{\rm mBZ} + \frac{\mu e}{\hbar} C \label{eq:PH11}
\end{align}
and
\begin{align}
	M_{\rm orb}(\mu) \approx -\frac{e}{2\hbar} C \left\langle a(\fvec k)-2\mu\right\rangle_{\rm mBZ}
	= -\frac{e}{2\hbar} C \left\langle a(\fvec k)\right\rangle_{\rm mBZ}  + \frac{\mu e}{\hbar} C \ .
	\label{eq:PH12}
\end{align}

\begin{figure}[htb]
	\begin{center}
		\fig{6.8in}{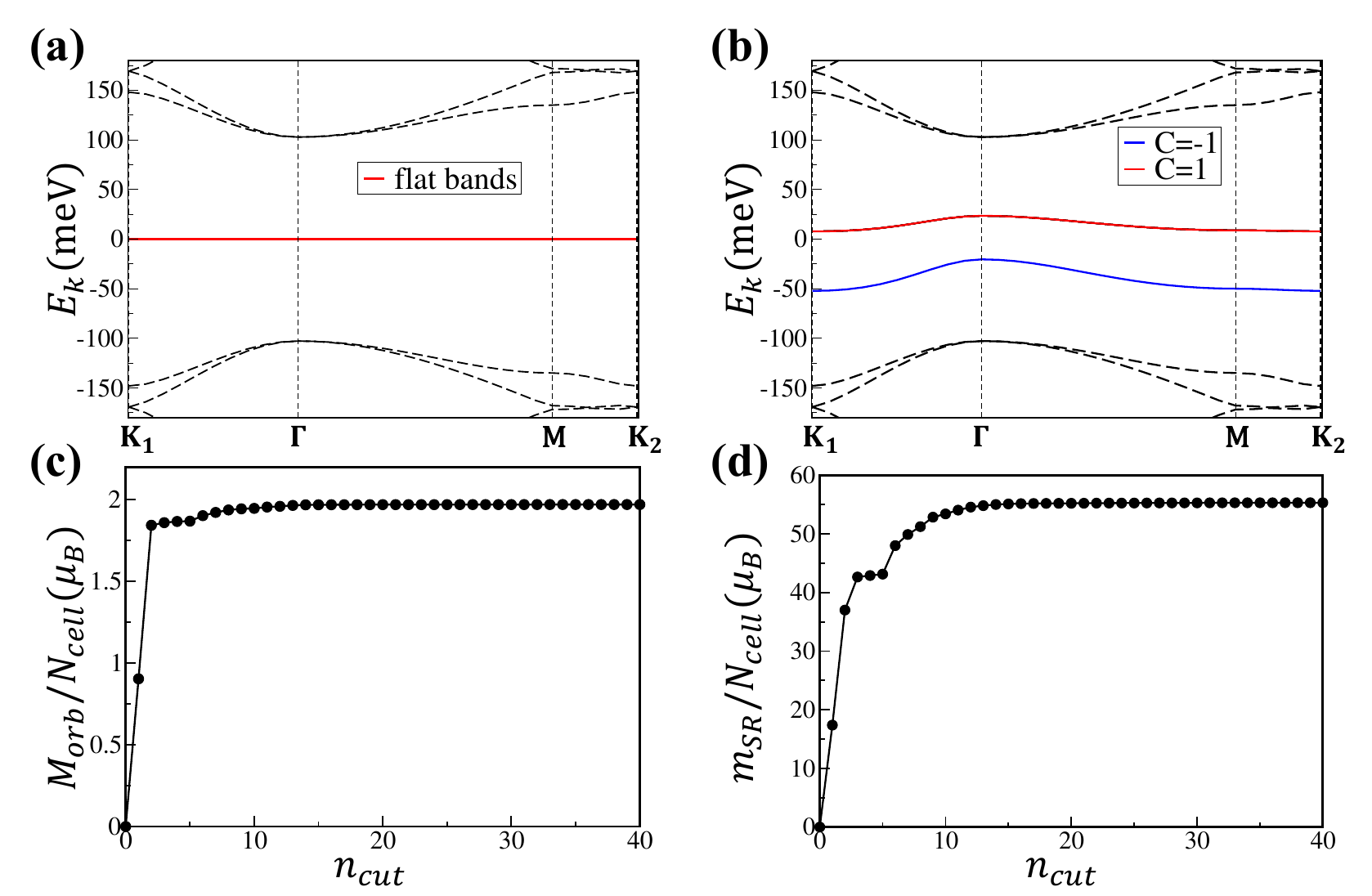}\caption{Band structures and orbital magnetization in the chiral limit, $u_0=0$, at filling $\nu=-3$. (a) Noninteracting BM band structure along the high-symmetry path. The two active bands (red) are exactly flat at charge neutrality, while the remote bands are shown as black dashed curves. (b) Self-consistent HF band structure of the Chern-insulating state. The active Chern bands with $C=-1$ and $C=1$ are shown in blue and red, respectively, whereas the frozen remote bands are shown as black dashed curves. (c,d) Total orbital magnetization $M_{\rm orb}/N_{\rm cell}$ (c) and self-rotation contribution $m_{\rm SR}/N_{\rm cell}$ (d), in units of $\mu_B$, as functions of the number of included remote-band pairs $n_{\rm cut}$. The chemical potential is fixed inside the HF gap at $\mu=-10\,\mathrm{meV}$, and a $30\times30$ momentum mesh is used. Both quantities vanish at $n_{\rm cut}=0$, when only the active bands are retained, and become finite upon including remote bands before rapidly converging with increasing $n_{\rm cut}$.
			\label{fig:chiral}}
	\end{center}
	\vskip-0.5cm
\end{figure}

\begin{figure}[htb]
	\begin{center}
		\fig{6.8in}{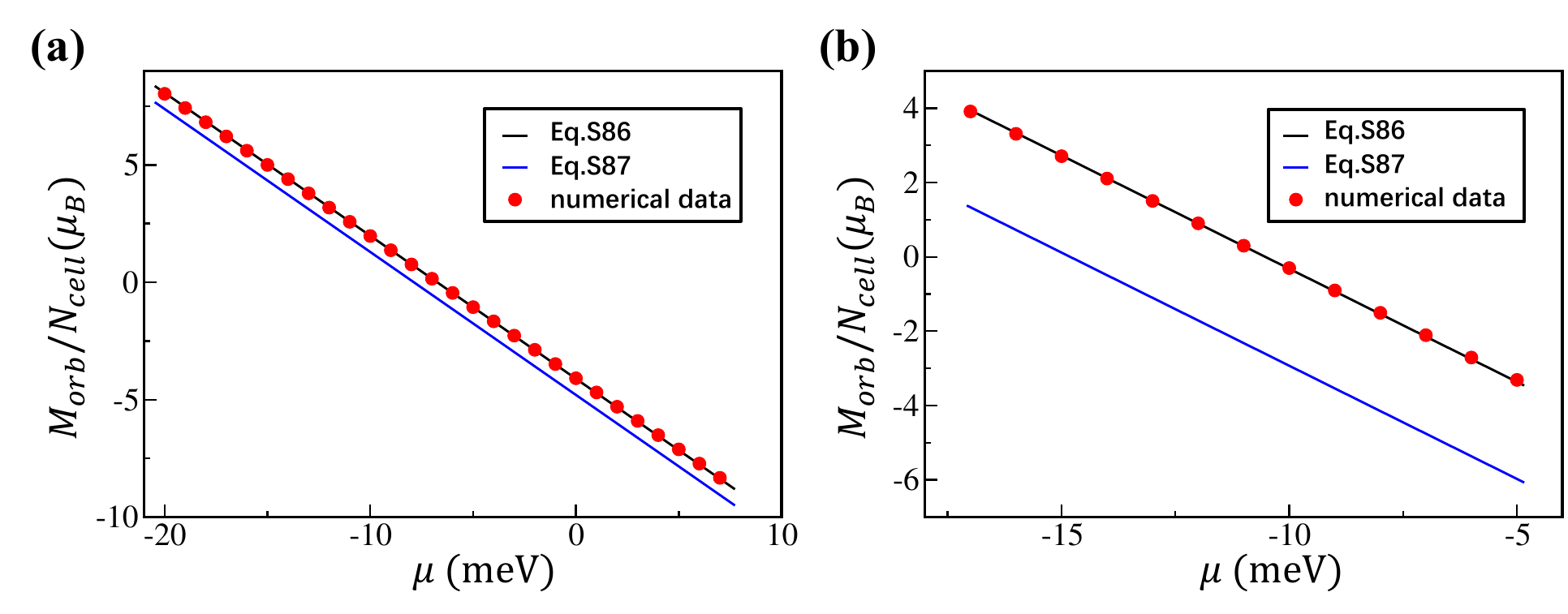}\caption{Comparison between the directly calculated orbital
		magnetization and the finite-chemical-potential expressions in Eqs.~\eqref{eq:PH11} and \eqref{eq:PH12} for the $n_{\rm R}^{\rm HF}=0$ Chern-insulating state at filling $\nu=-3$ with $C=-1$. Panels (a) and (b) show the chiral limit, $u_0=0$, and the realistic nonchiral case (the parameters used in the main text), respectively. The red circles denote the direct numerical results, the black curves represent the curvature-weighted expression in Eq.~\eqref{eq:PH11}, and the blue curves represent the factorized approximation in Eq.~\eqref{eq:PH12}. Each curve is restricted to chemical potentials inside the corresponding HF Chern gap. Both expressions reproduce the topological linear dependence on $\mu$ determined by $C=-1$. Equation~\eqref{eq:PH11} agrees closely with the numerical data in both cases, whereas Eq.~\eqref{eq:PH12} exhibits an approximately $\mu$-independent offset, particularly for the nonchiral parameters. The numerical calculations use a $30\times30$ momentum mesh and $n_{\rm cut}=30$.
			\label{fig:morb_ana}}
	\end{center}
	\vskip-0.5cm
\end{figure}

Figure~\ref{fig:morb_ana} provides a direct numerical test of Eqs.~\eqref{eq:PH11} and \eqref{eq:PH12}. In the chiral limit, the curvature-weighted expression in Eq.~\eqref{eq:PH11} agrees with the direct result to within $0.043\,\mu_B$ per moir\'e cell, consistent with the exact chiral-symmetry derivation up to finite-mesh effects. Remarkably, the same expression remains accurate to within $0.058\,\mu_B$ for the realistic nonchiral parameters, indicating that the corrections away from the chiral limit nearly cancel for this two-band HF state. By contrast, the factorized expression in Eq.~\eqref{eq:PH12} produces nearly $\mu$-independent offsets of approximately $0.70\,\mu_B$ and $2.62\,\mu_B$ in the chiral and nonchiral cases, respectively.

We emphasize that the gap between the active and remote bands in the chiral limit is approximately $100\mathrm{meV}$, nearly twice the value of $\sim50\mathrm{meV}$ obtained using the realistic parameters adopted in the main text and elsewhere in the Supplemental Material. A similar trend is observed for the interaction induced HF gap: it is approximately $30\,\mathrm{meV}$ in the chiral limit, compared with approximately $20\,\mathrm{meV}$ for the realistic parameters. Moreover, the purely active $(A-;A+)$ contribution vanishes  because of sublattice polarization in the chiral limit. Consequently, both $M_{\rm orb}$ and $m_{\rm SR}$ arise entirely from the mixed $(R-;A+)$ and $(A-;R+)$ sectors. Nevertheless, the magnitude of $M_{\rm orb}$ in the chiral limit remains approximately $70\%$ of its value for the realistic parameters, while $m_{\rm SR}$ remains nearly unchanged. These results further demonstrate that the orbital magnetization is governed primarily by the momentum space overlap between the active and remote bands, rather than by either the HF gap or the active-remote band gap alone.

\section{Extension of Hartree-Fock Calculation to Remote Bands}
\begin{figure}[H]
	\begin{center}
		\fig{6.2in}{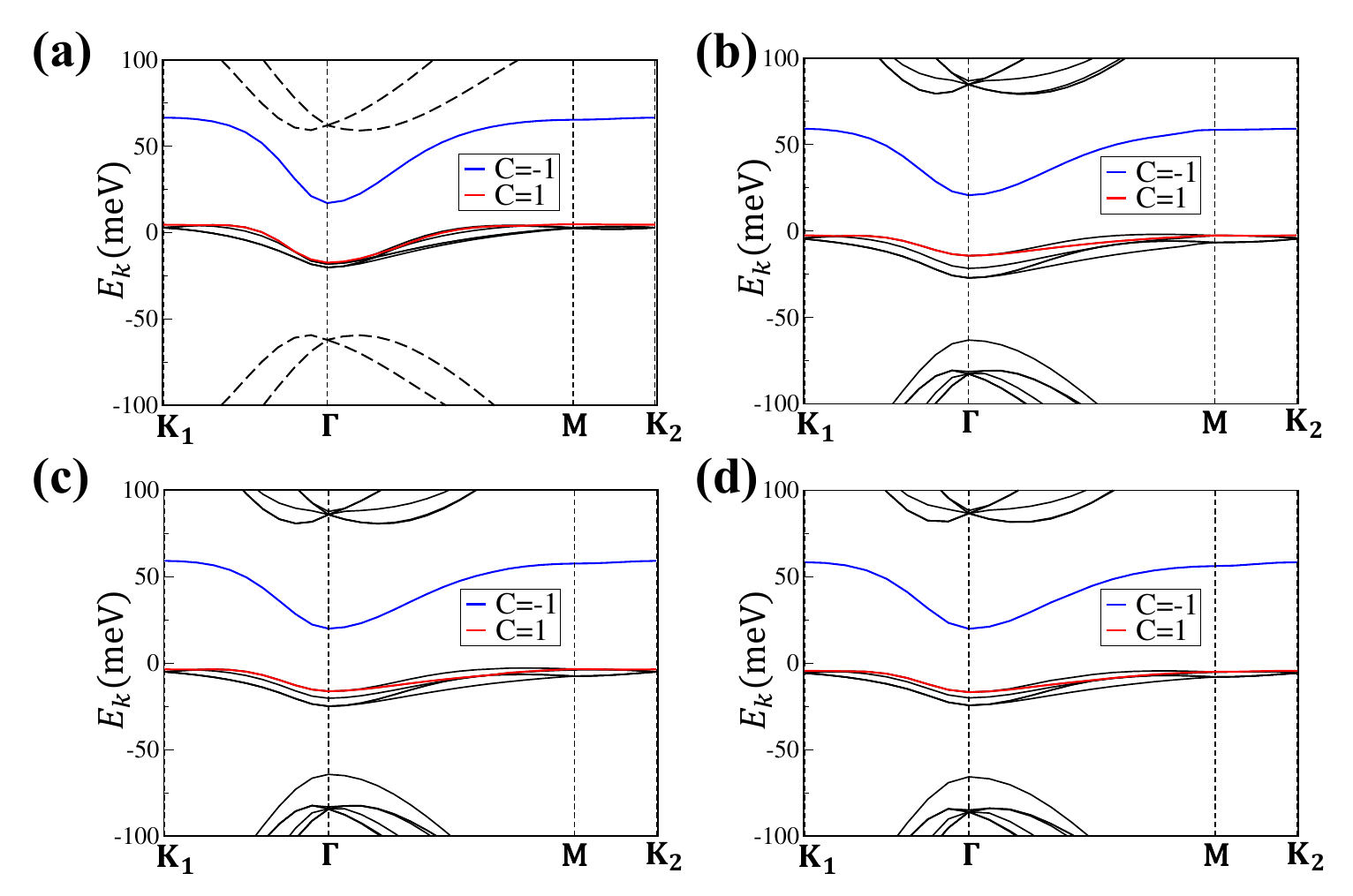}\caption{Evolution of the HF band structure of the Chern-insulating state at filling $\nu=3$ as remote bands are progressively incorporated into the self-consistent HF subspace. Panels (a)--(d) correspond to $n_{\rm R}^{\rm HF}=0$, $2$, $3$, and $4$, respectively, where $n_{\rm R}^{\rm HF}$ denotes the number of remote bands included on each side of charge neutrality for each spin and valley. Solid curves denote bands treated self-consistently. In panel (a), only the two active bands for each spin and valley are included in the HF calculation, while the black dashed curves show the two pairs of frozen remote bands closest to the active-band manifold. These remote bands are included self-consistently and therefore become solid curves in panels (b)--(d). The active Chern bands with $C=-1$ and $C=1$ are highlighted in blue and red, respectively, while the remaining bands are shown in black. 
			\label{fig:band_HF}}
	\end{center}
	\vskip-0.5cm
\end{figure}

To quantify the uncertainty introduced by the frozen-remote-band approximation, we extend the self-consistent HF calculation to include up to four remote bands both above and below the charge neutrality point for each spin and valley. We denote the number of remote bands included on each side by $n_{\rm R}^{\rm HF}$, with $n_{\rm R}^{\rm HF}=0$ corresponding to the frozen-remote-band approximation.  
The calculations for $n_{\rm R}^{\rm HF}=0$, $2$, and $3$ use a $30\times30$ momentum mesh, whereas that for $n_{\rm R}^{\rm HF}=4$ uses a $24\times24$ mesh because of its larger memory requirement. The resulting evolution of the HF band structure is shown in Fig.~\ref{fig:band_HF}.

\begin{figure}[H]
	\begin{center}
		\fig{6.2in}{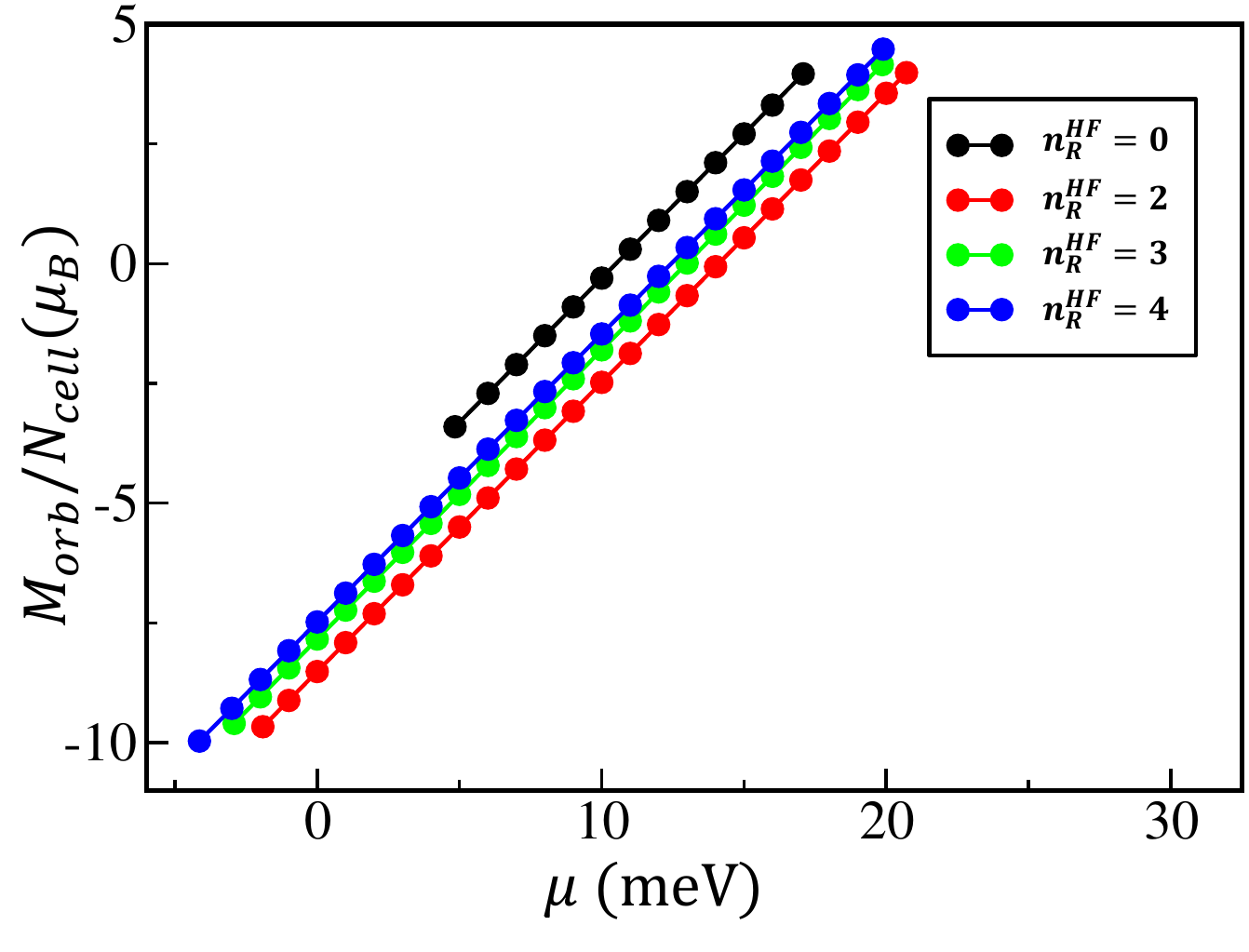}\caption{Total orbital magnetization $M_{\rm orb}/N_{\rm cell}$ as a function of the chemical potential $\mu$ inside the HF gap of the Chern-insulating state at filling $\nu=3$. The black, red, green and blue curves correspond to $n_{\rm R}^{\rm HF}=0$, $2$, $3$, and $4$, respectively. Each curve is plotted only between the valence-band maximum and the conduction-band minimum of its corresponding self-consistent HF spectrum. The linear dependence on $\mu$ and the nearly identical slopes for all four calculations are consistent with their common total Chern number $C=1$.
			\label{fig:Morb_nHF}}
	\end{center}
	\vskip-0.5cm
\end{figure}

As shown in Fig.~\ref{fig:Morb_nHF} and summarized in Table~\ref{tab:nhf_summary}, the orbital-magnetization quantities begin to approach convergence for $n_{\rm R}^{\rm HF}\gtrsim3$. Results for $n_{\rm R}^{\rm HF}=1$ are not shown because this truncation includes only one state from a pair of degenerate remote-band states at the $\Gamma$ point, thereby producing a substantial artificial breaking of $C_3$ symmetry. The magnitude of the self-rotation contribution, $|m_{\rm SR}|/N_{\rm cell}$, increases monotonically with $n_{\rm R}^{\rm HF}$, from $40.3\,\mu_B$ in the frozen-remote-band approximation to $49.6\,\mu_B$ at $n_{\rm R}^{\rm HF}=4$. This change corresponds to an increase of approximately $23\%$ relative to the frozen-band result.

A direct comparison of $M_{\rm orb}$ is more subtle because it depends linearly on the chemical potential within the insulating gap, while both the HF band structure and the gap boundaries vary with $n_{\rm R}^{\rm HF}$. At $\nu=3$, when the chemical potential is fixed at the bottom of the conduction band for each $n_{\rm R}^{\rm HF}$, $M_{\rm orb}/N_{\rm cell}$ takes the values $3.96$, $3.98$, $4.15$, and $4.48\,\mu_B$ for $n_{\rm R}^{\rm HF}=0$, $2$, $3$, and $4$, respectively. Thus, it changes by approximately $13\%$ between the frozen-remote-band result and $n_{\rm R}^{\rm HF}=4$. Alternatively, we fix $\mu=5.0\,\mathrm{meV}$, which lies inside the insulating gap for every value of $n_{\rm R}^{\rm HF}$ considered. Under this convention, $M_{\rm orb}/N_{\rm cell}$ is $-3.31$, $-5.50$, $-4.82$, and $-4.47\,\mu_B$ for $n_{\rm R}^{\rm HF}=0$, $2$, $3$, and $4$, respectively. Although the dependence is nonmonotonic, the closer agreement between the $n_{\rm R}^{\rm HF}=3$ and $4$ results indicates that $M_{\rm orb}$ is beginning to stabilize as additional remote bands are included in the HF calculation.

\begin{table}[H]
\caption{HF band edges and gaps, together with the converged orbital magnetization and self-rotation contribution, for the Chern-insulating state at filling $\nu=3$. Here, $E_{\rm VT}$ and $E_{\rm CB}$ denote the valence-band maximum and the bottom of the conduction band, respectively, and $\Delta_{\rm HF}=E_{\rm CB}-E_{\rm VT}$. The two $M_{\rm orb}$ columns give the values at the common chemical potential $\mu=5.0\,\mathrm{meV}$ and at the bottom of the conduction band, $\mu=E_{\rm CB}$, of each HF spectrum. The listed $M_{\rm orb}$ and $m_{\rm SR}$ values are converged with respect to the remote-band cutoff $n_{\rm cut}$ entering the orbital-magnetization calculation. Energies are in $\mathrm{meV}$, and magnetic moments are in $\mu_B$ per moir\'e unit cell.}
\label{tab:nhf_summary}
\begin{ruledtabular}
\begin{tabular}{ccccccc}
$n_{\rm R}^{\rm HF}$ & $E_{\rm VT}$ & $E_{\rm CB}$ & $\Delta_{\rm HF}$ & $M_{\rm orb}/N_{\rm cell}$ & $M_{\rm orb}/N_{\rm cell}$ & $m_{\rm SR}/N_{\rm cell}$ \\
 & $(\mathrm{meV})$ & $(\mathrm{meV})$ & $(\mathrm{meV})$ & $\mu=5.0\,\mathrm{meV}$ & $\mu=E_{\rm CB}$ & $(\mu_B)$ \\
 & & & & $(\mu_B)$ & $(\mu_B)$ & \\
\hline
$0$ & $4.834$ & $17.085$ & $12.251$ & $-3.31$ & $3.96$ & $-40.3$ \\
$2$ & $-1.918$ & $20.715$ & $22.633$ & $-5.50$ & $3.98$ & $-47.0$ \\
$3$ & $-2.929$ & $19.868$ & $22.796$ & $-4.82$ & $4.15$ & $-48.7$ \\
$4$ & $-4.144$ & $19.896$ & $24.040$ & $-4.47$ & $4.48$ & $-49.6$ \\
\end{tabular}
\end{ruledtabular}
\end{table}

\end{document}